\documentclass[twocolumn]{aastex631}

\newcommand{\dtemp}{d\textit{Temp}}

\usepackage{amsmath}

\newcommand\fakesection[1]{%
  \refstepcounter{section}%
  \addcontentsline{toc}{section}{\protect\numberline{\thesection}#1}%
  \sectionmark{#1}} 
  
\begin{document}

\defcitealias{Kostov_2019}{K19}
\defcitealias{Cloutier_2019}{C19}
\defcitealias{Demangeon_2021}{D21}

\title{Detailed Architecture of the L~98-59 System and Confirmation of a Fifth Planet in the Habitable Zone}

\correspondingauthor{Charles Cadieux}
\email{charles.cadieux.1@umontreal.ca}

\author[0000-0001-9291-5555]{Charles Cadieux}
\affiliation{Institut Trottier de recherche sur les exoplanètes (IREx), Université de Montréal, 1375 Ave Thérèse-Lavoie-Roux, Montréal, QC, H2V~0B3, Canada}

\author[0009-0005-6135-6769]{Alexandrine L'Heureux}
\affiliation{Institut Trottier de recherche sur les exoplanètes (IREx), Université de Montréal, 1375 Ave Thérèse-Lavoie-Roux, Montréal, QC, H2V~0B3, Canada}

\author[0000-0002-2875-917X]{Caroline Piaulet-Ghorayeb}
\altaffiliation{E. Margaret Burbridge Postdoctoral Fellow}
\affiliation{Department of Astronomy \& Astrophysics, University of Chicago, 5640 South Ellis Avenue, Chicago, IL 60637, USA}

\author[0000-0001-5485-4675]{Ren\'e Doyon}
\affiliation{Institut Trottier de recherche sur les exoplanètes (IREx), Université de Montréal, 1375 Ave Thérèse-Lavoie-Roux, Montréal, QC, H2V~0B3, Canada}
\affiliation{Observatoire du Mont-M\'egantic, Universit\'e de Montr\'eal, Montr\'eal H3C 3J7, Canada}

\author[0000-0003-3506-5667]{\'Etienne Artigau}
\affiliation{Institut Trottier de recherche sur les exoplanètes (IREx), Université de Montréal, 1375 Ave Thérèse-Lavoie-Roux, Montréal, QC, H2V~0B3, Canada}
\affiliation{Observatoire du Mont-M\'egantic, Universit\'e de Montr\'eal, Montr\'eal H3C 3J7, Canada}

\author[0000-0003-4166-4121]{Neil J. Cook}
\affiliation{Institut Trottier de recherche sur les exoplanètes (IREx), Université de Montréal, 1375 Ave Thérèse-Lavoie-Roux, Montréal, QC, H2V~0B3, Canada}

\author[0000-0002-2195-735X]{Louis-Philippe Coulombe}
\affiliation{Institut Trottier de recherche sur les exoplanètes (IREx), Université de Montréal, 1375 Ave Thérèse-Lavoie-Roux, Montréal, QC, H2V~0B3, Canada}

\author[0000-0001-6809-3520]{Pierre-Alexis Roy}
\affiliation{Institut Trottier de recherche sur les exoplanètes (IREx), Université de Montréal, 1375 Ave Thérèse-Lavoie-Roux, Montréal, QC, H2V~0B3, Canada}

\author[0000-0002-6780-4252]{David Lafreni\`ere}
\affiliation{Institut Trottier de recherche sur les exoplanètes (IREx), Université de Montréal, 1375 Ave Thérèse-Lavoie-Roux, Montréal, QC, H2V~0B3, Canada}

\author[0009-0003-2535-5641]{Pierrot Lamontagne}
\affiliation{Institut Trottier de recherche sur les exoplanètes (IREx), Université de Montréal, 1375 Ave Thérèse-Lavoie-Roux, Montréal, QC, H2V~0B3, Canada}

\author[0000-0002-3328-1203]{Michael Radica}
\affiliation{Department of Astronomy \& Astrophysics, University of Chicago, 5640 South Ellis Avenue, Chicago, IL 60637, USA}

\author[0000-0001-5578-1498]{Bj\"orn Benneke}
\affiliation{Department of Earth, Planetary, and Space Sciences, University of California, Los Angeles, CA USA}
\affiliation{Institut Trottier de recherche sur les exoplanètes (IREx), Université de Montréal, 1375 Ave Thérèse-Lavoie-Roux, Montréal, QC, H2V~0B3, Canada}

\author[0000-0003-0973-8426]{Eva-Maria Ahrer}
\affiliation{Max-Planck-Institut f\"{u}r Astronomie, K\"{o}nigstuhl 17, 69117 Heidelberg, Germany}

\author[0000-0002-7992-469X]{Drew Weisserman}
\affiliation{Department of Physics \& Astronomy, McMaster University, 1280 Main St W, Hamilton, ON, L8S 4L8, Canada}

\author[0000-0001-5383-9393]{Ryan Cloutier}
\affiliation{Department of Physics \& Astronomy, McMaster University, 1280 Main St W, Hamilton, ON, L8S 4L8, Canada}

\begin{abstract}

The L~98-59 system, identified by TESS in 2019, features three transiting exoplanets in compact orbits of 2.253, 3.691, and 7.451 days around an M3V star, with an outer 12.83-day non-transiting planet confirmed in 2021 using ESPRESSO. The planets exhibit a diverse range of sizes (0.8--1.6\,R$_{\oplus}$), masses (0.5--3\,M$_{\oplus}$), and likely compositions (Earth-like to possibly water-rich), prompting atmospheric characterization studies with HST and JWST. Here, we analyze 16 new TESS sectors and improve radial velocity (RV) precision of archival ESPRESSO and HARPS data using a line-by-line framework, enabling stellar activity detrending via a novel differential temperature indicator. We refine the radii of L~98-59\,b, c, and d to 0.837\,$\pm$\,0.019\,R$_{\oplus}$, 1.329\,$\pm$\,0.029\,R$_{\oplus}$, 1.627\,$\pm$\,0.041\,R$_{\oplus}$, respectively. Combining RVs with transit timing variations (TTV) of L~98-59\,c and d from TESS and JWST provides unprecedented constraints on the masses and eccentricities of the planets. We report updated masses of 0.46\,$\pm$\,0.11\,M$_{\oplus}$ for b, 2.00\,$\pm$\,0.13\,M$_{\oplus}$ for c, and 1.64\,$\pm$\,0.07\,M$_{\oplus}$ for d, and a minimum mass of 2.82\,$\pm$\,0.19\,M$_{\oplus}$ for e. We additionally confirm L~98-59\,f, a non-transiting super-Earth with a minimal mass of 2.80\,$\pm$\,0.30\,M$_{\oplus}$ on a 23.06-day orbit inside the Habitable Zone. The TTVs of L~98-59\,c and d ($<$\,3\,min, $P_{\rm TTV} = 396$\,days) constrain the eccentricities of all planets to near-circular orbits ($e \lesssim 0.04$). An internal structure analysis of the transiting planets reveals increasing water-mass fractions ($f_{\rm H_{2}O}$) with orbital distance, reaching $f_{\rm H_{2}O} \approx 0.16$ for L~98-59\,d. We predict eccentricity-induced tidal heating in L~98-59\,b with heat fluxes comparable to those of Io, potentially driving volcanic activity.
\end{abstract}

\section{Introduction} \label{sec:intro}

As we enter the eighth year of operation of TESS \citep{Ricker_2015} and the fourth year of JWST science \citep{Gardner_2023}, exoplanet research has slowly transitioned from hunting down a growing number of exoplanets to an era focused on the detailed characterization of a few keystone systems. Low-mass stellar hosts are of particular interest due to their intrinsic properties --- they are smaller, less massive, and cooler than our Sun --- offering optimal conditions for transit, secondary eclipse and radial velocity studies. Constraining the internal structure of terrestrial planets and super-Earths (1--10\,M$_{\oplus}$) requires precise mass measurements ($\lesssim$10\% error, \citealt{Plotnykov_2024}), a task largely feasible only for M-type stars with current instruments. The Habitable Zone around an M dwarf corresponds to orbital periods in the 10--30-day range, compared to 0.5--3\,years for Sun-like stars \citep{Kopparapu_2013}, which means most well-characterized temperate exoplanets orbit such a star. However, M dwarfs emit more extreme ultraviolet radiation (XUV; 10--130\,nm) due to their stronger magnetic activity, especially at a young age ($<$100\,Myr), and it remains unclear whether the atmospheres of small rocky planets can survive this environment and maintain habitable surface conditions (e.g., \citealt{Segura_2010, Luger-Barnes_2015, Louca_2023, Ridgway_2023}).

Multiplanetary systems offer a unique opportunity to study the outcomes of planetary formation and evolution within the same stellar environment. Planets around the same star spanning the radius valley \citep{Fulton_2017, Cloutier-Menou_2020} provide a way to test thermally-driven atmospheric mass loss mechanisms \citep{Owen_2017, Ginzburg_2018} over a shared stellar XUV flux history \citep{Owen_2020, VanWyngarden_2024}. Similarly, they allow for a comparison between stellar and planetary abundance ratios of refractory elements (e.g., Fe/Mg, Mg/Si) for planets formed with the same initial primordial materials \citep{Plotnykov_2020, Brinkman_2024}. Rocky planets in `multi' M-dwarf systems generally show lower core-mass fractions, with planet multiplicity increasing as the host star metallicity decreases \citep{Rodriguez_2023}. One hypothesis is that planet formation around metal-rich M dwarfs may favor giant planets in ‘single’ configurations, while lower metallicity (and less massive disks) could lead to multiple rocky planets in stable, compact, and coplanar arrangements \citep{Rodriguez_2023}.

\begin{figure}[ht!]
\centering
\hspace{-0.29cm}\includegraphics[width=0.99\linewidth]{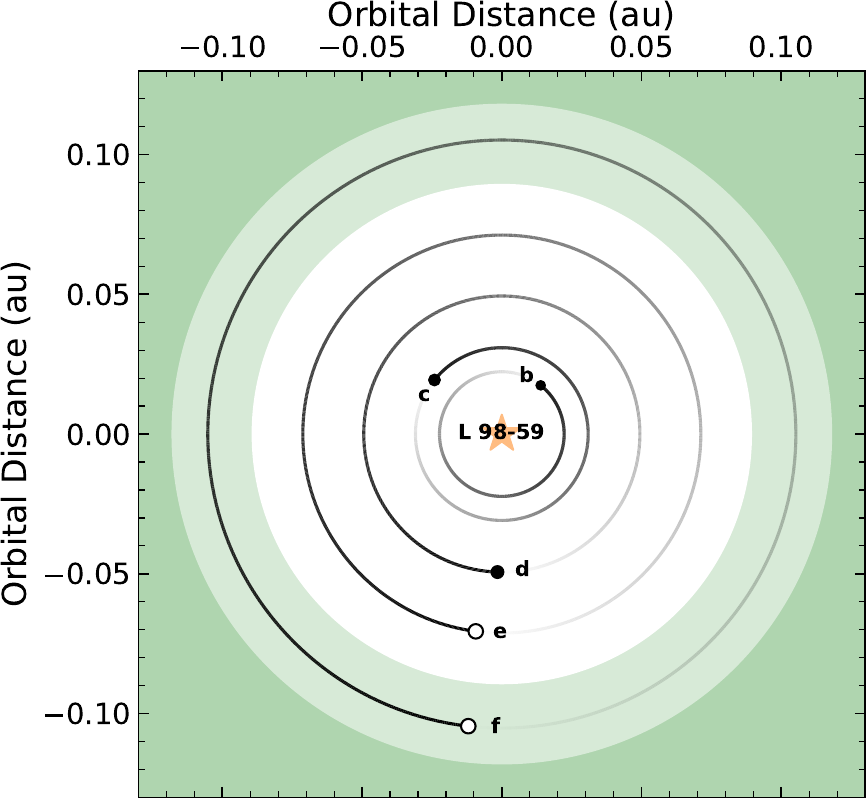}\\[0.25cm]
\includegraphics[width=1\linewidth]{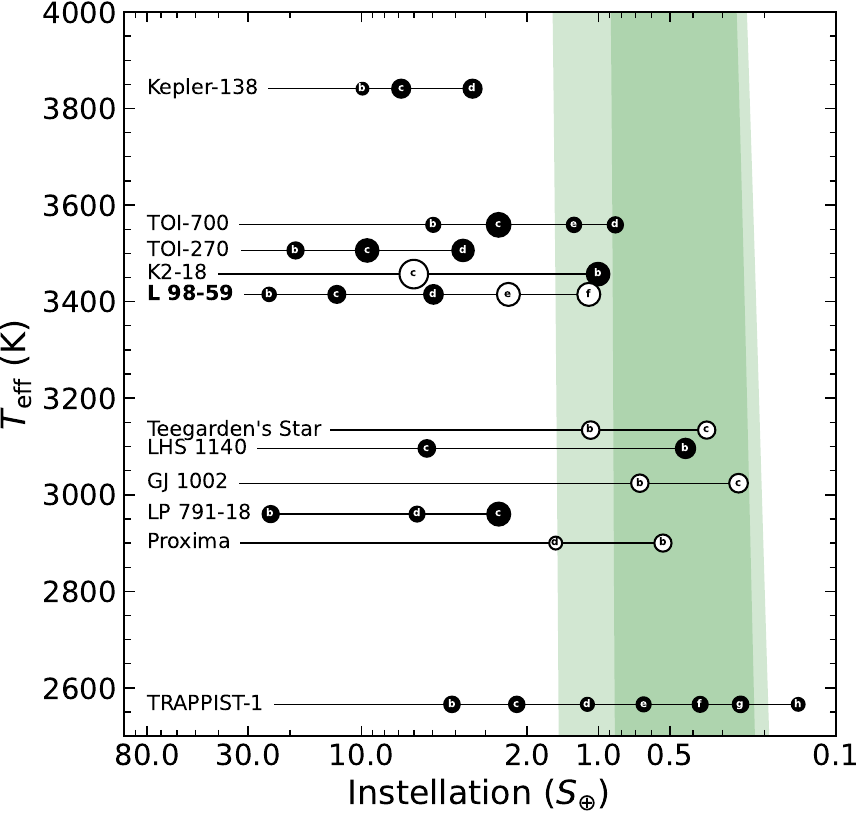}
  \caption{\textit{Top}: Above view of the L~98-59 planetary system on BJD = 2\,460\,000 assuming circular orbits. The line of sight toward Earth is to the right. The Habitable Zone defined in \citealt{Kopparapu_2013} is shown in green for runaway/maximum greenhouse (conservative) and pale green for early recent Venus/early Mars (optimistic). \textit{Bottom}: Comparative exoplanetology of M-dwarf multiplanetary systems ordered by stellar effective temperature and planetary instellation. An arbitrary shift of $+$100\,K in $T_{\rm eff}$ has been applied to \hbox{TOI-700} and Teegarden's star for clarity. In both panels, transiting (non-transiting) exoplanets are shown as filled (open) circles, with circle size proportional to planetary radius. For planets with only a minimum mass constraint, the radius is set to the 95$^{\rm th}$ percentile of the prediction from \texttt{spright} \citep{Parviainen_2024}.}
  \label{fig:orbit}
\end{figure}

Several well-studied multiplanetary systems exemplify the diversity of orbital architectures and atmospheric properties among planets orbiting the same star. One famous example is TRAPPIST-1 with seven transiting Earth-sized exoplanets around an ultra-cool M8 dwarf \citep{Gillon_2017}. The masses of the TRAPPIST-1 planets were constrained to an exquisite precision of 3--5\% from photo-dynamical modeling of transit timing variations (TTV, \citealt{Agol_2021}). Investigations of TRAPPIST-1 with JWST \citep{Gardner_2023} show strong evidence of stellar contamination in transmission data for TRAPPIST-1\,b~and~c \citep{Lim_2023, Radica_2025}, with eclipse photometry data showing a hot day-side for both planets consistent with no thick atmospheres \citep{Greene_2023, Ducrot_2024, Zieba_2023}. Another intriguing system is LHS~1140 \citep{Dittmann_2017, Ment_2019, Lillo-Box2020, Cadieux_2024a} with two super-Earths, including a water world candidate in the Habitable Zone (LHS~1140\,b) of this old ($>$5\,Gyr) and quiet M4V dwarf. Recent JWST transit observations of LHS~1140\,b have ruled out a low mean molecular weight around this planet, with a possible detection of an N$_2$-dominated atmosphere \citep{Damiano_2024, Cadieux_2024b}. A comparison of the orbital architecture of many known M-dwarf multiplanetary systems can be seen in the bottom panel of Figure~\ref{fig:orbit}.

L~98-59 is another keystone multiplanet system with ongoing transmission spectroscopy (GTO1201, GTO1224, GO2512, GO3942, GO4098) and eclipse photometry (GO3730) programs with JWST. In this paper, we analyze 25 sectors of TESS data along with archival ESPRESSO \citep{Pepe_2021} and HARPS \citep{Mayor_2003} radial velocities to refine the masses, radii, and orbital eccentricities of the L~98-59 planets, which are key parameters for accurate interpretation of JWST data. Our reanalysis of ESPRESSO and HARPS data with a line-by-line framework has improved the RV precision and provide strong evidence for a fifth outer non-transiting planet in the \hbox{L~98-59} system with a period of 23.06\,days and a minimum mass of $2.80 \pm 0.30$\,M$_{\oplus}$.

The previous findings on the L~98-59 planetary system are reviewed in Section~\ref{sec:L98-59}. The observations are described in Section~\ref{sec:observations} and the data analysis and results are presented in Section~\ref{sec:data}. In light of the updated masses, radii, and orbital eccentricities, the possible internal structures of the planets, potential for tidal heating, and timing of future secondary eclipses are discussed in Section~\ref{sec:discussion}. We end with a summary and concluding remarks in Section~\ref{sec:conclusions}.

\section{The L~98-59 Planetary System} \label{sec:L98-59}

L~98-59 is a relatively bright M3V thin disk star at 10.61\,pc \citep{Gaia_2023} with an estimated age of $\sim$5\,Gyr \citep{Engle_2023}. An overview of the stellar parameters is given in Table~\ref{table:stellarparams}.

The L~98-59 multiplanetary system is a key target in exoplanet research, with three transiting exoplanets with sizes ranging from 0.8 to 1.6\,R$_{\oplus}$, all suitable for atmospheric characterization with HST and JWST \citep{Pidhorodetska_2021}. The first three planets L~98-59\,b, c, and d with orbital periods of 2.253, 3.691, and 7.451\,days were revealed by TESS in a validation study by \citealt{Kostov_2019} (hereafter \citetalias{Kostov_2019}) using early primary mission data (sectors 2, 5, and 8) and a series of ground-based transit observations. The transit impact parameter of the three planets suggests that their orbits are nearly co-planar, with mutual inclination $\Delta i < 1^{\circ}$.

The trio of exoplanets was confirmed in a following study through mass measurement from a data set of 161 HARPS radial velocities (\citealt{Cloutier_2019}, hereafter \citetalias{Cloutier_2019}). Only an upper limit of 0.98\,M$_{\oplus}$ (95\%) for the mass of the inner planet b could be inferred from HARPS. The dynamical simulations presented by \citetalias{Cloutier_2019} show that the orbits must be nearly circular ($e < 0.1$) to maintain long-term stability. Based on the spectroscopic log\,$R^{\prime}_{HK}$ index and H$\alpha$ time series, \citetalias{Cloutier_2019} identified a periodicity of approximately 80 days attributed to the stellar rotation. The HARPS RVs were detrended from stellar activity using a Gaussian Process (GP) that modeled quasi-periodic signals at both the rotation period ($P_{\rm rot}$) and its first harmonic ($P_{\rm rot} / 2$).

\begin{deluxetable}{lcc} 
\setlength{\tabcolsep}{11pt}  
\tablewidth{\columnwidth} 
\tablecaption{L~98-59 stellar parameters}
\tablehead{
\colhead{Parameter} & \colhead{Value} & \colhead{Ref.}
}
\startdata
RA (J2000) & 08:18:07.621 & 1\\
DEC (J2000) & $-$68:18:46.805 & 1\\
$\mu_{\alpha} \cos \delta$ (mas\,yr$^{-1}$) & 94.794 $\pm$ 0.018 & 1\\
$\mu_{\delta}$ (mas\,yr$^{-1}$) & $-$340.084 $\pm$ 0.017 & 1\\
$\pi$ (mas) & 94.2664 $\pm$ 0.0155 & 1\\
Distance (pc) & 10.6082 $\pm$ 0.0017 & 1\\
$V$ (mag) & 11.685 $\pm$ 0.017 & 2\\
$K_{\rm s}$ (mag) & 7.101 $\pm$ 0.018 & 3\\
SpT & M3V & 4\\
$T_{\rm eff}$ (K) & 3415 $\pm$ 60 & 5\\
$\left[ {\rm Fe/H} \right]$ & $-$0.46 $\pm$ 0.26 & 5\\
Age (Gyr) &  4.94 $\pm$ 0.28 & 6\\
$M_{\star}$ (M$_{\odot}$) & 0.2923 $\pm$ 0.0067 & 7\\
$R_{\star}$ (R$_{\odot}$) & 0.3155 $\pm$ 0.0062 & 7\\
$\rho_{\star}$ (g\,$\cdot$\,cm$^{-3}$) & 13.0$^{+1.1}_{-0.9}$ & 7\\
log $g$ (dex) & 4.91 $\pm$ 0.02 & 8\\
$L_{\star}$ (L$_{\odot}$) & 0.0122 $\pm$ 0.0010 & 8\\
$P_{\rm rot}$ (days) & 77.5 $\pm$ 1.6 & 9\\
Optimistic HZ (au) & 0.090--0.237 & 10\\
Conservative HZ (au) & 0.118--0.224 & 10
\enddata
\tablecomments{(1) Gaia DR3 \citep{Gaia_2023}. (2) UCAC4 \citep{Zacharias_2013}. (3) 2MASS \citep{Skrutskie_2006}. (4) \cite{Kostov_2019}. (5) \cite{Demangeon_2021} from the spectral synthesis analysis.  (6) \cite{Engle_2023}. (7) This work using \cite{Mann_2015, Mann_2019} as prior and constrained by the transits. (8) This work derived from $M_{\star}$, $R_{\star}$, and $T_{\rm eff}$. (9) This work from \dtemp\ measurements. (10) \cite{Kopparapu_2013}.}
\label{table:stellarparams}
\end{deluxetable}

\vspace{-0.4cm}
The system was revisited in 2021 with 6 new sectors from TESS and 66 extreme precision RVs from ESPRESSO in addition to the existing HARPS data (\citealt{Demangeon_2021}, hereafter \citetalias{Demangeon_2021}). The revised radii and masses revealed that planet b ($R_{\rm p} = 0.85 \pm 0.05$\,R$_{\oplus}$, $M_{\rm p} = 0.40 \pm 0.16$\,M$_{\oplus}$) and planet c ($R_{\rm p} = 1.39 \pm 0.08$\,R$_{\oplus}$, $M_{\rm p} = 2.22 \pm 0.26$\,M$_{\oplus}$) have bulk densities suggestive of a rocky composition, but planet d ($R_{\rm p} = 1.52 \pm 0.11$\,R$_{\oplus}$, $M_{\rm p} = 1.94 \pm 0.28$\,M$_{\oplus}$) could be volatile-rich with water potentially totaling $\sim$30\% of the mass. The analysis of the RVs also confirmed L~98-59\,e, a non-transiting exoplanet with a period of 12.80\,days and an $M_{\rm p} \sin i = 3.06 \pm 0.34$. A tentative signal at 23.15\,days was also identified by \citetalias{Demangeon_2021}, but the four-planet model was favored by the data according to Bayesian model comparison. Both periodicities (12.80 and 23.15\,days) were undetected in the TESS light curve. With the discovery of at least one additional planet around L~98-59, \citetalias{Demangeon_2021} performed a new set of N body simulations to assess the stability of the system. The main conclusion of low eccentricities ($e < 0.1$) from \citetalias{Cloutier_2019} was confirmed. These simulations also showed that planet c and d, with period ratio of 2.019 close to 2:1, are not exactly in mean-motion resonance. An independent analysis of the ESPRESSO RVs by \cite{Rajpaul_2024} yielded consistent results with those of \citetalias{Demangeon_2021} with the exception that a more sophisticated stellar activity model using multi-dimensional GP was needed to robustly detect the small Keplerian signal of L~98-59\,b (0.56 $\pm$ 0.16\,m\,s$^{-1}$).

The semi-amplitudes ($K$) and the planet-to-star radius ratios ($R_{\rm p}/R_{\star}$) of the L~98-59 planets from previous studies are compared in Table~\ref{table:comparison}. While the sizes of the inner transiting planets have remained relatively consistent across studies, the changes in $K$ (and thus mass) primarily reflect improvements in RV data and stellar activity modeling. Figure~\ref{fig:orbit} illustrates the orbital architecture of the L~98-59 system and compares it to other multiplanetary systems around M dwarfs using data from the NASA Exoplanet Archive \citep{Akeson_2013}.

A first atmospheric reconnaissance of L~98-59\,b, c, and d in transmission was completed with the Wide Field Camera 3 (WFC3, 1.1--1.7\,$\mu$m) on the Hubble Space Telescope (HST) in program 15856 (PI: T. Barclay). The transmission spectrum of the sub-Earth L~98-59\,b from a set of five transits is mostly flat and consistent with no atmosphere or an atmosphere with high-altitude clouds or hazes (\citealt{Damiano_2022}, see also \citealt{Zhou_2022}). Similarly, a single transit with HST/WFC3 on L~98-59\,c (\citealt{Barclay_2023}; see also \citealt{Zhou_2023}) and on L~98-59\,d \citep{Zhou_2023} was insufficient to robustly constrain the presence of an atmosphere. Overall, HST observations have ruled out cloud-free hydrogen-rich atmospheres on planets b, c, or d, favoring either the absence of atmospheres or heavier atmospheres (e.g., H$_2$O- or CO$_2$-rich), with the latter scenario supported by atmospheric escape simulations \citep{Fromont_2024}. With all three planets located in the Venus-zone \citep{Ostberg_2023} and likely experiencing a runaway greenhouse state, \cite{Fromont_2024} show that for any initial water content, their atmosphere could accumulate significant quantities of oxygen through H$_2$O photolysis.

\begin{deluxetable}{lccc}
\setlength{\tabcolsep}{7.5pt}  
\tablecaption{Radial velocity semi-amplitudes and planet-to-star radius ratios for the L~98-59 planets from different studies}
\tablehead{\colhead{Parameter} & \colhead{\citetalias{Kostov_2019} and \citetalias{Cloutier_2019}} & \colhead{\citetalias{Demangeon_2021}} & \colhead{This work}}
\startdata \noalign{\vskip 3pt}
\multicolumn{4}{c}{\textit{L~98-59\,b} ($P = 2.2531$\,days)}\\[0.1cm]
$K$ (m\,s$^{-1}$) & $<1.03$ & 0.46$^{+0.20}_{-0.17}$ & 0.51$\pm$0.12\\
$R_{\rm p}/R_{\star}$ (\%) & 2.34$\pm$0.09 & 2.51$\pm$0.07 & 2.43$\pm$0.03 \\[0.1cm]
\hline \noalign{\vskip 3pt}
\multicolumn{4}{c}{\textit{L~98-59\,c} ($P = 3.6907$\,days)}\\[0.1cm]
$K$ (m\,s$^{-1}$) & 2.21$\pm$0.28 & 2.19$^{+0.17}_{-0.20}$ & 1.88$\pm$0.12\\
$R_{\rm p}/R_{\star}$ (\%) & 3.96$\pm$0.10  & 4.09$\pm$0.06 & 3.86$\pm$0.04\\[0.1cm]
\hline \noalign{\vskip 3pt}
\multicolumn{4}{c}{\textit{L~98-59\,d} ($P = 7.4507$\,days)}\\[0.1cm]
$K$ (m\,s$^{-1}$) & 1.61$\pm$0.36 & 1.50$^{+0.22}_{-0.19}$ & 1.22$\pm$0.05\\
$R_{\rm p}/R_{\star}$ (\%) & 4.62$\pm$0.33 & 4.48$\pm$0.10 & 4.72$\pm$0.08\\[0.1cm]
\hline \noalign{\vskip 3pt}
\multicolumn{4}{c}{\textit{L~98-59\,e} ($P = 12.828$\,days)}\\[0.1cm]
$K$ (m\,s$^{-1}$) & -- & 2.01$^{+0.16}_{-0.20}$ & 1.75$\pm$0.12\\[0.1cm]
\hline \noalign{\vskip 3pt}
\multicolumn{4}{c}{\textit{L~98-59\,f} ($P = 23.06$\,days)}\\[0.1cm]
$K$ (m\,s$^{-1}$) & -- & 1.37$^{+0.33}_{-0.44}$ & 1.43$\pm$0.15\\[0.1cm]
\enddata
\tablecomments{\citetalias{Kostov_2019} with TESS (3 sectors). \citetalias{Cloutier_2019} with 161 HARPS RVs. \citetalias{Demangeon_2021} with TESS (9 sectors), 161 HARPS RVs, and 66 ESPRESSO RVs. This work with TESS (21 sectors), 161 HARPS RVs, 66 ESPRESSO RVs, both improved with a line-by-line extraction, and transit timing from TESS and JWST.}
\label{table:comparison}
\end{deluxetable}

\vspace{-0.4cm}
Several ongoing (and future) JWST programs, with NIRSpec and NIRISS in transit spectroscopy and with MIRI in eclipse photometry, are expected to provide definitive answers about the presence and composition of atmospheres on the L~98-59 planets. Recent observations on L~98-59\,b with JWST NIRSpec/G395H (3--5\,$\mu$m) from a combination of four transits favor a SO$_2$-dominated atmosphere at 3.6$\sigma$ over no atmosphere \citep{Bello-Arufe_2025}. Two transits of the super-Earth L~98-59\,c with NIRSpec/G395H were analyzed in \cite{Scarsdale_2024} showing no clear signatures of molecular absorption. Given the sensitivity of this data set, L~98-59\,c either has no atmosphere, a muted transmission spectrum from opaque clouds/hazes or a high mean molecular weight atmosphere ($\gtrsim$10\,g\,mol$^{-1}$). For L~98-59\,d, \cite{Gressier_2024} revealed hints of SO$_2$ in its transmission spectrum derived from a single transit with NIRSpec/G395H (see also retrieval analysis by \citealt{Banerjee_2024}).

If confirmed, the detection of SO$_2$ on L~98-59\,b and d would indicate disequilibrium processes in their atmospheres, as SO$_2$ typically forms through photochemistry or volcanic activity. The potential water-rich nature of L~98-59\,d could facilitate production of SO$_2$ in the upper atmosphere through photochemical reactions \citep{Yang_2024}. On planet b, a sulfur-rich atmosphere could instead suggest active volcanism driven by tidal interactions between the planets due to nonzero eccentricity, akin to Jupiter’s moon Io. According to \cite{Seligman_2024}, the L~98-59 system is the most likely to sustain tidally-induced volcanism, estimating that 5--10 transits with JWST are required to formally detect volcanic activity tracers (e.g., H$_2$S and SO$_2$) in the atmospheres of L~98-59\,b, c, and d. While the eccentricities of the planets remain weakly constrained\,---\,0.103$^{+0.117}_{-0.045}$ for b, 0.103$^{+0.045}_{-0.058}$ for c, and 0.074$^{+0.057}_{-0.046}$ for d (\citetalias{Demangeon_2021})\,---\,these values suggest that eccentricity-driven tidal heating is plausible.

Highly irradiated rocky planets are vulnerable to photoevaporation, making atmospheric retention unlikely, as described by the ‘cosmic shoreline’ concept \citep{Zahnle_2017}. However, continuous volcanic outgassing from tidal dissipation in compact systems could maintain secondary atmospheres on such planets. For L~98-59\,b, an internal heating per unit mass of at least eight times that of Io is required to retain an atmosphere \citep{Bello-Arufe_2025}. Recent hints of atmospheres on the L~98-59 planets from JWST make it one of the best system to search for secondary atmospheres on small rocky worlds around M dwarfs.

\section{Observations} \label{sec:observations}

\subsection{TESS Photometry} \label{sec:tess}

L~98-59 (TOI-175, $T = 9.412$) is a near Southern Continuous Viewing Zone star that was observed at a 2-minute cadence with TESS in 25 sectors\footnote{Sectors 2, 5, 8, 9, 10, 11, 12, 28, 29, 32, 35, 36, 37, 38, 39, 61, 62, 63, 64, 65, 69, 87, 88, 89, and 90} between August 2018 and April 2025. We collected the \texttt{PDCSAP} light curves (\citealt{Smith_2012}; \citealt{Stumpe_2012}, \citeyear{Stumpe_2014}) of L~98-59 produced by the TESS Science Processing Operations Center (SPOC, \citealt{Jenkins_2016}) at NASA Ames Research Center\footnote{Publicly available on the Mikulski Archive for Space Telescopes: \href{https://archive.stsci.edu/tess/}{\texttt{archive.stsci.edu/tess/}}}. 

The \texttt{PDCSAP} light curves have been corrected for both instrumental systematics and for flux dilution from neighboring stars within a few TESS pixels (21$\arcsec$ per pixel). Two of these nearby stars have been identified as eclipsing binaries (EB) in \citetalias{Kostov_2019}: TIC 307210817 (80$\arcsec$ NW, $\Delta T = 4.184$) with a period of 10.43\,days and TIC 307210845 (54$\arcsec$ SSE, $\Delta T = 6.630$) with a period of 1.0492\,days. Folding the light curves of all the sectors to the periods of the EBs, we do not detect any contamination from the 10.43-day EB in the \texttt{PDCSAP} data, but we clearly detect shallow eclipses of approximately 200\,ppm from TIC 307210817 every 1.0492\,days. The eclipse depths also vary from sector to sector depending on the position of TIC 307210817 relative to the default \texttt{PDCSAP} aperture mask. The EB eclipses with a 1.0492-day periodicity frequently coincide with the transits of L~98-59\,b, c, and d with respective transit depths of 600, 1600, and 2000\,ppm (\citetalias{Demangeon_2021}). To avoid any bias in transit depth measurements, we tested two approaches: (1) masking all photometric data during the apparent 1.6-hour eclipse and (2) modeling the eclipse as if it was a transit (see Sect.~\ref{sec:transit_fit}), scaling its depth for each sector, and correcting the affected photometric points by removing the shallow eclipse. For both approaches, the radius of the planets was unchanged so we decided to work with the EB-cleaned (method 2) light curve for the rest of the analysis.

We applied the following corrections to the out-of-transit light curve of each sector: (1) a smoothed light curve was generated from a rolling mean with a window size of 11 data points, (2) outlying fluxes that deviated by more than 3.5 times the median absolute deviation (MAD) from the smoothed curve were discarded, (3) stellar flares were identified and removed when the smoothed curve exceeded 5 MAD from the median flux, (4) the cleaned light curve was normalized by the median flux, and (5) a Gaussian Process (GP) regression was applied to the normalized per-sector light curve following a procedure similar to that outlined by \citetalias{Kostov_2019} and \citetalias{Demangeon_2021}, which we briefly summarize here.

We used a simple harmonic oscillator (SHO) kernel in \texttt{celerite2} (\citealt{celerite1_2017}; \citeyear{celerite2_2018}) to model residual instrumental or stellar activity correlated noise with a GP. As in \citetalias{Kostov_2019} and \citetalias{Demangeon_2021}, the SHO term is critically damped with a quality factor $Q$ fixed to $\sqrt{0.5}$ (no periodicity). We allow for a baseline flux $f_0$ and fit for an amplitude $\sigma_{\rm phot}$ and a coherence timescale $\tau$ of the variations. We also include an extra white noise term $\sigma_{\rm jitter}$ added in quadrature to the diagonal of the covariance matrix to account for uncorrelated excess jitter. The parameter optimization was done using a Bayesian framework and the MCMC sampling package \texttt{emcee} \citep{Foreman-Mackey_2013}. Wide log-uniform priors were used for all parameters, respectively $\mathcal{LU} \left(0.5, 1.5\right)$ for $f_0$, $\mathcal{LU} \left(1, 10^6\right)$\,ppm for $\sigma_{\rm phot}$, $\mathcal{LU} \left(0.2, 20\right)$\,days for $\tau$, and $\mathcal{LU} \left(1, 10^6\right)$\,ppm for $\sigma_{\rm jitter}$, with 100 walkers, 3000 steps, and a burn-in of 500 steps. 

The resulting GP models for all sectors are shown in Figure~\ref{fig:tess_gp}. The light curves were finally corrected by dividing the fluxes, both in- and out-of-transit, by the mean prediction from the corresponding GP model. Only the corrected in-transit photometric points are later used in the analysis. 
This work adds 16 new TESS sectors compared to the analysis presented in \citetalias{Demangeon_2021} and 22 from \citetalias{Kostov_2019}.

\subsection{ESPRESSO Radial Velocity}

We obtained publicly accessible ESPRESSO observations of L~98-59 from the ESO archive \citep{Delmotte_2006}. This data set, described in detail in \citetalias{Demangeon_2021}, contains 66 spectra taken between November 2018 and March 2020 as part of the ESPRESSO GTO program (ID: 102.C-0744, 1102.C-0958, and 1104.C-0350). Rather than directly using the velocities from \citetalias{Demangeon_2021} extracted with a cross-correlation function (CCF), we instead performed a custom line-by-line (LBL) analysis starting from the extracted orber-by-order spectra (\texttt{S2D}) processed with the ESPRESSO pipeline (version 2.2.1)\footnote{\href{https://www.eso.org/sci/software/pipelines/espresso/}{\texttt{eso.org/sci/software/pipelines/espresso/}}}. We follow the LBL method of \cite{Artigau_2022} integrated in the \texttt{LBL} package v0.65\footnote{\href{https://github.com/njcuk9999/lbl}{\texttt{github.com/njcuk9999/lbl}}} compatible with many echelle spectrographs (e.g., ESPRESSO, HARPS, NIRPS, SPIRou). The LBL method, like other template matching codes (e.g., \citealt{Anglada-Escude_2012}; \citealt{Astudillo-Defru_2017}; \citealt{Zechmeister_2018}; \citealt{Silva_2022}), fully extracts the RV information from a stellar spectrum, while providing extra resilience against outlying spectral features (e.g., telluric residuals, cosmic rays, and hot pixels).

The \texttt{S2D} spectra are first telluric corrected in \texttt{LBL} by adjusting TAPAS atmospheric models \citep{Bertaux_2014}. This correction is based on the first step of the telluric correction in \texttt{APERO} \citep{Cook_2022} and is similar to the model-based method developed by \cite{Allart_2022} for ESPRESSO that significantly improves the RV precision by up to $\sim$25\% for M dwarfs. A template spectrum of L~98-59 is then built by taking the median of all 66 telluric-corrected spectra. Line-by-line velocities are calculated for $\sim$22\,000 individual spectral lines by matching the residuals between the observed spectrum and the template with the wavelength derivative of the template \citep{Bouchy_2001}. Similarly, \texttt{LBL} also computes stellar temperature perturbations --- the differential temperature indicator (\dtemp) --- from the projection of the template in $T_{\rm eff}$ space using a library of high-resolution spectra from HARPS archival data spanning multiple spectral types \citep{Artigau_2024}. The \dtemp\ indicator has been demonstrated to be at least as effective as the differential line width (dLW; \citealt{Zechmeister_2018}) or CCF FWHM for stellar activity detrending. The error-weighted average of all per-line RV and \dtemp\ are finally compiled for each epoch following the outlier identification method described in Appendix B of \citealt{Artigau_2022}.

The replacement of the fiber-link of ESPRESSO in June 2019 produced an RV offset \citep{Pepe_2021}. During the analysis, we consider the data taken before and after this upgrade as two instruments labeled ``ESPRESSOpre" and ``ESPRESSOpost". We note a significant median SNR difference between pre- and post-fiber change data, with the pre data having a SNR per pixel $\sim$20\% lower. Four epochs were flagged as $>$5$\sigma$ outliers\footnote{ESPRESSO clipped epochs (BJD\,$-$\,2457000):~1562.6476, 1637.4621, 1858.6204, and 1924.6451} using an iterative sigma-clipping algorithm, and were thus removed. The final RVs and \dtemp\ are listed in Table~\ref{table:rv}. The median RV uncertainty with LBL is 0.26\,m\,s$^{-1}$ or about a factor three smaller than the CCF RVs (0.80\,m\,s$^{-1}$) presented in \citetalias{Demangeon_2021}. A comparable improvement over CCF was also obtained in the reanalysis of the ESPRESSO data of the M4.5V dwarf LHS 1140 \citep{Cadieux_2024a}.

\subsection{HARPS Radial Velocity}

The HARPS data set presented in \citetalias{Cloutier_2019} and \citetalias{Demangeon_2021} was reanalyzed using the LBL algorithm and integrated in this study. A total of 166 L~98-59 spectra were collected with HARPS between October 2018 and April 2019 (ID: 1102.C-0339, 0102.C-0525, and 0102.D-0483), overlapping with the ESPRESSOpre campaign. The extracted spectra were retrieved from the ESO archive as processed data reduced with the HARPS pipeline v3.5.

In previous studies, the RVs of L~98-59 were calculated with a template matching method \citep{Astudillo-Defru_2017}, achieving a median precision of 2.00\,m\,s$^{-1}$ (\citetalias{Cloutier_2019}). Our LBL RV extraction, applied to the same HARPS data set, improves this precision to 1.29\,m\,s$^{-1}$, despite both methods being conceptually similar. A comparable gain was reported by \cite{Radica_2022}, who found that LBL outperformed template matching for CARMENES \citep{Quirrenbach_2018} observations of K2-18. Reprocessing the HARPS data with LBL also provides \dtemp\ time series for stellar activity modeling.

At some epochs, two consecutive exposures with HARPS were taken. We work with the per-night binned velocities and end up with a time series of 106 RVs with a median uncertainty of 1.10\,m\,s$^{-1}$, after removing two $>$5$\sigma$ outlying measurements\footnote{HARPS clipped epochs (BJD\,$-$\,2457000):~1447.8036 and 1503.7951}. The final RV and \dtemp\ measurements are reported in Table~\ref{table:rv}.

\subsection{JWST Transit Timing}

We combined precise transit timing measurements of L~98-59\,c and L~98-59\,d from JWST to constrain a global RV and TTV model. We excluded the HST mid-transit time for L~98-59\,c, reported in Table~2 of \citet{Barclay_2023}, due to a systematic offset of approximately two minutes in $t_0$ likely because the transit was partial, capturing only the ingress.

Two consecutive transits of L~98-59\,c were observed with JWST/NIRSpec and reported in \citet{Scarsdale_2024}. For our analysis, we adopted $t_0$ values calculated as the average of the mid-transit times reported in their Table~2 for two reductions of the NRS1 detector: $2460135.10507 \pm 3\cdot10^{-5}$ and $2460138.79577 \pm 3\cdot10^{-5}$\,BJD. We applied a correction of 0.5\,days to these times to account for an apparent mix-up between MJD and BJD in \cite{Scarsdale_2024}. For L~98-59\,d, we used the mid-transit time $t_0 = 2460121.11252 \pm 7\cdot10^{-5}$\,BJD from \cite{Gressier_2024} derived from the white light curve analysis of a single transit with JWST/NIRSpec. 

Finally, we included three additional transit times obtained with JWST/NIRISS and JWST/NIRSpec as part of programs GTO1201 and GO4098 that will be presented in a forthcoming publication (Benneke et al.\ in prep.). These new measurements are $2459941.79320 \pm 5\cdot10^{-5}$ and $2460381.88699 \pm 4\cdot10^{-5}$\,BJD for L~98-59\,d, and $2460736.68798 \pm 4\cdot10^{-5}$\,BJD for L~98-59\,c.

\section{Data Analysis \& Results} \label{sec:data}

\subsection{Transit Fit} \label{sec:transit_fit}

Since 2018, TESS has observed 260 transits of L~98-59\,b, 154 transits of c, and 79 transits of d. Due to the extensive TESS data set with more than 55\,000 in-transit photometric points, we opted for a sequential analysis that consists of first fitting the transits and then using the best-fit parameters in the subsequent RV and TTV analyses.

We used the \texttt{juliet} package \citep{Espinoza_2018} to fit \texttt{batman} transit models \citep{Kreidberg_2015} with an implementation of the \texttt{nautilus} nested sampling package \citep{Lange_2023}. \texttt{nautilus} is well suited for high-dimensional and large data set problems, requiring often an order of magnitude fewer likelihood evaluations to accurately estimate Bayesian evidence and parameter posteriors compared to standard \citep{Skilling_2006} or dynamic \citep{Higson_2019} nested sampling. \texttt{nautilus} enhances the efficiency of importance nested sampling \citep{Feroz_2019} by using multiple neural networks to guide the sampler toward the bulk of the distribution.

The orbital elements for planet $k=\{ \textrm{b, c, d}\}$ are the period $P_{k}$, the time of inferior conjunction $t_{0,k}$, the scaled semi-major axis $a_{k}/R_{\star}$, the eccentricity $e_{k}$, and the argument of periastron $\omega_{k}$. In a system with two or more planets, such as L~98-59, it is convenient to fit for a common stellar density $\rho_{\star}$ instead of a $a_{k}/R_{\star}$ for each planet. Here, the stellar density is fitted in logarithmic form ($\ln \rho_{\star}$) to avoid reported biases for planets transiting at a high impact parameter \citep{Gilbert_2022}. One must still be cautious of photo-eccentric bias in $\rho_{\star}$ if the orbits are incorrectly assumed to be circular (\citealt{Kipping_2010}, \citeyear{Kipping_2014}; \citealt{Dawson_2012}; \citealt{Van-Eylen_2015}). This bias can reach 30\% systematic error in $\rho_{\star}$ even for relatively small eccentricities of $e \sim 0.1$. 
Here, we fix $e_{k} = 0$ and $\omega_{k} = 90^{\circ}$ for the transit fit, but allow for free eccentricity in our joint RV--TTV analysis presented in Section~\ref{sec:joint_fit} that confirms negligible level of photo-eccentric bias.

The transit impact parameter $b_{k}$ and the planet-to-star radius ratio $R_{\textrm{p,}k} / R_{\star}$ are sampled using the $r_{1, k}$ and $r_{2, k}$ parameter transformation outlined in \cite{Espinoza_2019} that only allow for physically plausible $b_{k}$ and $R_{\textrm{p,}k} / R_{\star}$ combination. The limb-darkening effect in the TESS bandpass is modeled using the quadratic parameters $q_1$ and $q_2$ which take values between 0 and 1 \citep{Kipping_2013}. An additional photometric jitter term $\sigma_{\rm TESS}$ is also included in the fit.

A normal prior $\mathcal{N}\left(9.50, 0.09^2\right)$ (SI units) is adopted for $\ln \rho_{\star}$ based on stellar mass ($M_{\star} = 0.2926 \pm 0.0069$\,M$_{\odot}$) and radius ($R_{\star} = 0.3139 \pm 0.0093$\,R$_{\odot}$) estimates from empirical calibrations of M dwarfs with absolute ${K_{\rm s}}$ magnitude \citep{Mann_2015, Mann_2019}. Wide uniform priors are used for the other transit parameters (see Table~\ref{table:transit_fit}). The nested sampling is performed with $100\times n_{\rm dim}$ live points, with $n_{\rm dim} = 16$ the number of free parameters. As recommended by \cite{Lange_2023} for more accurate Bayesian evidence and posterior distributions in \texttt{nautilus}, we discarded all samples at the end of the exploration phase and drew new ones to construct a sample with an effective size of at least $n_{\rm eff} = 10\,000$.

\begin{figure*}[ht!]
\centering
\minipage{0.2196\textwidth}
\includegraphics[width=1\linewidth]{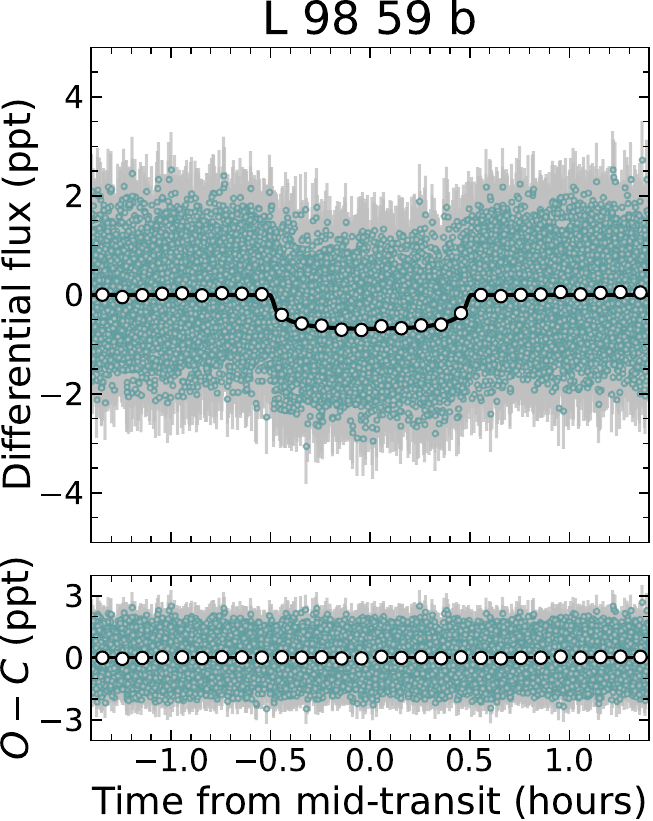}
\endminipage
\hfill
\minipage{0.19\textwidth}
\vspace{0.01cm}
\includegraphics[width=1\linewidth]{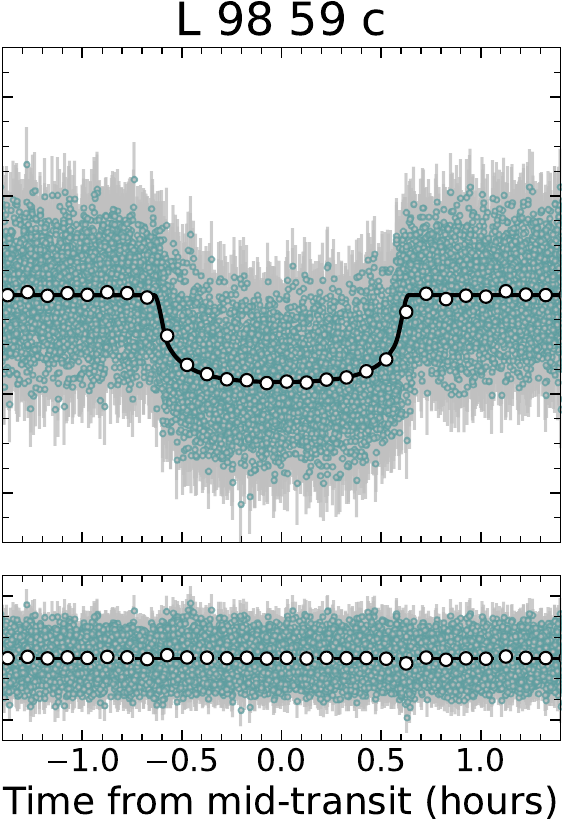}
\endminipage
\hfill
\minipage{0.19\textwidth}
\vspace{0.01cm}
\includegraphics[width=1\linewidth]{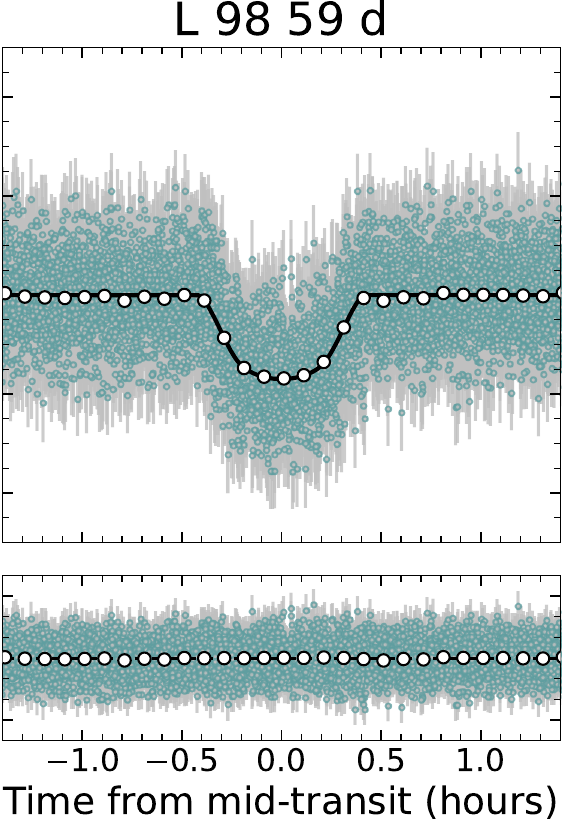}
\endminipage
\hfill
\minipage{0.1895\textwidth}
\vspace{-1.17cm}
\includegraphics[width=1\linewidth]{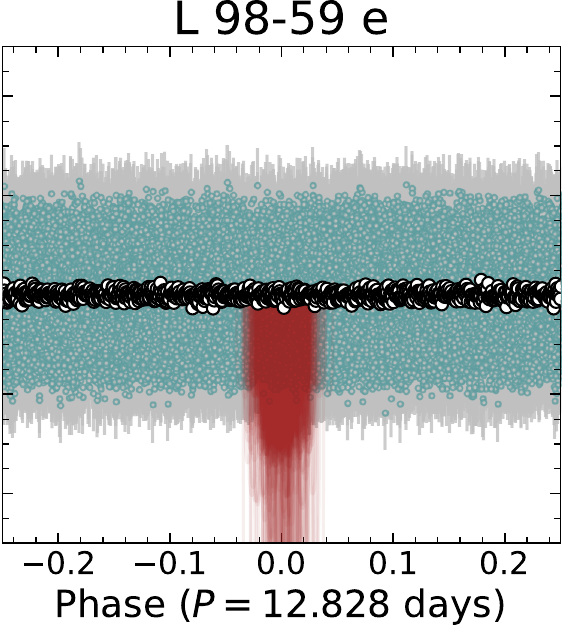}
\endminipage
\hfill
\minipage{0.1895\textwidth}
\vspace{-1.185cm}
\includegraphics[width=1\linewidth]{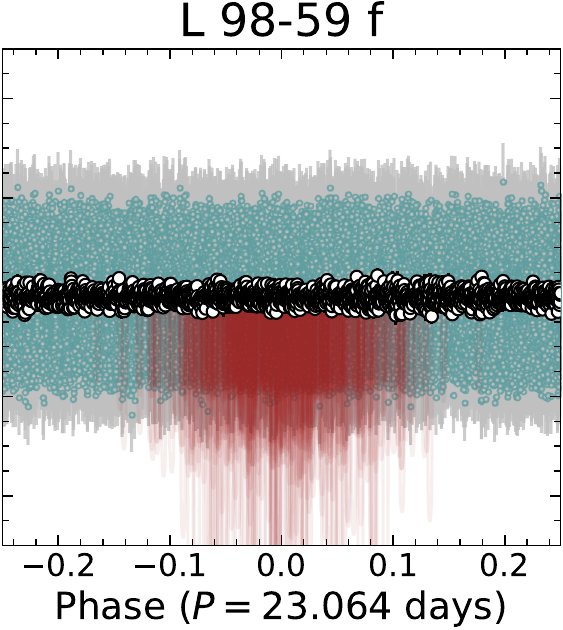}
\endminipage
\\[0.2cm] 
\minipage{0.2196\textwidth}
\includegraphics[width=1\linewidth]{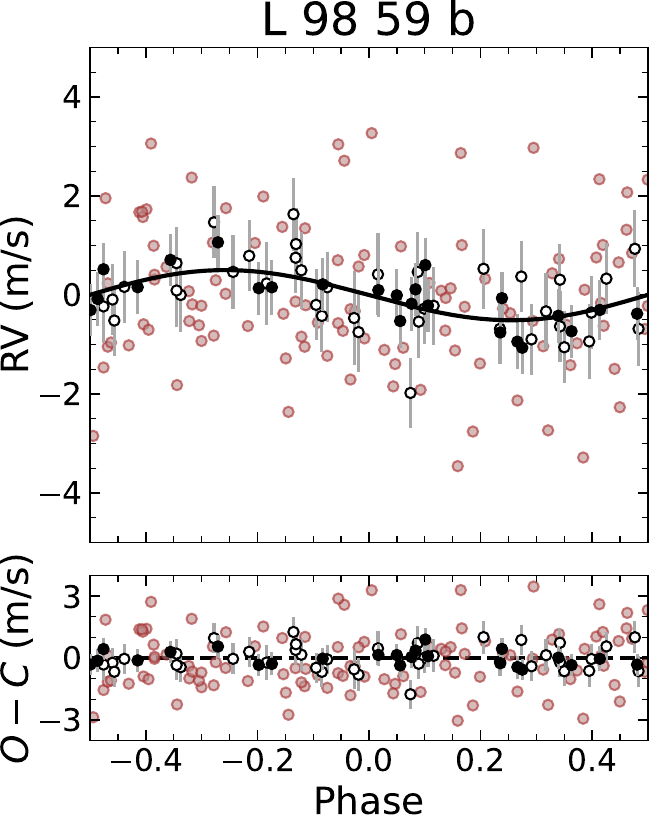}
\endminipage
\hfill
\minipage{0.19\textwidth}
\includegraphics[width=1\linewidth]{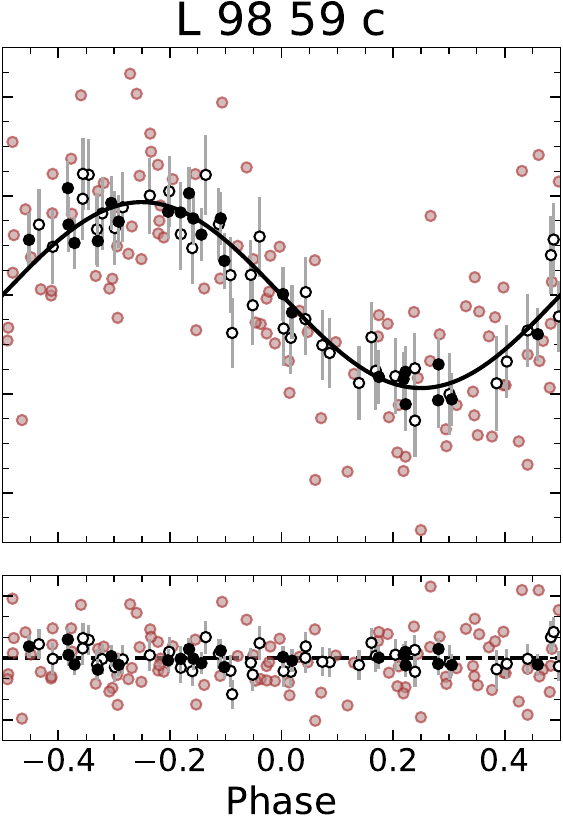}
\endminipage
\hfill
\minipage{0.19\textwidth}
\includegraphics[width=1\linewidth]{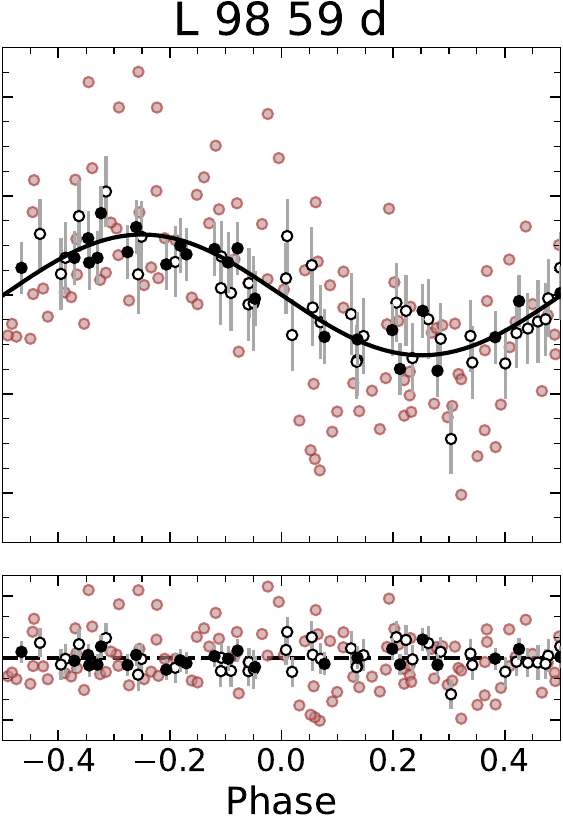}
\endminipage
\hfill
\minipage{0.19\textwidth}
\includegraphics[width=1\linewidth]{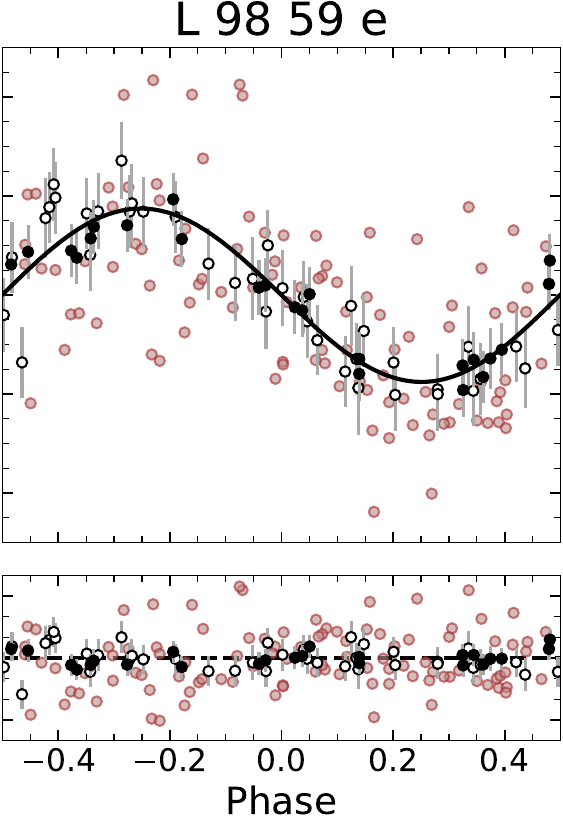}
\endminipage
\hfill
\minipage{0.19\textwidth}
\includegraphics[width=1\linewidth]{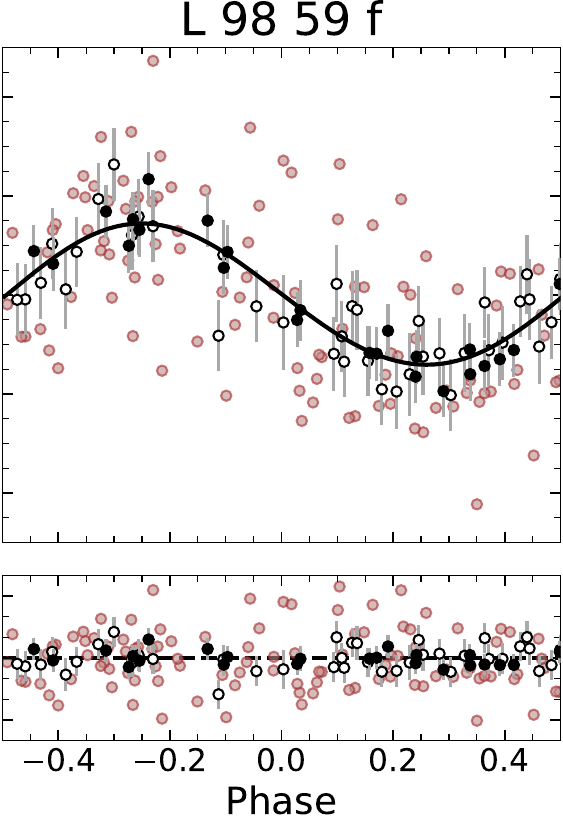}
\endminipage
\caption{Family portrait of the five planets in the L~98-59 system showing their transit (or absence of transit) observed by TESS in the top panels and their Doppler signature from HARPS (red circles) and ESPRESSO (white/black circles for pre/post data) in the bottom panels. Only the error bars for ESPRESSO are shown to improve visibility. Black curves represent best-fit models from the transit analysis (top) and joint RV–TTV fit (bottom, using fixed orbital inclinations, see Sect.~\ref{sec:joint_fit}). For L~98-59\,e and f, 1000 predicted transit curves are shown in red using posterior samples from the RV--TTV fit and radii estimated from their minimum masses using \texttt{spright} \citep{Parviainen_2024}. Planets e and f do not transit L~98-59.}
\label{fig:transit_rv}
\end{figure*}

The results of the transit fit are reported as the median, 16$^{\rm th}$ and 84$^{\rm th}$ percentiles of the posteriors in Table~\ref{table:transit_fit}. The best-fit transit models for L~98-59\,b, c, and d are shown in the top panels of Figure~\ref{fig:transit_rv}. These transits, observed at distinct impact parameters of $0.51^{+0.04}_{-0.05}$, $0.41^{+0.05}_{-0.07}$, and $0.921^{+0.007}_{-0.007}$ for L~98-59\,b, c, and d, respectively, provide only marginal additional constraints on the stellar density ($\ln \rho_{\star} = 9.47 \pm 0.08$). Taking this $\rho_{\star}$ from the transit fit and following Appendix C.2 of \cite{Cadieux_2024a}, we update $M_{\star}$ to 0.2923 $\pm$ 0.0067\,M$_{\odot}$ and $R_{\star}$ to 0.3155 $\pm$ 0.0062\,R$_{\odot}$ for L~98-59, which we adopt to derive the planetary parameters.

\begin{deluxetable}{lcr}
\tablecaption{Transit parameters from TESS data}
\tablehead{\colhead{Parameter} & \colhead{Prior} & \colhead{Posterior}}
\startdata
\multicolumn{3}{c}{\textit{Stellar parameters}} \\
$\ln \rho_{\star}$ & $\mathcal{N}\left(9.50, 0.09^2\right)$ & 9.47$\pm$0.08 \\[0.1cm]
\multicolumn{3}{c}{\textit{L~98-59\,b}} \\
$P_{\rm b}$ (days) & $\mathcal{U}$(\citetalias{Demangeon_2021} $\pm$ 0.01) & 2.2531140(4) \\
$t_{\rm 0,b}$ (\footnotesize{BJD\,-\,2458360}) & $\mathcal{U}$(\citetalias{Demangeon_2021} $\pm$ 0.01) & 6.1706(2) \\
$r_{\rm 1,b}$ & $\mathcal{U}\left(0, 1\right)$ & 0.67$^{+0.03}_{-0.03}$ \\
$r_{\rm 2,b}$ & $\mathcal{U}\left(0, 1\right)$ & 0.0243(3) \\[0.1cm]
\multicolumn{3}{c}{\textit{L~98-59\,c}} \\
$P_{\rm c}$ (days) & $\mathcal{U}$(\citetalias{Demangeon_2021} $\pm$ 0.01) & 3.6906764(4) \\
$t_{\rm 0,c}$ (\footnotesize{BJD\,-\,2458360}) & $\mathcal{U}$(\citetalias{Demangeon_2021} $\pm$ 0.01) & 7.2730(2) \\
$r_{\rm 1,c}$ & $\mathcal{U}\left(0, 1\right)$ & 0.61$^{+0.03}_{-0.04}$ \\
$r_{\rm 2,c}$  & $\mathcal{U}\left(0, 1\right)$ & 0.0385(4) \\[0.1cm]
\multicolumn{3}{c}{\textit{L~98-59\,c}} \\
$P_{\rm d}$ (days) & $\mathcal{U}$(\citetalias{Demangeon_2021} $\pm$ 0.01) & 7.450729(2) \\
$t_{\rm 0,d}$ (\footnotesize{BJD\,-\,2458360}) & $\mathcal{U}$(\citetalias{Demangeon_2021} $\pm$ 0.01) & 2.7400(3) \\
$r_{\rm 1,d}$ & $\mathcal{U}\left(0, 1\right)$ & 0.947$^{+0.004}_{-0.005}$ \\
$r_{\rm 2,d}$ & $\mathcal{U}\left(0, 1\right)$ & 0.0472(8) \\[0.1cm]
\multicolumn{3}{c}{\textit{Instrumental parameters}} \\
$q_{\rm 1,TESS}$ & $\mathcal{U}\left(0, 1\right)$ & 0.78$^{+0.14}_{-0.15}$ \\
$q_{\rm 2,TESS}$ & $\mathcal{U}\left(0, 1\right)$ & 0.08$^{+0.07}_{-0.05}$ \\
$\sigma_{\rm TESS}$ (ppm) & $\mathcal{LU}$(1, 1000) & 5$^{+10}_{-3}$ \\
\enddata
\tablecomments{Priors on $P$ and $t_0$ are centered on the values from Table 3 of \citetalias{Demangeon_2021}.}
\label{table:transit_fit}
\end{deluxetable}

\subsection{Transit Timing Variations} \label{sec:ttv}

With five planets inside 0.1\,au (see Fig.~\ref{fig:orbit}), the compact system around L~98-59 is conducive to TTV caused by mutual gravitational interactions between the planets. The characteristic period of TTVs for a pair of planets with a period ratio $P_{2}/P_{1}$ close to the ratio ($m/n$) between two integers $m$ and $n$ is given by \citep{Agol_2018}:
\begin{equation}
    P_{\rm TTV} = 1 / |m / P_2 - n/P_1 |
\end{equation}

For example, planets L~98-59\,c and d with a period ratio ($P_{\rm d} / P_{\rm c} = 2.019$) near the 2:1 mean-motion resonance ($m=2$ and $n=1$) would produce TTVs with a super-period $P_{\rm TTV} = 396.32 \pm 0.01$\,days. The six years of TESS data provide a sufficiently long baseline to potentially detect this super-period.

We employed the \texttt{TTVFast} code \citep{Deck_2014} adapted in Python\footnote{\href{https://github.com/simonrw/ttvfast-python}{\texttt{github.com/simonrw/ttvfast-python}}} to simulate TTVs in the L~98-59 system over the timescale of the TESS data. The numerical integrations are done on time steps of 0.1\,days which respect the recommendation of integration steps $<$1/20 of the shortest orbital period. \texttt{TTVFast} takes as input the stellar mass $M_{\star}$ and seven parameters per planet: $M_{\textrm{p,} k}$, $P_{k}$, $e_{k}$, $\omega_{k}$, $i_{k}$, $\Omega_{k}$, $M_{0,k}$. Here, $M_{\textrm{p,} k}$ is the planet mass, $i_{k}$ the orbital inclination, $\Omega_{k}$ the longitude of ascending node set arbitrarily to 0$^{\circ}$, and $M_{0,k}$ is the mean anomaly at a reference epoch taken as the start of the simulation (2\,458\,345\,BJD). The $M_{0,k}$ are derived from $t_{0,k}$ and $P_{k}$, where $k$ represents up to five planets ($k=\{ \textrm{b, c, d, e, f}\}$). Non-transiting planets such as L~98-59\,e can gravitationally perturb the orbits of the inner transiting planets, potentially inducing detectable TTVs, especially during closest separation where short-timescale perturbations --- known as ``chopping'' signals --- can arise. Using \texttt{TTVFast} and the parameters in Table~3 of \citetalias{Demangeon_2021} for their five-planet model, small TTVs are predicted in this system, with amplitudes of 0.3\,min for L~98-59\,b, 3\,min for L~98-59\,c, and 6\,min for L~98-59\,d.

We examined the TESS data for evidence of TTVs, focusing our search on L~98-59\,c and d due to the shallow transit (600\,ppm) and TTV signal likely below 1 minute for L~98-59\,b. Only full transits of L~98-59\,c and d with a sufficient baseline of at least one transit duration $t_{14}$ before ingress and after egress were analyzed. The TESS data contains many double and triple transits, which were corrected by removing the transit signals of the `contaminating' planets using \texttt{batman} models generated with the parameters listed in Table~\ref{table:transit_fit}. Following this cleaning step, we are left with a set of isolated transits of planets c and d.

For each isolated transit, we measured any timing deviation $\delta t$ from the constant period relation $t_{\textrm{c}, k} = t_{0, k} + E P_k$, where $t_{\textrm{c}, k}$ is the time of inferior conjunction at epoch number $E$ for planet $k$. The $\delta t$ are determined by shifting in time a \texttt{batman} model at a high temporal resolution of 0.0001\,days ($\sim$10\,sec) to minimize the chi-square. The $\delta t$ were limited to $\pm30$\,min around the linear ephemeris prediction. The fitting was carried out using MCMC sampling with \texttt{emcee} (100 walkers, 2000 steps, 500 burn-in), yielding both the best-fit $\delta t$ values and their uncertainties. The resulting TTVs display clear sinusoidal variations with a super-period consistent with 396\,days expected for the gravitational perturbation between L~98-59\,c and d (Fig.~\ref{fig:ttv}). We report the transit timing measurements in Table~\ref{table:ttv} that we use in Section~\ref{sec:joint_fit} to fit a TTV model.

\subsection{Stellar Activity Modeling}

The archival ESPRESSO and HARPS observations of L~98-59 serve as a good benchmark data set to test the \dtemp\ formalism of \cite{Artigau_2024}. The \dtemp\ indicator showed an almost perfect anti-correlation ($-0.922$) with the small-scale magnetic field for the active M1V star AU Mic observed with the near-infrared spectropolarimeter SPIRou \citep{Donati_2020}, implying that \dtemp\ traces the same underlying physics. The \dtemp\ sequence of Barnard's star (M4V), also observed with SPIRou, demonstrates the power of this indicator in constraining the rotation period of this quiet star ($P_{\rm rot} = 153 \pm 3$\,days), a result in line with polarimetry ($P_{\rm rot} = 136 \pm 13$\,days, \citealt{Donati_2023}).

Here, the ESPRESSO and HARPS \dtemp\ time series of L~98-59 were modeled with a common GP using \texttt{george} \citep{Ambikasaran_2015}, with a quasi-periodic covariance kernel (\citealt{Haywood_2014}; \citealt{Rajpaul_2015}; \citealt{Stock_2023}):
\begin{equation}
\label{eq:quasi-per}
k_{i,j} = A^2 \exp \left[ -\frac{|t_i - t_j|^2}{2 \ell^2} - \Gamma \sin^2 \left( \frac{\pi | t_i - t_j|}{P_{\rm rot}} \right) \right]
\end{equation}
where $|t_i - t_j|$ is the time lag between data $i$ and $j$, $A$ is the amplitude of the GP, $\ell$ is the coherence timescale, $\Gamma$ is a dimensionless scaling factor of the periodic component of the GP, and $P_{\rm rot}$ is the stellar rotation period. Constant offsets $c_{\scalebox{0.6}{\rm HARPS}}$, $c_{\rm pre}$ and $c_{\rm post}$ between instruments were fitted. White noise terms $\sigma_{\scalebox{0.6}{\rm HARPS}}$, $\sigma_{\rm pre}$, and $\sigma_{\rm post}$ were also added in quadrature to the covariance function above. 

In both \citetalias{Demangeon_2021} and \cite{Rajpaul_2024}, the rotation period of L~98-59 was not well constrained by a quasi-periodic GP regression of the CCF FWHM time series, yielding $P_{\rm rot} = 33^{+43}_{-19}$\,days and $P_{\rm rot} = 50\pm20$\,days, respectively, inferred from the same data set. We thus first run a model with wide log uniform prior $\mathcal{LU} \left(30, 120\right)$\,days for $P_{\rm rot}$ and $\mathcal{LU} \left(30, 1000\right)$\,days for $\ell$. Other priors are the same for all models, namely $\mathcal{LU} \left(0.1, 10\right)$ for $\Gamma$ following \cite{Stock_2023}, $\mathcal{LU} \left(0.01, 10\right)$\,K for $A$, $\sigma_{\scalebox{0.6}{\rm HARPS}}$, $\sigma_{\rm pre}$, $\sigma_{\rm post}$, and finally $\mathcal{U} \left(-10, 10\right)$\,K for $c_{\scalebox{0.6}{\rm HARPS}}$, $c_{\rm pre}$, and $c_{\rm post}$. The posterior distribution is sampled with \texttt{emcee} using 100 walkers for 25\,000 steps, rejecting the first 5000 steps.

This first exploratory run converges to a well-defined $P_{\rm rot} = 76.7^{+2.5}_{-2.3}$\,days with the 1-$\sigma$ uncertainty widened by secondary peaks (unlikely) near 50 and 100\,days, capturing only a negligible fraction of the probability distribution. This period of 76.7\,days is consistent with the periodicity of $\sim$80\,days identified by \citetalias{Cloutier_2019} in various activity indices, most prominently in the H$\alpha$ time series and from the star's log\,$R^{\prime}_{HK}$. With the \dtemp\ indicator also appearing to be sensitive to the rotational signal, a second run more guided toward the 80-day periodicity was attempted. This time, we restricted the prior on $P_{\rm rot}$ to $\mathcal{LU} \left(65, 95\right)$\,days and $\ell$ to $\mathcal{LU} \left(80, 1000\right)$\,days, informed by \cite{Stock_2023} that this timescale parameter correlates with the lifetime of surface magnetic regions (e.g., spots, plages) that usually survive for multiple rotations.

\begin{figure}[ht!]
\centering
\includegraphics[width=1\linewidth]{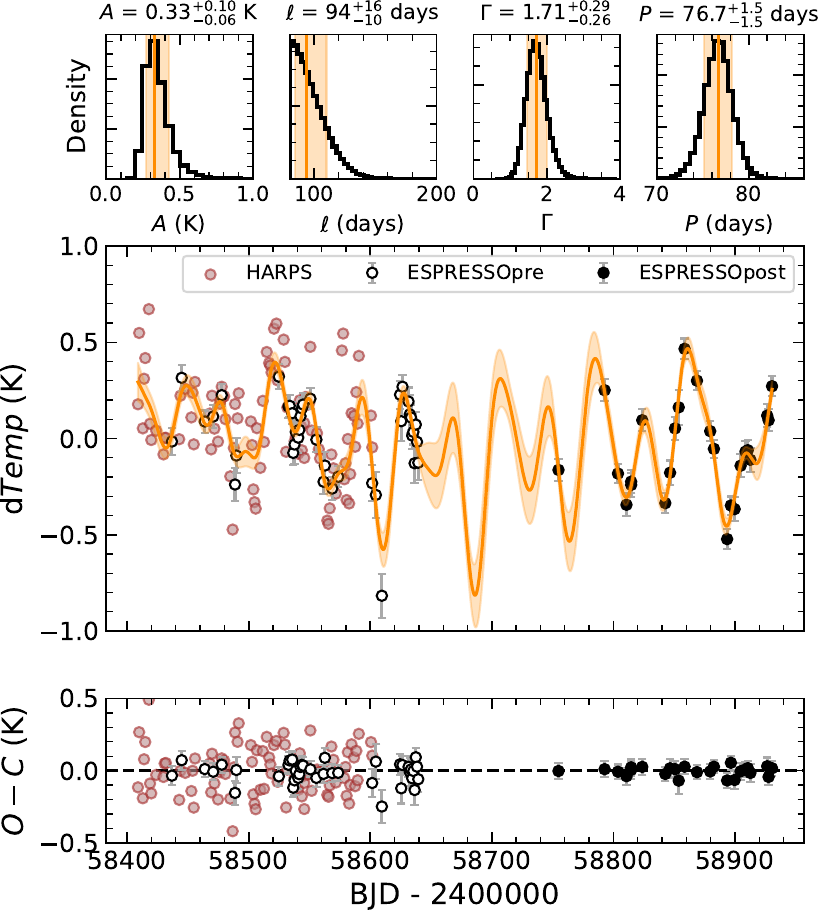}
\caption{Sub-Kelvin differential temperature measurements of L~98-59 from ESPRESSO and HARPS, with only the error bars for ESPRESSO displayed for clarity. The posterior distributions on the hyperparameters for a quasi-periodic GP regression are shown above with the median GP prediction shown with the orange curve and 1$\sigma$ uncertainty envelope. The \dtemp\ indicator reveal that L~98-59 is a relatively quiet star with a rotational period of 76.7$\pm1.5$\,days.}
\label{fig:dtemp}
\end{figure}

The results of this fit are shown in Figure~\ref{fig:dtemp}. The \dtemp\ indicator enables the first robust determination of L~98-59’s rotation period ($P_{\rm rot} = 76.7 \pm 1.5$\,days). The residuals of our GP model have a dispersion of 0.17\,K for HARPS and 0.05\,K for ESPRESSO, which is equivalent to sub-mmag precision photometry over the 1.5-year sequence assuming a $T_{\rm eff}^4$ dependency of flux. This opens the door to adapting the $FF^{\prime}$ method of \cite{Aigrain_2012} for estimating RV jitter by using \dtemp\ instead of photometry. The offsets $c_{\rm pre} = -0.06 \pm 0.15$\,K and $c_{\rm post} = 0.02 \pm 0.18$\,K are both consistent with 0\,K. There is thus no evidence that the fiber change of ESPRESSO resulted in a systematic bias in \dtemp, or by extension systematic offset in spectral line contrast. The posterior distributions for $\ell$, $\Gamma$, and $P_{\rm rot}$ constrained by \dtemp\ (shown in Fig.~\ref{fig:dtemp}) serve as priors for the GP activity modeling of the radial velocities.

\subsection{RV Statistical Confirmation of L~98-59\,f} \label{sec:rv_analysis}

The improved ESPRESSO and HARPS RVs and the new stellar rotation constraints from \dtemp\ motivate a reanalysis of the RV data to characterize the Keplerian signals of the L~98-59 planets and to investigate evidence for a fifth planet at 23\,d, proposed by \citetalias{Demangeon_2021}. We modeled stellar activity with a GP trained on the simultaneous \dtemp\ time series and compared the Bayesian evidence ($\mathcal{Z}$) for models with three, four, and five planets on circular orbits. This analysis provides statistical confirmation of L~98-59\,f in the RV data, supporting a five-planet architecture in the joint RV--TTV fit presented in the next Section~\ref{sec:joint_fit}.

For the activity GP, we use a quasi-periodic kernel (Eq.~\ref{eq:quasi-per}) with a log uniform prior $\mathcal{LU}\left(10^{-3}, 5 \right)$\,m\,s$^{-1}$ for the amplitude $A$ and approximate the \dtemp\ hyperparameter posteriors (Fig.~\ref{fig:dtemp}) using analytical forms to serve as priors: a truncated normal distribution $\mathcal{TN}(85, 20^2, 80, 1000)$ for $\ell$, and normal distributions $\mathcal{N}(1.71, 0.29^2)$ for $\Gamma$ and $\mathcal{N}(76.7, 1.5^2)$ for $P_{\rm rot}$. Here, $\mathcal{TN}(\mu, \sigma^2, a, b)$ denotes a normal distribution with mean $\mu$, variance $\sigma^2$, defined between $a$ and $b$. We use \texttt{radvel} \citep{Fulton_2018} to generate RV Keplerian models for each planet $k$ described by a semi-amplitude $K_k$ and four orbital elements ($P_{k}$, $t_{0,k}$, $e_k$, $\omega_k$). All models assume circular orbits ($e_k = 0$, $\omega_k = 90^\circ$). For L~98-59 b, c, and d, $P_{k}$ and $t_{0,k}$ are constrained using a normal prior based on the median and 16$^{\rm th}$--84$^{\rm th}$ percentile values from the transit fit (Table~\ref{table:transit_fit}). For planet e and f, we adopt broad uniform priors for $P_{k}$ based on the periodogram of the raw RV (Fig.~\ref{fig:full_rv_fit}), namely $\mathcal{U}\left(8, 18\right)$ and $\mathcal{U}\left(18, 35\right)$\,days, and use uniform priors on $t_{0,k}$ to cover all phases. We also consider per-instrument RV offsets ($\gamma_{\scalebox{0.6}{\rm HARPS}}$, $\gamma_{\rm pre}$ and $\gamma_{\rm post}$) and extra white noise terms ($\sigma_{\scalebox{0.6}{\rm HARPS}}$, $\sigma_{\rm pre}$ and $\sigma_{\rm post}$). We use \texttt{nautilus} with $100\times n_{\rm dim}$ live points to sample the posterior distributions, with $n_{\rm dim}$ varying between 19 and 25 depending on the number of planets. Again, we follow the suggested \texttt{nautilus} setup of performing exploration and sampling phases to achieve unbiased distributions with at least $n_{\rm eff} = 10\,000$ effective samples \citep{Lange_2023}.

The resulting log Bayesian evidences are reported in Table~\ref{table:model_comparison}. Compared to \citetalias{Demangeon_2021}, where a fifth planet lowered the $\ln \mathcal{Z}$ by 0.81 and 0.4 depending on the activity modeling, our analysis robustly confirms L~98-59\,f with a $\Delta \ln \mathcal{Z} = 11.34$ over the four-planet solution, which corresponds to a strong detection according to \cite{Trotta2008} or equivalently a 5.1$\sigma$ detection following \cite{Benneke_2012}. The retrieved Keplerian signature of L~98-59\,f has a period (23.08 $\pm$ 0.07\,days) and a semi-amplitude (1.43 $\pm$ 0.17\,m\,s$^{-1}$) fully consistent with the candidate planet reported by \citetalias{Demangeon_2021}. Injection-recovery simulations, detailed in Appendix~\ref{sec:sensitivity_map}, quantify potential signal suppression introduced by the activity GP over periods 1--500\,days and confirm that the combined ESPRESSO and HARPS data are sensitive to a 1.43\,m\,s$^{-1}$ signal at 23\,days.

\begin{deluxetable}{lcc}
\tablecaption{Bayesian model comparison of the RV-only analysis of L~98-59}
\tablehead{\colhead{Model} & \colhead{$\ln \mathcal{Z}$} & \colhead{$\Delta \ln \mathcal{Z}$}}
\startdata
Activity GP + 3 planets & $-391.51$ & $-35.86$\\
Activity GP + 4 planets & $-355.65$ & 0\\
Activity GP + 5 planets & $-344.31$ & $11.34$\\
\enddata
\label{table:model_comparison}
\tablecomments{The uncertainty on $\ln \mathcal{Z}$ estimated from repeated runs was typically 0.01.}
\end{deluxetable}

\subsection{Joint RV--TTV Fit} \label{sec:joint_fit}

\begin{figure*}[ht!]
\centering
\includegraphics[width=1\linewidth]{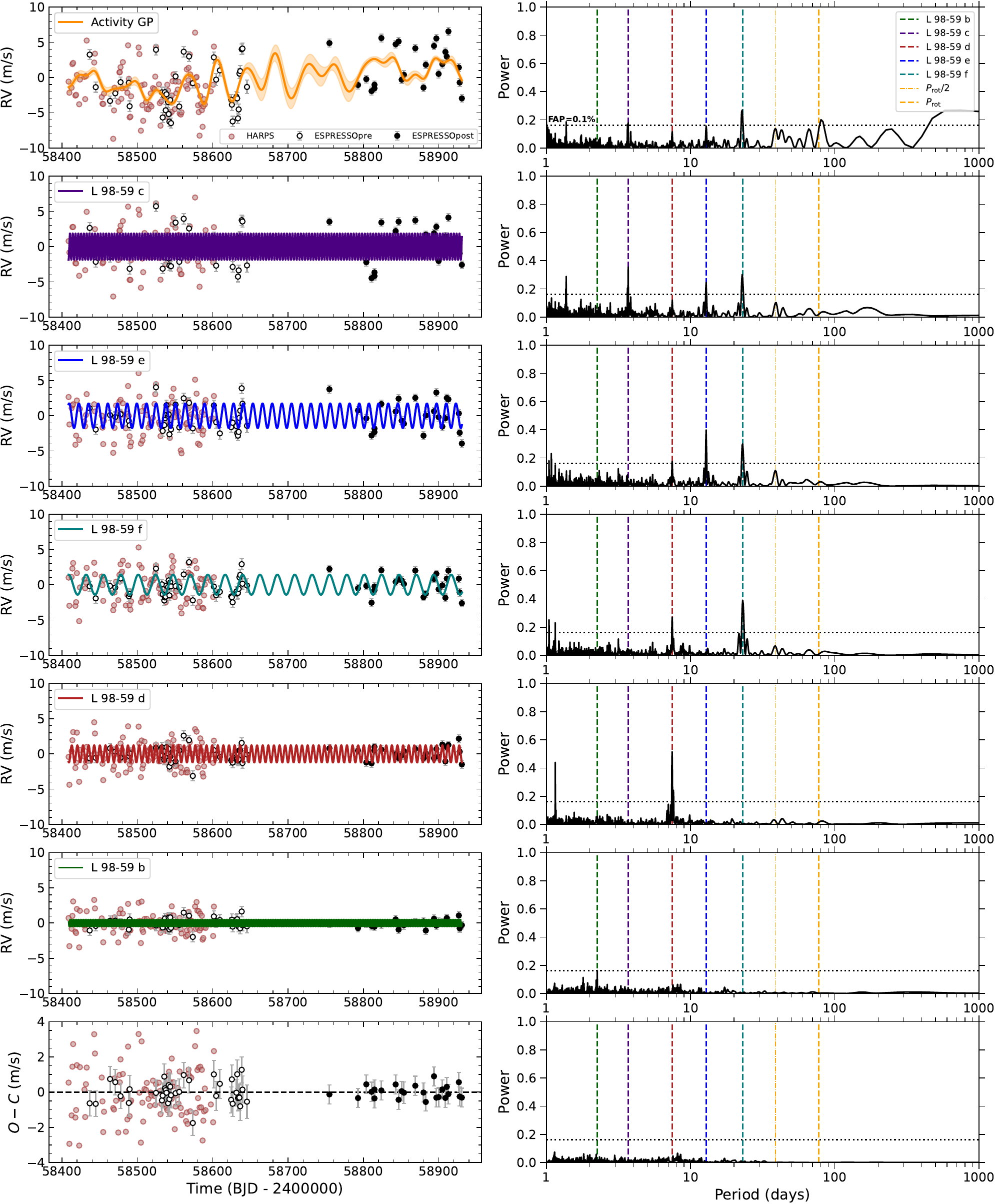}
\caption{Summary of the RV component from the joint RV--TTV fit. \textit{Left panels}:~Radial velocity of L~98-59 from ESPRESSO and HARPS decomposed into many components (activity and planets). For clarity, only the error bars from ESPRESSO are shown. The first row shows the raw RVs with systemic offsets removed. Each subsequent row sequentially subtracts one component at a time in decreasing order of variance (activity GP, then planets c, e, f, d, and b). \textit{Right panels}: Lomb-Scargle periodograms computed on the combined data at each step. The false alarm probability (FAP) of 0.1\% is shown with a dotted horizontal line.}
\label{fig:full_rv_fit}
\end{figure*}

\begin{figure*}[ht!]
\centering
\includegraphics[width=1\linewidth]{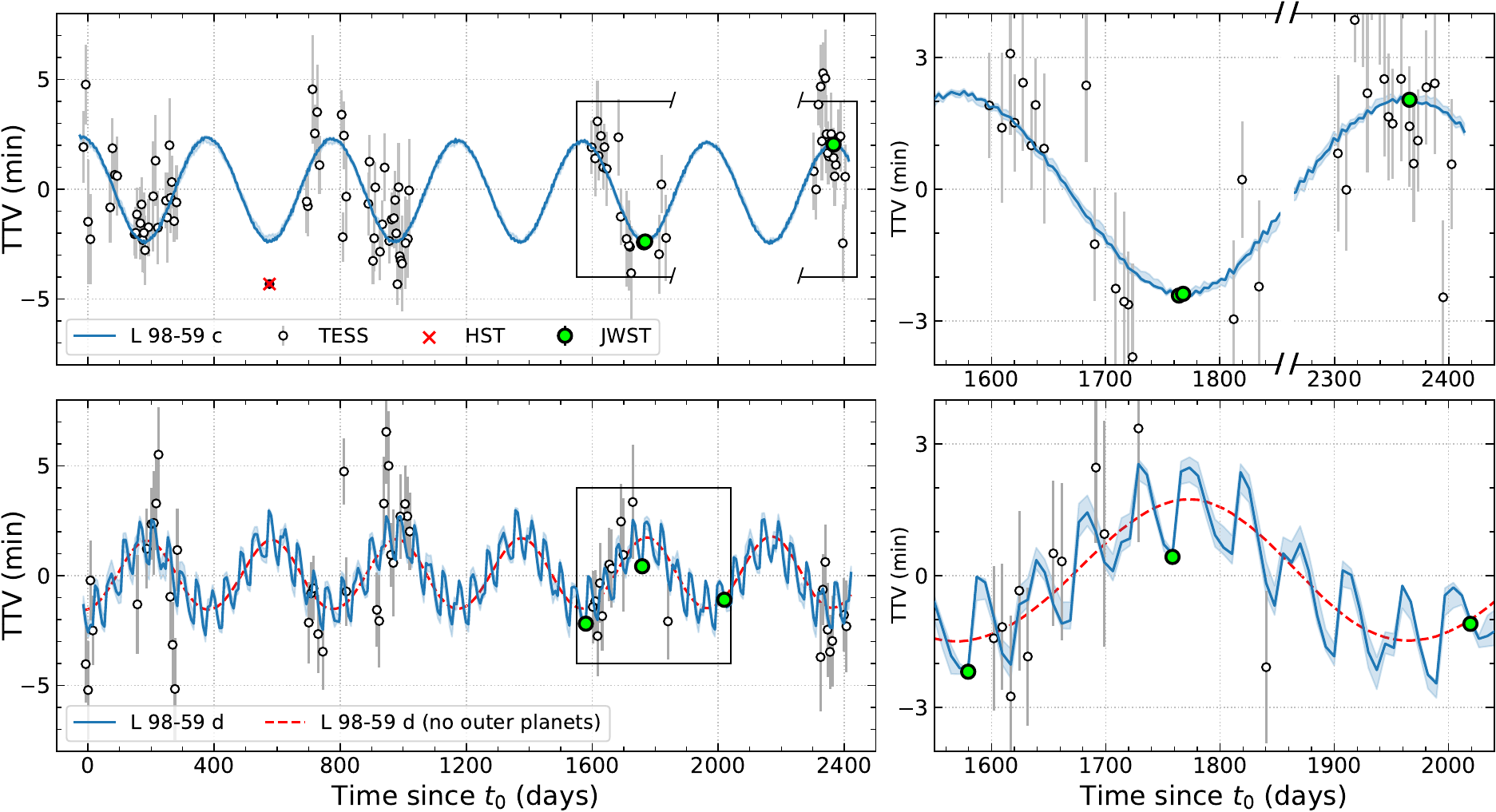}
\caption{Transit timing variations of L~98-59\,c (top) and L~98-59\,d (bottom) from TESS (white circles), HST (red cross), and JWST (green circles) with a zoomed-in region shown on the right. The HST partial transit of L~98-59\,c was not included in the fit. The blue curve depicts the median TTV model from the joint RV--TTV fit using a fixed inclination for all planets. A 68\% confidence envelope is shown in lighter blue and was determined from 1000 random solutions drawn from the posterior distribution. The red dashed curve shows a TTV model without the outer non-transiting planets L~98-59\,e and f.}
\label{fig:ttv}
\end{figure*}

We conducted a joint analysis of L~98-59's RV from ESPRESSO and HARPS, combined with TTVs of planets c and d from TESS and recent transit timing measurements from JWST. The global fit implements RV Keplerian models with \texttt{radvel}, RV stellar activity GP using \texttt{george}, and TTV modeling with \texttt{TTVFast}, following the same methodology outlined in the previous Sections. 

The RV and TTV models share common planetary and orbital parameters ($P_{k}$, $t_{0,k}$, $K_k$, $e_k$, $\omega_k$, $i_k$). Again, we use a quasi-periodic GP trained on the simultaneous \dtemp \ indicator to correct for stellar activity in the RV data. We adopt the same priors as in the RV-only analysis, with the exceptions that we now fix $P_{k}$ and $t_{0,k}$ for L~98-59\,b, c, and d to their value inferred from the transit fit (Table~\ref{table:transit_fit}) to limit the number of free parameters, and now allow eccentric orbits for all planets. Instead of directly sampling $e_k$ and $\omega_k$, we fit $\sqrt{e_k} \cos \omega_k$ and $\sqrt{e_k} \sin \omega_k$, each with uniform prior between $-\sqrt{0.3}$ and $\sqrt{0.3}$, forcing $e_k < 0.3$ to prevent orbit crossings.

Transit timing variations scale with the absolute masses of the planets ($M_{\textrm{p,}k}$). We relate $K_k$ with $M_{\textrm{p,}k}$ using the standard semi-amplitude equation (e.g., \citealt{Lovis_2010}):
\begin{equation} 
K_k = \left( \frac{2 \pi G}{P_k} \right)^{1/3} \frac{M_{\textrm{p,}k} \sin i_k}{\left( M_{\star} +  M_{\textrm{p,}k} \right)^{2/3}} \left( 1 - e_k^2 \right)^{-1/2}
\label{eq:semi-amplitude}
\end{equation}
assuming $M_{\textrm{p,}k} \ll M_{\star}$ and adding $M_{\star}$ as a parameter in the global RV--TTV model with a prior $\mathcal{N}\left(0.2923, 0.0067^2\right)$\,$M_\odot$. For L~98-59\,b, c, and d, the inclinations $i_k$ are fixed to the median values derived from the transit fit, respectively 88.08, 88.88, and 88.44$^{\circ}$. For L~98-59\,e and f, we explored two models: (1) assuming their orbits are co-planar with the inner planets ($i_{\rm e} = i_{\rm f} = 88.47^{\circ}$), and (2) letting their inclinations vary freely between 30$^{\circ}$ and 90$^{\circ}$.

The joint RV--TTV model has $n_{\rm dim} = 30$ free parameters, with two more when the orbital inclinations of L~98-59\,e and f are allowed to deviate from co-planarity. We use \texttt{nautilus} to explore the prior volume and to sample the posterior until reaching an effective sample size $n_{\rm eff} = 10\,000$. The tightly packed configuration of the L~98-59 system causes TTV predictions from \texttt{TTVFast} to be highly sensitive to even small eccentricities, which affected the sampling efficiency because highly improbable orbital configurations were attributed equal prior probability. To mitigate this and speed up the nested sampling, we restricted the priors on $P_{\rm e}$ and $P_{\rm f}$ to $\mathcal{U}\left(12, 14 \right)$ and $\mathcal{U}\left(20, 25 \right)$\,days, respectively, and imposed upper limits of $e_{\rm b} < 0.2$, $e_{\rm c} < 0.02$, $e_{\rm d} < 0.05$, $e_{\rm e} < 0.1$, and $e_{\rm f} < 0.25$. These constraints were determined after an initial exploratory run with 1000 live points ($n_{\rm live}$), using the same parameterization of $\sqrt{e_k} \cos \omega_k$ and $\sqrt{e_k} \sin \omega_k$. The posteriors from this run showed that the probability density outside these bounds was effectively zero.


To account for the complexity and high dimensionality of the posterior, the final fits were performed with $n_{\rm live} = 200 \times n_{\rm dim}$ to ensure thorough exploration of the parameter space and consistent log Bayesian evidence across repeated runs. We note that the nested sampling occasionally struggled to locate high-likelihood peaks when constructing new bounds, which drove down the sampling efficiency from 0.30--0.45 to below 0.01. To remedy this problem, we tuned two parameters from the default \texttt{nautilus} configuration. First, we reduced the ellipsoid-splitting threshold factor from $\beta = 100$ to $\beta = 10$ to allow tighter ellipsoids around the sharp likelihood peaks, thereby improving the sampling efficiency (see \citealt{Lange_2023} for details). Second, we adopted a slightly more complex neural network setup by doubling the number of neurons in each hidden layer from (100, 50, 20) to (200, 100, 40). This modification improved the network’s ability to estimate the likelihood surface and accelerated the inference process by reducing the number of computationally expensive likelihood evaluations. We emphasize that this issue affected only a small number of bounds, where the sampler struggle to converge toward the narrow set of TTV solutions consistent with the extremely precise JWST transit times.

The fixed-inclination model yielded the highest Bayesian evidence ($\Delta \ln \mathcal{Z} = 5.0$ compared to free inclinations), suggesting the added complexity of allowing $i_{\rm e}$ and $i_{\rm f}$ to vary was not statistically justified. From the free-inclination fit, we report 2$\sigma$ lower limits of $i_{\rm e} > 80^\circ$ and $i_{\rm f} > 74^\circ$. These lower limits arise because the RVs primarily constrain $K_{\rm e}$ and $K_{\rm f}$, and lower inclinations imply higher true masses for planets e and f, which in turn produce stronger TTVs incompatible with the data. We adopt the prefered fixed-inclination solution and list the priors and posteriors (16$^{\rm th}$, 50$^{\rm th}$, and 84$^{\rm th}$ percentiles) in Table~\ref{table:full_model_params}.

The phase-folded Keplerian signals for all five planets are shown in Figure~\ref{fig:transit_rv}. Using the posterior samples on $P_{\rm e}$, $P_{\rm f}$ and on $t_{\rm 0,e}$, $t_{\rm 0,f}$ with realistic transit depth predictions, we show in the top panel of Figure~\ref{fig:transit_rv} the absence of transit of L~98-59\,e and f in the TESS data within one year of the RV observations, an interval chosen to minimize ephemeris uncertainty and with sufficient photometric coverage to detect transits if they occurred. A detailed summary of the RV component of the joint RV--TTV fit is presented in Figure~\ref{fig:full_rv_fit}, where each row highlights the signals (activity and Keplerian) identified in the data. The median TTV solution is plotted on top of the TTV measurements in Figure~\ref{fig:ttv}. With a precision of a few seconds in transit timing, JWST detects deviations from a pure 396-day sinusoidal function for L~98-59\,d, which can be attributed to the influence of the outer, non-transiting planets L~98-59\,e and f.

\begin{figure}
\vspace{1cm}
    \centering
\includegraphics[width=1\linewidth]{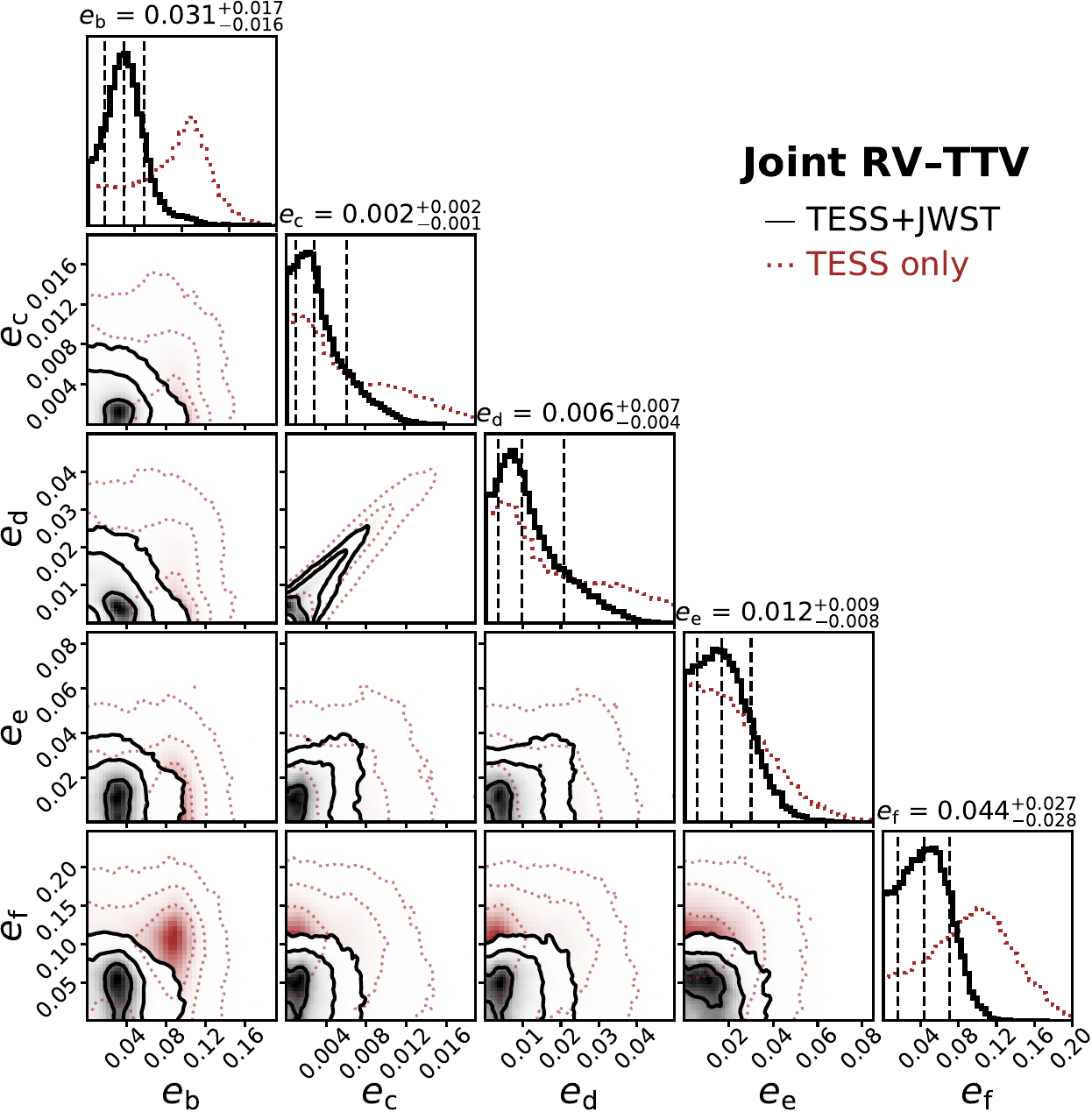}
\caption{Marginalised posterior distributions of the orbital eccentricities of the L~98-59 planets from the joint TTV--RV fit (fixed inclinations) using the full dataset (black) and TTVs from TESS only (red). The 2D posteriors show 1-, 2-, and 3$\sigma$ contours.}
    \label{fig:ecc_post}
\end{figure}

The posterior distributions on the orbital eccentricities are presented in Figure~\ref{fig:ecc_post}. To illustrate the relative contributions from TESS and JWST, we show in red the eccentricity constraints from a joint RV--TTV fit using TESS timing only. A comparison with an RV-only analysis is also presented in Figure~\ref{fig:dist_rv_vs_rvttv}. The long baseline provided by TESS is essential to constrain the 396-day periodicity. The JWST observations, on the other hand, narrowed the range of allowed TTV solutions, pushing the eccentricity of all planets toward zero. We also note an interesting feature in the $e_{\rm c}$--$e_{\rm d}$ 2D posteriors of Figure~\ref{fig:ecc_post}. The data disfavor to 2$\sigma$ the possibility of both planets having zero eccentricity. Instead, the data are consistent with either one of the two planets having some eccentricity, or both planets exhibiting highly correlated eccentricities.

\begin{deluxetable*}{lccccc}
\tablecaption{Main planetary parameters derived from the transit and joint RV--TTV analyses}
\tablehead{\colhead{Parameter}
& \colhead{L~98-59\,b} & \colhead{L~98-59\,c} & \colhead{L~98-59\,d} & \colhead{L~98-59\,e} & \colhead{L~98-59\,f}}
\startdata
\multicolumn{6}{c}{\textit{Orbital parameters}}\\[0.1cm]
$P$ (days) &  2.2531140 $\pm$ 4$\cdot10^{-7}$ & 3.6906764 $\pm$ 4$\cdot10^{-7}$ & 7.450729 $\pm$ 2$\cdot10^{-6}$ & 12.8278 $\pm$ 0.0018 & 23.064 $\pm$ 0.055 \\
$t_{\rm 0}$ (TBJD$^*$) & 1366.17056 $\pm$ 0.00024 & 1367.27303 $\pm$ 0.00015 & 1362.74002 $\pm$ 0.00025 & 1438.48 $\pm$ 0.15 & 1437.84 $\pm$ 0.75\\
$a$ (au) & 0.0223 $\pm$ 0.0007 & 0.0309 $\pm$ 0.0010 & 0.0494 $\pm$ 0.0016 & 0.0712 $\pm$ 0.0022 & 0.1052 $\pm$ 0.0033\\
$i$ ($^{\circ}$) & 88.08$^{+0.23}_{-0.20}$ & 88.88$^{+0.21}_{-0.17}$ & 88.44$^{+0.05}_{-0.05}$ & $>80$$^{\dagger}$ & $>74$$^{\dagger}$\\
$e$ & 0.031$^{+0.017}_{-0.016}$ & 0.002$^{+0.002}_{-0.001}$ & 0.006$^{+0.007}_{-0.004}$ & 0.012$^{+0.009}_{-0.008}$ & 0.044$^{+0.027}_{-0.028}$\\[0.15cm]
\multicolumn{6}{c}{\textit{Transit parameters}}\\[0.1cm]
$b$ & 0.51$^{+0.04}_{-0.05}$ & 0.41$^{+0.05}_{-0.07}$ & 0.921$^{+0.007}_{-0.007}$ & -- & --\\
$\delta$ (ppm) & 589 $\pm$ 14 & 1486 $\pm$ 30 & 1445 $\pm$ 47 & -- & -- \\
$t_{14}$ (hours) & 1.01 $\pm$ 0.03 & 1.28 $\pm$ 0.04 & 0.84 $\pm$ 0.04 & -- & -- \\[0.15cm]
\multicolumn{6}{c}{\textit{Physical parameters}}\\[0.1cm]
$R_{\rm p}$ (R$_{\oplus}$) & 0.837 $\pm$ 0.019 & 1.329 $\pm$ 0.029 & 1.627 $\pm$ 0.041 & 1.42$^{+0.38}_{-0.08}$$^{\times}$ & 1.43$^{+0.38}_{-0.10}$$^{\times}$\\
$M_{\rm p}$ (M$_{\oplus}$) & 0.46 $\pm$ 0.11 & 2.00 $\pm$ 0.13 & 1.64 $\pm$ 0.07 & 2.82 $\pm$ 0.19$^{\raisebox{0.5ex}{\scalebox{0.55}{$\square$}}}$ & 2.80 $\pm$ 0.30$^{\raisebox{0.5ex}{\scalebox{0.55}{$\square$}}}$\\
$\rho$ (g\,$\cdot$\,cm$^{-3}$) & 4.3$^{+1.1}_{-1.0}$ & 4.7$^{+0.5}_{-0.4}$ & 2.2$^{+0.2}_{-0.2}$ & 5.2$^{+1.3}_{-2.6}$ & 5.0$^{+1.7}_{-2.4}$\\
$S$ (S$_{\oplus}$) & 24.5 $\pm$ 2.5 & 12.7 $\pm$ 1.3 & 4.97 $\pm$ 0.52 & 2.40 $\pm$ 0.25 & 1.10 $\pm$ 0.11\\
$T_{\textrm{eq},A_{\rm B} = 0}$ (K) & 620 $\pm$ 13 & 526 $\pm$ 11 & 416 $\pm$ 9 & 347 $\pm$ 7 & 285 $\pm$ 6\\
\enddata
\tablecomments{$^*\textrm{TBJD} = \textrm{BJD} - 2\,457\,000$. $^{\dagger}$2$\sigma$ lower limit with a maximum inclination for planet e of 88.82$^{\circ}$ and for planet f of 89.20$^{\circ}$ constrained by the absence of transit. $^{\times}$Radius prediction from empirical mass--radius relation \citep{Parviainen_2024}. $^{\raisebox{0.5ex}{\scalebox{0.55}{$\square$}}}$Minimum mass $M_{\rm p} \sin i$.}
\label{table:derivedparams}
\end{deluxetable*}

\section{Discussion} \label{sec:discussion}

\subsection{Updating the Parameters of the L~98-59 Planets}

The combined results from our transit and joint RV--TTV analyses are summarized in Table~\ref{table:derivedparams}. We revise the masses and radii of L~98-59\,b, c, and d to 0.46 $\pm$ 0.11\,M$_{\oplus}$ and 0.837 $\pm$ 0.019\,R$_{\oplus}$, 2.00 $\pm$ 0.13\,M$_{\oplus}$ and \hbox{1.329 $\pm$ 0.029\,R$_{\oplus}$}, and 1.64 $\pm$ 0.07\,M$_{\oplus}$ and 1.627 $\pm$ 0.041\,R$_{\oplus}$, respectively. For the non-transiting planets L~98-59\,e and f, we measure minimum masses of 2.82 $\pm$ 0.19\,M$_{\oplus}$ and 2.80 $\pm$ 0.30\,M$_{\oplus}$. The absence of stronger TTVs observed in the system constrains the true masses of planets e and f to be at most 3.18\,M$_{\oplus}$ and 3.28\,M$_{\oplus}$ at 2$\sigma$ (free-inclination model), placing both planets in the super-Earth mass regime.

Our new mass and radius measurements reduce the relative uncertainty by about 50\% for all planets compared to the latest work (\citetalias{Demangeon_2021}). One notable improvement is the tighter orbital eccentricity constraints. Taking the example of L~98-59\,c, \citetalias{Demangeon_2021} measured \hbox{$e_{\rm c} = 0.103^{+0.045}_{-0.058}$} from RV data only, while our joint analysis of improved RV and TTVs yields $e_{\rm c} = 0.002^{+0.002}_{-0.001}$. This work demonstrates that long-term transit monitoring from TESS coupled with extreme precision transit timing from JWST atmospheric work can be used to constrain orbital eccentricity in compact systems, even with TTV signals below a few minutes.

\begin{figure}[ht!]
\centering
\includegraphics[width=1\linewidth]{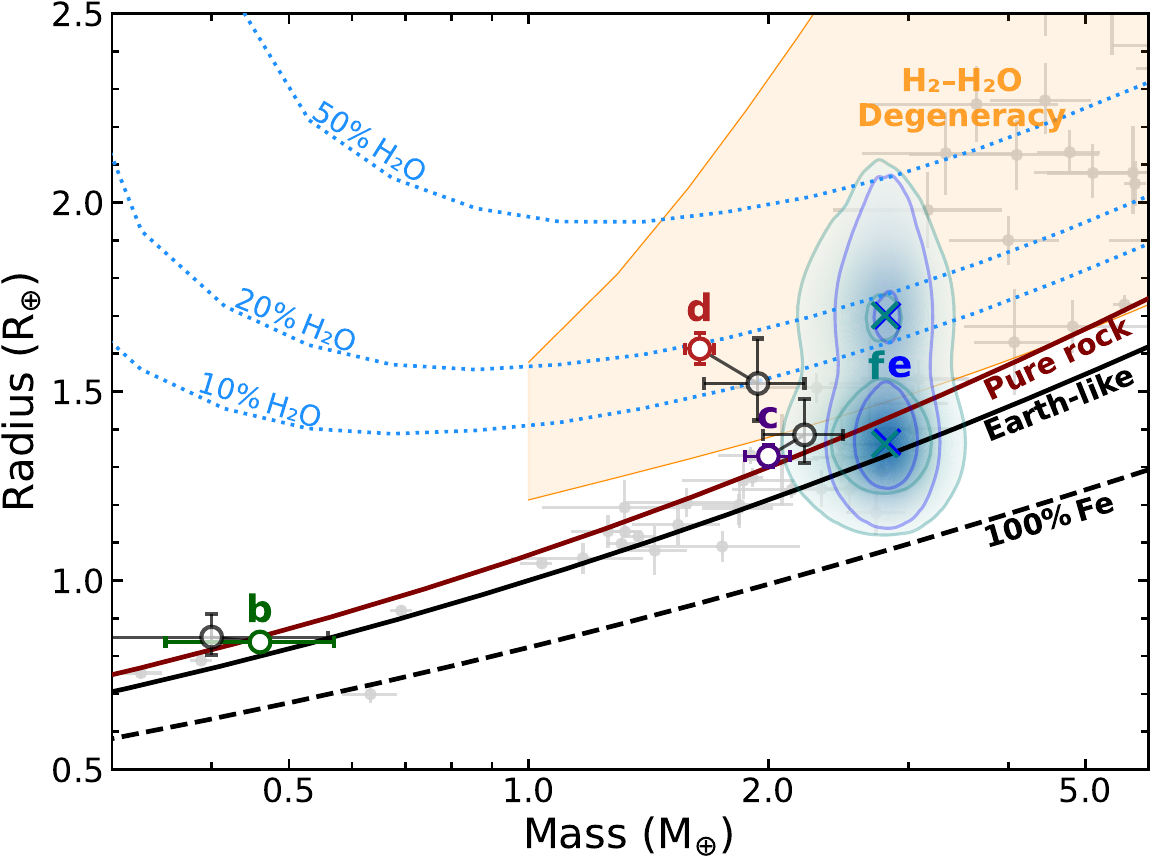}
\caption{Mass--radius constraints on the L~98-59 planets (b: green, c: purple, d: red, e: blue, f: teal) with other exoplanets around M dwarfs in the background (gray points). The previous values from \citetalias{Demangeon_2021} are shown in black attached to revised measurements. For L~98-59\,e and f, 1- and 2-$\sigma$ confidence contours are shown with the radius estimated using \texttt{spright} \citep{Parviainen_2024}  and with a cross symbol showing the two modes (rocky and volatile-rich). Rocky composition curves from \cite{Zeng_2019} are drawn from pure iron to pure silicate rocks. Water-rich interior models at $T = 400$\,K from \cite{Aguichine_2021} are shown with dotted blue curves. The orange region delimits a degeneracy (H$_2$- or H$_2$O-rich) in composition \citep{Rogers_2023}.}
\label{fig:mr}
\end{figure}

The updated masses and radii of the L~98-59's planets are plotted in a $M$--$R$ diagram with different pure composition curves in Figure~\ref{fig:mr}. For L~98-59\,e and f, we used \texttt{spright} \citep{Parviainen_2024}, a Bayesian radius--density--mass relation for small planets, to predict planetary radii of 1.42$^{+0.38}_{-0.08}$\,R$_{\oplus}$ and 1.43$^{+0.38}_{-0.10}$\,R$_{\oplus}$, respectively, given their minimum masses. The exoplanet population around M dwarfs with relative mass and radius precision below 30\% is also shown in the background of Figure~\ref{fig:mr} using data from the NASA Exoplanet Archive. The orange region corresponds to the interval of plausible H/He-rich sub-Neptunes after 5\,Gyr of thermal evolution and atmospheric photoevaporation \citep{Rogers_2023}. This region is also consistent with migrated steam worlds (H$_2$O-rich) according to the synthetic population of water worlds by \cite{Burn_2024}.  

The planets around L~98-59 are seemingly showing diverse compositions, which we explore in more details in the next Section~\ref{sec:internal_structure}. L~98-59\,b is a sub-Earth planet with a density consistent with an Earth-like interior, in accordance with previous results from \citetalias{Demangeon_2021}. L~98-59\,c is particularly interesting because our revision of the mass and radius, consistent within 1$\sigma$ with \citetalias{Demangeon_2021}, confirms that the planet is underdense compared to the Earth-like track. An iron-depleted rocky interior or a thick volatile-rich envelope (e.g., steam) are two possibilities to explain the density of planet c. The main difference from \citetalias{Demangeon_2021} concerns L~98-59\,d, for which we measure a lower density by 1.5$\sigma$. This places the planet in a strongly degenerate region of the mass--radius diagram, where its composition could be equally explained by either a volatile-rich (e.g., H$_2$O) or gas-rich (H$_2$) envelope on top of a rocky interior. In Figure~\ref{fig:mr}, no other M~dwarfs planets are found near L~98-59\,d, making it the 1.6-R$_{\oplus}$ planet with the lowest density ($\rho = 2.2 \pm 0.2$\,g\,$\cdot$\,cm$^{-3}$) currently known. 

The masses and predicted radii for the two non-transiting planets e and f encompass two modes: one corresponding to a rocky composition with a peak at $R_{\rm p} \approx 1.36$\,R$_{\oplus}$ and another indicative of a volatile-rich nature near $R_{\rm p} \approx 1.69$\,R$_{\oplus}$. However, the low density of L~98-59\,d suggesting a significant amounts of water/gas accreted during formation, may point to a similar composition for the outer planets. The second mode at a larger radius appears as a more likely scenario for L~98-59\,e and f.

\subsection{Internal Structure Analysis} \label{sec:internal_structure}

\begin{figure*}[ht!]
\centering
\minipage{0.25\textwidth}
\vspace{-1.9cm}
\includegraphics[width=1\linewidth]{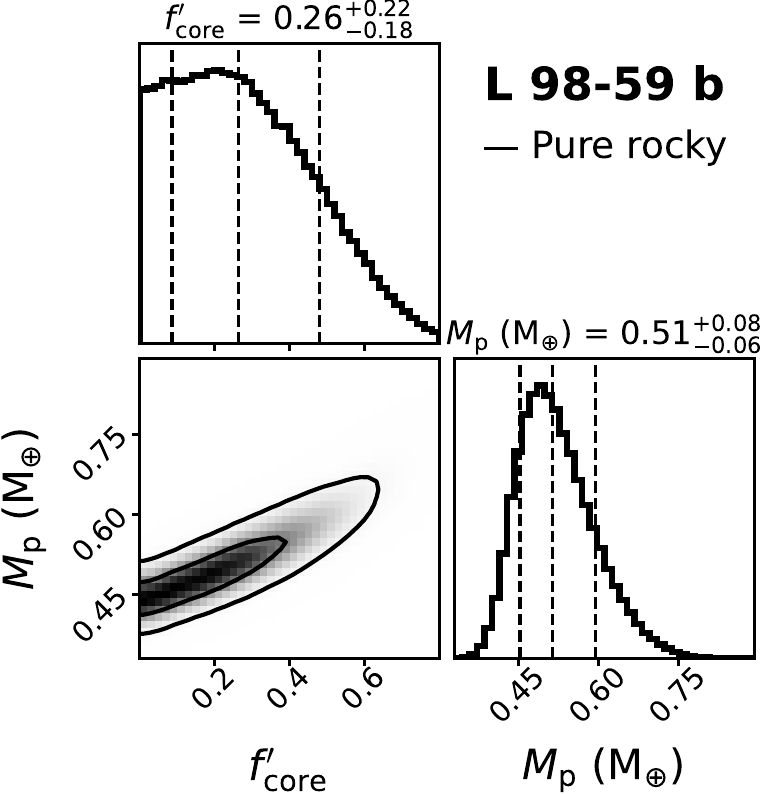}
\endminipage
\hfill
\minipage{0.36\textwidth}
\includegraphics[width=1\linewidth]{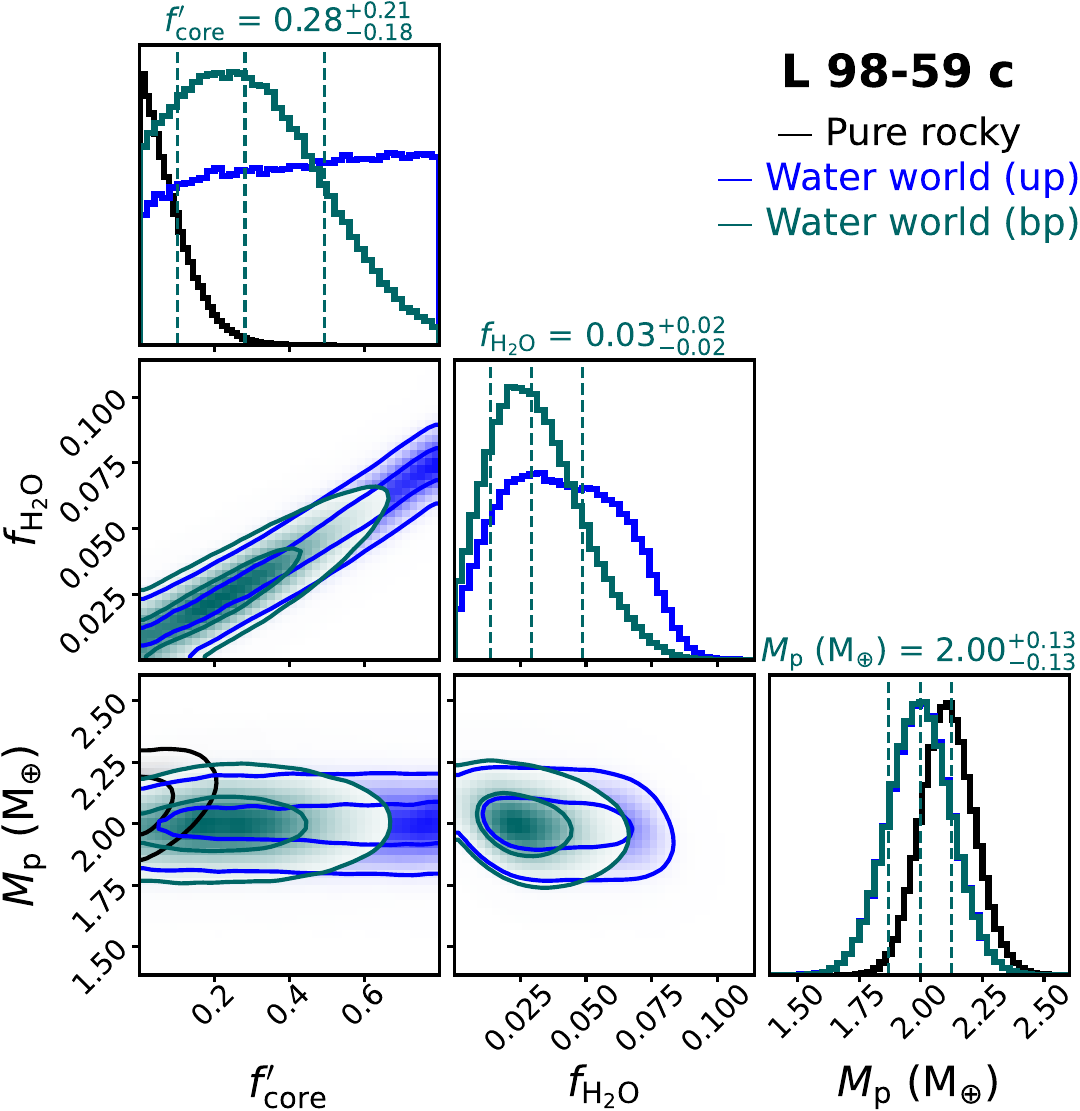}
\endminipage
\hfill
\minipage{0.36\textwidth}
\includegraphics[width=1\linewidth]{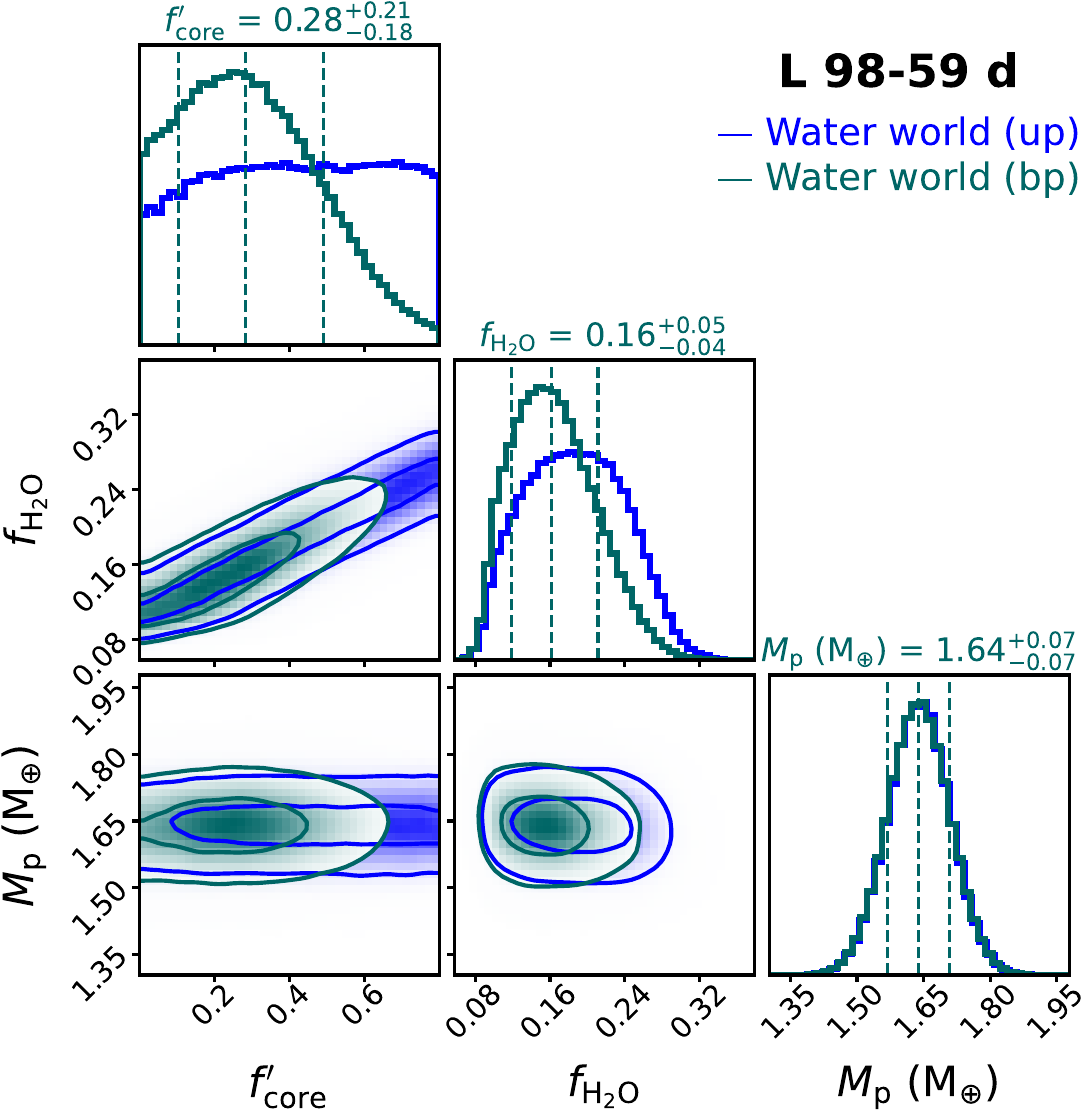}
\endminipage
\caption{Marginalised 1D and 2D posterior distributions of internal structure parameters for L~98-59\,b (left), c (center), and d (right) derived with \texttt{smint} \citep{Piaulet_2023}. The 2D posteriors display the 1- and 2-$\sigma$ contours. The analysis constrains the core- and water-mass fractions ($f^{\prime}_{\rm core}$, $f_{\rm H_{2}O}$) using priors on planetary mass ($M_{\rm p}$) and irradiation temperature ($T_{\rm irr}$; not shown), while requiring consistency with the observed planetary radius ($R_{\rm p}$). The sub-Earth L~98-59\,b is assumed to be purely rocky ($f_{\rm H_{2}O} = 0$). We explored a pure rocky model for L~98-59\,c and water-rich models for L~98-59\,c and d with either a uniform prior on $f^{\prime}_{\rm core}$ (`up') or using $f^{\prime}_{\rm core}$ from planet b as a prior (`bp'). L~98-59\,b has a $f^{\prime}_{\rm core}$ consistent with a wide range of iron content in the interior (from coreless to super-Mercury), with a mode close to Earth's value ($f^{\prime}_{\rm core} = 0.325$, \citealt{Wang_2018}). L~98-59\,c likely has a small $f_{\rm H_{2}O} \sim 0.03$ to explain its mass and radius. For L~98-59\,d, we infer a statistically significant $f_{\rm H_{2}O}$ = 0.16--0.18 irrespective of the choice of prior.}
\label{fig:interior}
\end{figure*}

We focus on the transiting planets L~98-59\,b, c, and d with exquisite precision of 3\% for the radius and 4--24\% for the mass to constrain their core-mass fraction ($f_{\rm core}$) and their water-mass fraction ($f_{\rm H_{2}O}$). Transmission spectroscopy with HST/WFC3 and JWST/NIRSpec ruled out clear hydrogen and helium atmospheres for the three planets \citep{Damiano_2022, Barclay_2023, Zhou_2023, Scarsdale_2024, Gressier_2024, Banerjee_2024, Bello-Arufe_2025}. Due to their proximity to L~98-59, the planets have been subjected to substantial cumulative XUV radiation over the lifetime of the system ($\sim$5\,Gyr), leading to atmospheric escape and efficient hydrogen loss \citep{Fromont_2024}. Based on these previous results and their positions in Figure~\ref{fig:mr}, we assume an Earth-like composition for L~98-59\,b and explore the possibility that L~98-59\,c and d are steam worlds with a significant water-mass fractions ($f_{\rm H_{2}O} \gtrsim$1\%).

Tentative detections of sulfur species on L~98-59\,b and d suggest atmospheric mean molecular weights ($\mu$) around 64 (pure SO$_2$) and 10 (H$_2$S, SO$_2$ with H$_2$ as background gas), respectively for the two planets, based on the maximum a posteriori retrieval results of \cite{Bello-Arufe_2025, Gressier_2024, Banerjee_2024}. We neglected an atmosphere for L~98-59\,b but have applied a fixed radius correction associated with a potential atmosphere for L~98-59\,d equivalent to five atmospheric scale heights ($\sim$300 km), representing approximately 3\% of the radius.

We made use of the \texttt{smint} code \citep{Piaulet_2023} that couples an interpolator of planetary interior models of rocky \citep{Zeng_2016} and irradiated ocean planets \citep{Aguichine_2021} with the MCMC sampler \texttt{emcee}. The planets are modeled with three layers: an iron core, a silicate rock mantle, and a water layer on top. The water envelope is expected to be in vapor/supercritical phase as the planets all have a $T_{\rm eq} > 400$\,K \citep{Aguichine_2021, Turbet_2023}. The planetary radius produced by the models for a given mass is compared to the observed radius. The mass, treated as a model parameter, is constrained by a Gaussian prior based on our RV--TTV fit (Table~\ref{table:derivedparams}). For the water-rich models, we apply a Gaussian prior on the irradiation temperature $T_{\rm irr}$ by taking $T_{\rm eq}$ and its uncertainty from Table~\ref{table:derivedparams} calculated with a Bond albedo $A_{\rm B} = 0$. In \texttt{smint}, the core-mass fraction is sampled using $f^{\prime}_{\rm core}$ that denotes the iron content (by mass) in the core+mantle interior only. The total planetary core-mass fraction is given by $f_{\rm core} = f^{\prime}_{\rm core} (1 - f_{\rm H_{2}O})$. For a pure rocky planet ($f_{\rm H_{2}O} = 0$), $f^{\prime}_{\rm core} = f_{\rm core}$. We limit $f^{\prime}_{\rm core}$ to 0.8 for the maximum iron enrichment \citep{Plotnykov_2020}.

The following interior scenarios were considered:
\begin{itemize}
\setlength\itemsep{0em}
    \item A rocky planet (free $f^{\prime}_{\rm core}$, $f_{\rm H_{2}O} = 0$) for L~98-59\,b and c
    \item A water world with uniform prior on $f^{\prime}_{\rm core}$ of $\mathcal{U} \left(0, 0.8\right)$ for L~98-59\,c and d, i.e., the uniform prior case `up'
    \item A water world with an informed prior on $f^{\prime}_{\rm core}$ using the core-mass fraction of planet b for L~98-59\,c and d, i.e., the b as a prior case `bp'
\end{itemize}
For both the `up' and `bp' models, the prior on $f_{\rm H_{2}O}$ is uniform between 0 and 1. Our MCMC analysis uses 100 walkers per model, running for 10\,000 steps with a burn-in of 2\,000. The posterior results of the interior modeling of L~98-59\,b, c, and d are summarized in Figure~\ref{fig:interior}. The posteriors on $T_{\rm irr}$ for the water-rich models (not shown in Fig.~\ref{fig:interior}) were equal to the priors.

L~98-59\,b, with about half the mass of the Earth (0.46 $\pm$ 0.11\,M$_{\oplus}$), is consistent with a $f^{\prime}_{\rm core} = 0.26^{+0.22}_{-0.18}$, spanning a range from a coreless planet to a super-Mercury (see Fig.~\ref{fig:interior}), with the mode of the distribution peaking near Earth's core-mass fraction  ($f_{\rm core} = 0.325 \pm 0.003$, \citealt{Wang_2018}). Under the hypothesis that all the planets around L~98-59 formed with a similar $f^{\prime}_{\rm core}$, a scenario supported both theoretically (e.g., \citealt{Bond_2010}, \citealt{Thiabaud_2015}, \citealt{Unterborn_2016}) and empirically (e.g., \citealt{Dorn_2017, Bonsor_2021, Brinkman_2024}), the inferred distribution on $f^{\prime}_{\rm core}$ from L~98-59\,b can serve as prior information for models incorporating a water-rich layer (see below, the `bp' case).

For planet c, the pure rocky planet scenario has a $f^{\prime}_{\rm core}$ that peaks at 0, indicating a coreless (iron-deficient) planet. This is also illustrated in Figure~\ref{fig:mr}: the mode of the mass--radius constraints for L~98-59\,c lies slightly above the pure Mg-Si rocky curve for planets entirely made of mantle material. This scenario cannot be firmly excluded with current density measurement, but it would require a lower $f^{\prime}_{\rm core}$ for planet c compared to planet b, with the mass $M_{\rm p}$ of planet c deviating at 1\,$\sigma$ away from the observed values (see Fig.~\ref{fig:interior}). Alternatively, the density of L~98-59\,c can be explained by a small $f_{\rm H_{2}O} \sim 0.03$ in both the `up' and `bp' cases. We infer a 2$\sigma$ upper limit of 0.08 for the fraction of water in L~98-59\,c (`up' model). Here, we modeled the water as a steam/super-critical envelope atop a rocky interior, but similar bulk densities can be obtained if the water is instead sequestered within the core and the mantle, as proposed by \citet{Luo_2024}.

The location of L~98-59\,d in the mass--radius diagram (Fig.~\ref{fig:mr}) makes it highly improbable that the planet is composed entirely of rocks and metals. Our water-rich analyses result in a statistically significant $f_{\rm H_{2}O} = 0.19^{+0.06}_{-0.06}$ for `up' and $f_{\rm H_{2}O} = 0.16^{+0.05}_{-0.04}$ for `bp' runs. These results suggest that L~98-59\,d is a strong water world candidate, comparable to several recent discoveries of similar planets around M dwarfs (e.g., \citealt{Luque_2021}, \citealt{Cadieux_2022}, \citealt{Luque_2022}, \citealt{Piaulet_2023}, \citealt{Cherubim_2023}, \citealt{Cadieux_2024a}, \citealt{Osborne_2024}). Water worlds are predicted by planet population synthesis models that incorporate the latest equation of state for water, accounting for all possible phases, and atmospheric mass loss due to photoevaporation after disk dispersal \citep{Venturini_2024}. In this paradigm, rocky super-Earths are formed \textit{in situ} inside the ice line, while sub-Neptunes are assembled \textit{ex situ} before migrating inward to appear today with an volatile-rich envelope inflated by high irradiation.

\subsection{Potential for Tidally-Induced Magma Oceans} \label{sec:volcanism}

In this section, we discuss the potential for subsurface magma oceans on L~98-59\,b and c driven by tidal heating. The planet L~98-59\,d is excluded from this analysis because of its substantial volatile mass fraction (e.g., H$_2$O) that could reach $\sim$20\%. We use the open-source code \texttt{melt}\footnote{\href{https://github.com/cpiaulet/melt}{\texttt{github.com/cpiaulet/melt}}} to compute tidal heating supported by a forced eccentricity. This code has previously been applied to assess the interior state of LP~791-18\,d \citep{peterson_temperate_2023} and follows a prescription originally developed to study tidal heating on Io \citep{fischer_thermal-orbital_1990,moore_tidal_2003,henning_tidally_2009,dobos_viscoelastic_2015}, as well as the {TRAPPIST-1} planets \citep{barr_interior_2018}. The energy balance calculations for tidally-heated rocky planets are detailed in Appendix~\ref{tidal_heating_methods} and briefly summarized here.

We compute the tidal energy dissipation flux $F_\mathrm{tidal}$ and the convective flux $F_\mathrm{conv}$ for a range of mantle temperatures $T_\mathrm{mantle}$ solving for the equilibrium where $F_\mathrm{conv} = F_\mathrm{tidal}$. An equilibrium at $T_\mathrm{mantle}$ higher than the solidus temperature of a pure rock composition \citep{solomatov_scaling_2000} suggests a partially molten mantle layer under the planet's surface. To account for model uncertainty, we find solutions for several values of rock viscosity parameterized by the melt fraction coefficient $B$ spanning a experimentally-constrained range between 10 and 40 (details in Appendix~\ref{tidal_heating_methods}). The model takes as input the planet mass, radius, orbital period, and orbital eccentricity taken from Table~\ref{table:derivedparams}. Here, we adopt the maximum a posteriori eccentricity value ($e_{\rm b} = 0.031$, $e_{\rm c} = 0.001$), as it provides a more conservative estimate compared to upper limits, since higher eccentricities generally yield higher mantle temperatures. 

\begin{figure}
\centering
\includegraphics[width=1\linewidth]{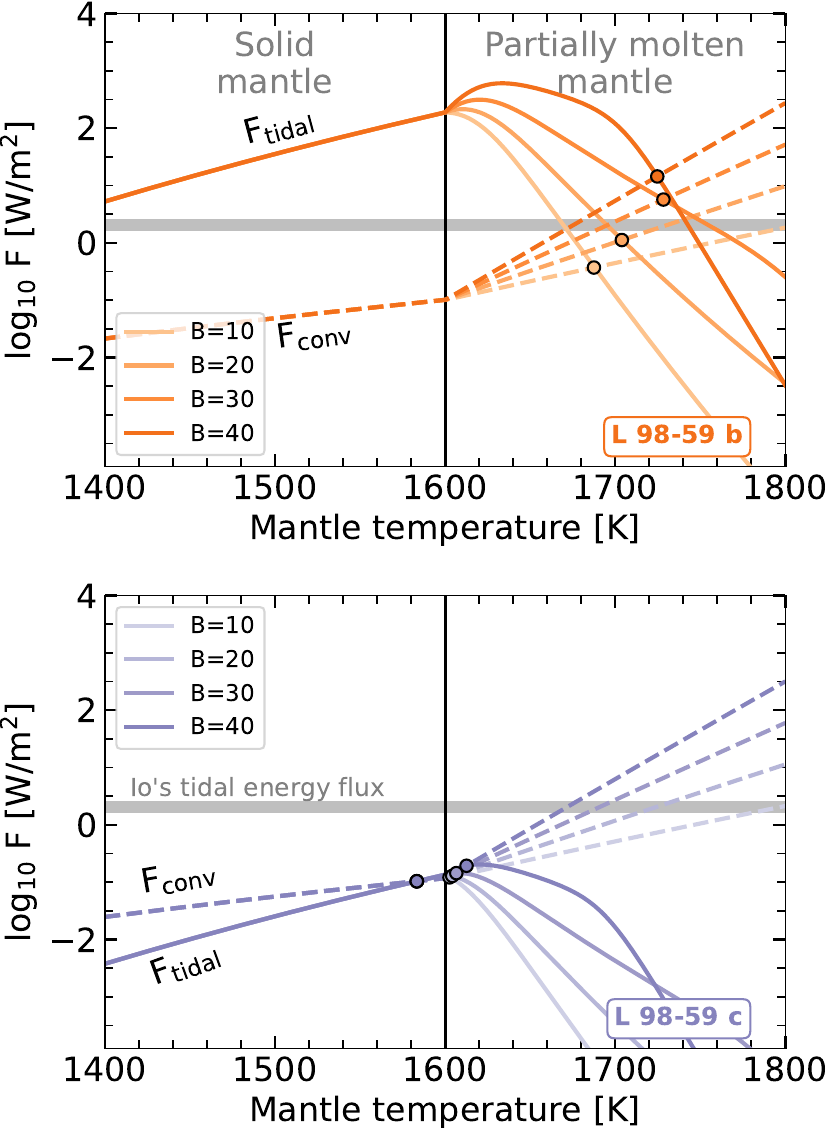}
\caption{\label{fig:energy_balance_tides} Internal energy balance inferred with \texttt{melt} \citep{peterson_temperate_2023} for L~98-59 b (top) and c (bottom) as a result of tidal heating. We show the tidal heating flux per unit surface area (solid lines) and the convective heat flux (dashed lines) as a function of mantle temperature, with the equilibrium points shown as circles, colored according to the value of the melt fraction coefficient $B$ adopted in the model (methods in Appendix~\ref{tidal_heating_methods}). The solidus temperature is shown as the vertical black line. The equilibrium happens beyond the solidus for L~98-59\,b, at similar heat fluxes to those inferred for Jupiter's moon Io, implying a potential molten magma layer under the planets' surfaces which could be conducive to active volcanic activity. The lower eccentricity-driven tidal flux for planet c results in equilibrium mantle temperatures consistent with a solidified mantle or marginally beyond the solidus. Under these conditions, we cannot conclude that L~98-59\,c has a partially molten mantle layer.}
\end{figure}

Beyond the solidus, our model considers only the mantle viscosity as a source of tidal heat flux. However, recent modeling demonstrated that the molten rock that acts as a fluid also contributes to the mantle heat flux \citep{farhat_tides_2025, Gkouvelis_2025}. This leads to an additional heat source that increases with the mantle temperature beyond the solidus, (while our model lead to a decreasing tidal response with increasing temperature, see Fig.~\ref{fig:energy_balance_tides}). The tidal fluxes we infer can therefore be considered as lower limits for L~98-59 b and c.

We find distinct thermal outcomes for planets b and c, as shown in Figure~\ref{fig:energy_balance_tides}. For L~98-59\,b, the stable equilibrium mantle temperature is in the range 1687--1724\,K. These temperatures lie beyond the solidus, which implies a partially molten magma layer (melt fraction between 22 and 31\%), which may be conducive to active volcanism. Specifically, the corresponding heat surface fluxes (0.4 to 14\,W/m$^2$) are about one order of magnitude larger than the Earth's surface heat flux \citep{davies_earths_2010}, and similar to those estimated for Io, with the range of values from \citealp{veeder_io_2012} also shown in Figure~\ref{fig:energy_balance_tides}.

In contrast, L~98-59\,c receives a much weaker tidal forcing, with $F_{\rm tidal}$ roughly 1800 times lower than for planet b, primarily due to its smaller eccentricity. The energy balance calculations yield two equilibrium points on either side of the solidus temperature (see \citealt{barr_interior_2018} for a detailed description of the two equilibrium states). One solution corresponds to a solid mantle but is considered unstable, as the convective heat flux cannot balance $F_{\rm tidal}$ if the latter increases. The second equilibrium point is stable, but occurs at a temperature only marginally exceeding the solidus ($T_\mathrm{mantle} \approx 1610$\,K). These results are highly sensitive to eccentricity: adopting $e_{\rm c} = 0.004$ (84$^{\rm th}$ percentile) in a separate simulation (not shown) yields a single stable equilibrium past the solidus, producing a partially molten layer with a melt fraction 9--16\%. Independent simulations by \citet{Nicholls_2025} similarly indicate that the transition between sustaining an eccentricity-driven subsurface magma ocean and complete solidification occurs at $e_{\rm c} \approx 0.003$. The current uncertainties in both eccentricity and interior modeling prevent us from confidently establishing whether L~98-59\,c hosts a partially molten mantle.

The main conclusion of this exercise is to confirm the plausibility of strong geologic activity on L~98-59\,b and c accounting for our updated eccentricity constraints, a result inline with the previous calculations of \cite{Quick_2020, Seligman_2024, Bello-Arufe_2025, Nicholls_2025}. We reiterate that this conclusion relies on the assumption that both planets are purely rocky, neglecting any volatile layer. While this is a reasonable approximation for L~98-59\,b, the lower density of L~98-59\,c consistent with 3--5\% H$_2$O by mass could affect the energy balance calculation presented here.

\subsection{A New Habitable Zone Exoplanet in the Vicinity of the Earth} \label{sec:L9859f}

The newly confirmed exoplanet L~98-59\,f with a period of $23.064 \pm 0.055$\,days and a minimal mass of 2.80 $\pm$ 0.30\,M$_{\oplus}$ orbits the Habitable Zone of L~98-59 according to \cite{Kopparapu_2013}. Our joint RV–TTV fit with free eccentricity yields $e_{\rm f} = 0.044^{+0.027}_{-0.028}$, although the data are equally consistent with a circular orbit. On average, L~98-59\,f receives an Earth-like bolometric instellation of $1.10 \pm 0.11$\,S$_{\oplus}$, with this median value potentially varying between 1.01 and 1.20\,S$_{\oplus}$ over one orbit if the planet follows the eccentric solution. Because its M3V host star emits proportionally more infrared light than a Sun-like star and atmospheric greenhouse gases (e.g., H$_2$O, CO$_2$) absorb infrared radiation efficiently, the runaway greenhouse threshold shifts to lower bolometric flux levels for planets around M dwarfs, pushing the inner edge of the Habitable Zone farther from the star. For this reason, the super-Earth L~98-59\,f can be considered in the `optimistic' Habitable Zone as it would require particular conditions (e.g., cooling effects from reflective clouds) to maintain liquid water on its surface. The exact definition of an Habitable Zone remains imprecise, as it depends on many factors including the planetary mass \citep{Kopparapu_2014}, 3D effects from tidal locking \citep{Yang_2013, Kopparapu_2016}, rotation rate \citep{Yang_2014}, or initial conditions conducive to formation of oceans early on \citep{Turbet_2023}. A precise definition may only emerge once surface liquid water is inferred on multiple planets through consistency between atmospheric observables (e.g., transmission and emission spectra, phase curves) and 3D global climate models.

At 10.61\,pc, L~98-59 currently ranks as the 12$^{\rm th}$ closest system with an exoplanet in the Habitable Zone according to the NASA Exoplanet Archive, following well-known nearby multiplanetary systems such as Proxima Centauri \citep{Anglada-Escude_2016}, Teegarden’s Star \citep{Zechmeister_2019}, and GJ~1002 \citep{Suarez-Mascarno_2023}, with the particularity of being the only system with inner transiting planets. The proximity to L~98-59 means the orbital distance of planet f ($0.1052 \pm 0.0033$\,au) equals to a projected separation of 9.9 $\pm$ 0.3\,mas --- or a maximum of 10.4 $\pm$ 0.3\,mas at apoastron if an eccentric orbit is robustly confirmed. This makes L~98-59\,f barely inaccessible for reflected-light atmospheric characterization with ANDES \citep{Marconi_2024} on the upcoming ESO/ELT, given the minimal angular separations requirements (2$\lambda/D$) of 10.6\,mas at 1000\,nm and 16.92\,mas at 1600\,nm \citep{Palle_2023}. Follow-up study of the atmosphere of this Habitable Zone super-Earth may thus be only feasible with future space-based facilities such as the Habitable World Observatory \citep{Gaudi_2020} or the LIFE interferometric mission \citep{Quanz_2022}.

\subsection{Secondary Eclipse Ephemerides} \label{sec:secondary_eclipse}

Secondary eclipse observations of the inner planets in the L 98-59 system, combined with transmission spectroscopy, offer the most effective approach for detecting atmospheres and constraining their dayside temperatures, which may be influenced by tidal heating in the planets' interiors. An important consideration when observing secondary eclipses is ensuring a sufficiently long baseline, particularly when the orbital eccentricity is unconstrained. A faint eclipse signal could either indicate a genuinely small eclipse depth or that the eclipse occurred outside the observation window due to the incorrect assumption of a circular orbit.

Here, the results of the RV--TTV fit are used to estimate future timing of secondary eclipses for L~98-59\,b, c, and d. The time of secondary eclipse is given by:
\begin{equation}
\setlength\itemsep{0em}
    t_{\textrm{sec}, k} = t_{\textrm{c}, k} + P_k / 2 + \underbrace{\frac{2 e_k \cos \omega_k}{\pi} P_k}_{\delta_{\textrm{ecc}, k}} + \delta_{\textrm{travel}, k} + \delta_{\textrm{ETV}, k}
    \label{eq:eclipse_timing}
\end{equation}
where $t_{\textrm{c}, k} = t_{0, k} + EP_k$ is the transit timing (i.e., inferior conjunction) for constant period $P_k$ at epoch number $E$, $\delta_{\textrm{ecc}, k}$ is a deviation term proportional to $P_k$ for an elliptical orbit \citep{Winn_2010}, $\delta_{\textrm{travel} , k} \approx 2a_k/c$ is a delay to account for light travel time \citep{Huber_2017} and $\delta_{\textrm{ETV}, k}$ is an extra time difference for eclipse timing variations (ETV) in compact system. Entering the posteriors of the joint RV--TTV fit into Equation~\ref{eq:eclipse_timing}, we infer $\delta_{\rm ecc,b} = -6^{+31}_{-40}$\,min, $\delta_{\rm ecc,c} = -1^{+4}_{-5}$\,min, and $\delta_{\rm ecc,d} = 25^{+27}_{-26}$\,min. The $\delta_{\textrm{travel} , k}$ for planet b, c, and d are approximately 22\,s, 31\,s, and 49\,s, respectively. Because we expect ETV to be similar to TTV, we assume deviations of $\delta_{\rm ETV,b} = 0$\,min, $\delta_{\rm ETV,c} = 0\pm2$\,min, and $\delta_{\rm ETV,d} = 0\pm2$\,min, where the uncertainty is an approximate 1\,$\sigma$ dispersion (Fig.~\ref{fig:ttv}). 

We report the following relations to plan future eclipse observations for the three inner planets around L~98-59:
\begin{equation}
\begin{split}
    & (t_{\rm sec,b} - 2460500) = (0.9919 + EP_{\rm b}) \pm (0.0249) \ \textrm{BJD}\\
    & (t_{\rm sec,c} - 2460500) = (2.3290 + EP_{\rm c}) \pm (0.0034) \ \textrm{BJD}\\
    & (t_{\rm sec,d} - 2460500) = (4.8425 + EP_{\rm d}) \pm (0.0184) \ \textrm{BJD}
\end{split}
\end{equation}
with $E$ the epoch number and $P_{\rm b} = 2.2531140$\,days, $P_{\rm c} = 3.6906764$\,days, and $P_{\rm d} = 7.450729$\,days. The uncertainty in eclipse timing due to errors in the orbital period (scaling with $E$) is several orders of magnitude smaller than the uncertainty from $\delta_{\rm ecc}$ and has been neglected in the relations above. The $e_k \cos \omega_k$ constraints from our RV--TTV fit provide good timing precision of about half an hour for L~98-59\,b and d, and of only a few minutes for L~98-59\,c, allowing the scheduling of secondary eclipse observations with JWST. The relations above can serve as priors on the secondary eclipse mid-time in the analysis of upcoming JWST observations.

\section{Summary \& Conclusion} \label{sec:conclusions}

In this paper, we revisited the multiplanetary system L~98-59 with four known small exoplanets, including three transiting the M3V host. We modeled the transits from the most up-to-date TESS data, including 16 new sectors, and performed a line-by-line analysis of archival ESPRESSO and HARPS data to enhance the radial velocity (RV) precision. A search for transit timing variations (TTV) of L~98-59\,c and L~98-59\,d over six years of TESS data revealed a 396-day periodicity with an amplitude of a few minutes, mainly caused by the mutual gravitational interaction between the two planets near a 2:1 orbital resonance. A joint analysis of the ESPRESSO+HARPS RVs along with the TTVs from TESS and from recent JWST observations delivers the most precise constraints yet on the global architecture of the L~98-59 system. The key findings of this study are outlined below.

{
\begin{itemize}
    \item We report updated planetary radii and masses of 0.837\,$\pm$\,0.019\,R$_{\oplus}$ and 0.46\,$\pm$\,0.11\,M$_{\oplus}$ for L~98-59\,b, 1.329\,$\pm$\,0.029\,R$_{\oplus}$ and 2.00\,$\pm$\,0.13\,M$_{\oplus}$ for L~98-59\,c, 1.627\,$\pm$\,0.040\,R$_{\oplus}$ and 1.64\,$\pm$\,0.07\,M$_{\oplus}$ for L~98-59\,d, and a minimal mass of 2.82\,$\pm$\,0.19\,M$_{\oplus}$ for the non-transiting L~98-59\,e.
    \item The rotation period of the star L~98-59 is constrained for the first time ($P_{\rm rot } = 76.7 \pm 1.5$\,days) using a novel spectroscopic activity indicator, the \dtemp \ \citep{Artigau_2024}.
    \item Using a Gaussian process informed by \dtemp \ to model stellar activity in the RV data, we robustly confirm (5.1$\sigma$) the non-transiting exoplanet L~98-59\,f, first announced as a candidate in \cite{Demangeon_2021}, with a minimum mass of 2.80\,$\pm$\,0.30\,M$_{\oplus}$.
    \item This new planet f is the outermost in the L~98-59 system, with an orbital period of 23.064\,$\pm$\,0.055\,days and a distance of 0.1052\,$\pm$\,0.0033\,au, placing it in the optimistic Habitable Zone with an instellation level similar to Earth (1.10\,$\pm$\,0.12\,S$_{\oplus}$).
    \item The absence of stronger TTVs recorded by TESS and JWST constrains the eccentricities of all planet to near-circular: 0.031$^{+0.017}_{-0.016}$, 0.002$^{+0.002}_{-0.001}$, 0.006$^{+0.007}_{-0.004}$, 0.012$^{+0.009}_{-0.008}$, and 0.044$^{+0.027}_{-0.028}$, respectively for planets b to f. Additionally, a TTV model assuming L~98-59\,e and f are co-planar with the inner planets is statistically preferred. The free-inclination fit yields 2$\sigma$ lower limits of $i_{\rm e} > 80^{\circ}$ and $i_{\rm f} > 74^{\circ}$.
    \item The updated orbital parameters now constrain the timing of future secondary eclipse observations to within approximately 30\,min for L~98-59\,b and d, and just 5\,min for L~98-59\,c.
    \item We modeled the internal structures of the inner planets. The sub-Earth L~98-59\,b most likely has an Earth-like composition, while the super-Earths L~98-59 c and d may host H$_2$O steam/supercritical envelopes to explain their lower densities. Using planet b’s core-mass fraction as a prior ($f_{\rm core}=0.26^{+0.22}_{-0.18}$), our analysis yields water-mass fractions $f_{\rm H_{2}O} = 0.03^{+0.02}_{-0.02}$ for c and $f_{\rm H_{2}O} = 0.16^{+0.05}_{-0.04}$ for d.
    \item The apparent compositional gradient in the L~98-59 planets, with $f_{\rm H_{2}O}$ increasing with orbital distance, suggests formation beyond the ice line, where water ice is more abundant, followed by inward migration to their current orbits.
    \item The compact nature of the L~98-59 system is conducive to tidal heating in the planetary interiors from nonzero eccentricity. Tidal heating calculations in L~98-59\,b and c reveal potential for subsurface magma oceans and strong volcanism for planet b, with surface heat fluxes comparable to those of Jupiter's moon Io.
\end{itemize}
}

The atmospheric characterization of the L~98-59 planets is already underway with JWST, using both transmission and emission spectroscopy. Our revised planetary and orbital parameters will be essential to accurately interpret the existing and upcoming observations. L~98-59 is emerging as a benchmark system for super-Earths around M dwarfs, offering a rare opportunity to constrain interior and atmospheric properties for planets within a single planetary system. Such comprehensive characterization is key to advancing our understanding of planet formation and evolution around low-mass stars.

\fakesection{Acknowledgments}
\vspace{1cm}

We thank the anonymous referee whose constructive report improved the clarity and quality of the manuscript.

This study uses public ESPRESSO data under GTO program IDs 1102.C-0744, 1102.C-0958, and 1104.C-0350 (PI: F.\ Pepe) and HARPS data taken under program IDs 1102.C-0339 (PI: X. Bonfils), 0102.C-0525 (PI: M., Diaz), and 0102.D-0483 (PI: Z. Berdinas).

This work is partly supported by the Natural Science and Engineering Research Council of Canada, the Canadian Space Agency and the Trottier Family Foundation through their support of the Trottier Institute for Research on Exoplanets (IREx). This work benefited from support of the Fonds de recherche du Québec – Nature et technologies (FRQNT), through the Center for Research in Astrophysics of Quebec.

We acknowledge the use of public TESS Alert data from pipelines at the TESS Science Office and at the TESS Science Processing Operations Center. Resources supporting this work were provided by the NASA High-End Computing (HEC) Program through the NASA Advanced Supercomputing (NAS) Division at Ames Research Center for the production of the SPOC data products. This paper includes data collected by the TESS mission that are publicly available from the Mikulski Archive for Space Telescopes (MAST).

This work is based on observations made with the NASA/ESA/CSA JWST under programs GTO1201 (PI: D.\ Lafrenière), GTO1224 (PI: S. Birkmann), GO2512 (PI: N. Batalha), and GO4098 (PI: B.\ Benneke).

A.L'H. acknowledges support from the FRQNT under file \#349961. C.P.-G acknowledges support from the E. Margaret Burbidge Prize Postdoctoral Fellowship from the Brinson Foundation.

\facilities{ESO/ESPRESSO, ESO/HARPS, NASA/TESS, NASA/JWST}

\software{\texttt{Astropy} \citep{Astropy_2018}; \texttt{batman} \citep{Kreidberg_2015}; \texttt{celerite2} (\citealt{celerite1_2017}; \citeyear{celerite2_2018}; \texttt{emcee} \citep{Foreman-Mackey_2013}; \texttt{george} \citep{Ambikasaran_2015}; \texttt{juliet} \citep{Espinoza_2019}; \texttt{LBL} \citep{Artigau_2022}; \texttt{matplotlib} \citep{Hunter_2007}; \texttt{melt} \citep{peterson_temperate_2023}; \texttt{nautilus} \citep{Lange_2023}; \texttt{NumPy} \citep{Harris_2020}; \texttt{radvel} \citep{Fulton_2018}; \texttt{SciPy} \citep{Virtanen_2020}; \texttt{smint} \citep{Piaulet_2023}; \texttt{spright} \citep{Parviainen_2024}; \texttt{TTVFast} \citep{Deck_2014}}

\bibliography{L9859}{}

\begin{thebibliography}{}
\expandafter\ifx\csname natexlab\endcsname\relax\def\natexlab#1{#1}\fi
\providecommand{\url}[1]{\href{#1}{#1}}
\providecommand{\dodoi}[1]{doi:~\href{http://doi.org/#1}{\nolinkurl{#1}}}
\providecommand{\doeprint}[1]{\href{http://ascl.net/#1}{\nolinkurl{http://ascl.net/#1}}}
\providecommand{\doarXiv}[1]{\href{https://arxiv.org/abs/#1}{\nolinkurl{https://arxiv.org/abs/#1}}}

\bibitem[{{Agol} \& {Fabrycky}(2018)}]{Agol_2018}
{Agol}, E., \& {Fabrycky}, D.~C. 2018, in Handbook of Exoplanets, ed. H.~J. {Deeg} \& J.~A. {Belmonte}, 7, \dodoi{10.1007/978-3-319-55333-7_7}

\bibitem[{{Agol} {et~al.}(2021){Agol}, {Dorn}, {Grimm}, {Turbet}, {Ducrot}, {Delrez}, {Gillon}, {Demory}, {Burdanov}, {Barkaoui}, {Benkhaldoun}, {Bolmont}, {Burgasser}, {Carey}, {de Wit}, {Fabrycky}, {Foreman-Mackey}, {Haldemann}, {Hernandez}, {Ingalls}, {Jehin}, {Langford}, {Leconte}, {Lederer}, {Luger}, {Malhotra}, {Meadows}, {Morris}, {Pozuelos}, {Queloz}, {Raymond}, {Selsis}, {Sestovic}, {Triaud}, \& {Van Grootel}}]{Agol_2021}
{Agol}, E., {Dorn}, C., {Grimm}, S.~L., {et~al.} 2021, Planet. Sci. J., 2, 1, \dodoi{10.3847/PSJ/abd022}

\bibitem[{{Aguichine} {et~al.}(2021){Aguichine}, {Mousis}, {Deleuil}, \& {Marcq}}]{Aguichine_2021}
{Aguichine}, A., {Mousis}, O., {Deleuil}, M., \& {Marcq}, E. 2021, \apj, 914, 84, \dodoi{10.3847/1538-4357/abfa99}

\bibitem[{{Aigrain} {et~al.}(2012){Aigrain}, {Pont}, \& {Zucker}}]{Aigrain_2012}
{Aigrain}, S., {Pont}, F., \& {Zucker}, S. 2012, \mnras, 419, 3147, \dodoi{10.1111/j.1365-2966.2011.19960.x}

\bibitem[{{Akeson} {et~al.}(2013){Akeson}, {Chen}, {Ciardi}, {Crane}, {Good}, {Harbut}, {Jackson}, {Kane}, {Laity}, {Leifer}, {Lynn}, {McElroy}, {Papin}, {Plavchan}, {Ram{\'\i}rez}, {Rey}, {von Braun}, {Wittman}, {Abajian}, {Ali}, {Beichman}, {Beekley}, {Berriman}, {Berukoff}, {Bryden}, {Chan}, {Groom}, {Lau}, {Payne}, {Regelson}, {Saucedo}, {Schmitz}, {Stauffer}, {Wyatt}, \& {Zhang}}]{Akeson_2013}
{Akeson}, R.~L., {Chen}, X., {Ciardi}, D., {et~al.} 2013, \pasp, 125, 989, \dodoi{10.1086/672273}

\bibitem[{{Allart} {et~al.}(2022){Allart}, {Lovis}, {Faria}, {Dumusque}, {Sosnowska}, {Figueira}, {Silva}, {Mehner}, {Pepe}, {Cristiani}, {Rebolo}, {Santos}, {Adibekyan}, {Cupani}, {Di Marcantonio}, {D'Odorico}, {Gonz{\'a}lez Hern{\'a}ndez}, {Martins}, {Milakovi{\'c}}, {Nunes}, {Sozzetti}, {Su{\'a}rez Mascare{\~n}o}, {Tabernero}, \& {Zapatero Osorio}}]{Allart_2022}
{Allart}, R., {Lovis}, C., {Faria}, J., {et~al.} 2022, \aap, 666, A196, \dodoi{10.1051/0004-6361/202243629}

\bibitem[{{Ambikasaran} {et~al.}(2015){Ambikasaran}, {Foreman-Mackey}, {Greengard}, {Hogg}, \& {O'Neil}}]{Ambikasaran_2015}
{Ambikasaran}, S., {Foreman-Mackey}, D., {Greengard}, L., {Hogg}, D.~W., \& {O'Neil}, M. 2015, IEEE Transactions on Pattern Analysis and Machine Intelligence, 38, 252, \dodoi{10.1109/TPAMI.2015.2448083}

\bibitem[{{Anglada-Escud{\'e}} \& {Butler}(2012)}]{Anglada-Escude_2012}
{Anglada-Escud{\'e}}, G., \& {Butler}, R.~P. 2012, \apjs, 200, 15, \dodoi{10.1088/0067-0049/200/2/15}

\bibitem[{{Anglada-Escud{\'e}} {et~al.}(2016){Anglada-Escud{\'e}}, {Amado}, {Barnes}, {Berdi{\~n}as}, {Butler}, {Coleman}, {de La Cueva}, {Dreizler}, {Endl}, {Giesers}, {Jeffers}, {Jenkins}, {Jones}, {Kiraga}, {K{\"u}rster}, {L{\'o}pez-Gonz{\'a}lez}, {Marvin}, {Morales}, {Morin}, {Nelson}, {Ortiz}, {Ofir}, {Paardekooper}, {Reiners}, {Rodr{\'\i}guez}, {Rodr{\'\i}guez-L{\'o}pez}, {Sarmiento}, {Strachan}, {Tsapras}, {Tuomi}, \& {Zechmeister}}]{Anglada-Escude_2016}
{Anglada-Escud{\'e}}, G., {Amado}, P.~J., {Barnes}, J., {et~al.} 2016, \nat, 536, 437, \dodoi{10.1038/nature19106}

\bibitem[{{Artigau} {et~al.}(2022){Artigau}, {Cadieux}, {Cook}, {Doyon}, {Vandal}, {Donati}, {Moutou}, {Delfosse}, {Fouqu{\'e}}, {Martioli}, {Bouchy}, {Parsons}, {Carmona}, {Dumusque}, {Astudillo-Defru}, {Bonfils}, \& {Mignon}}]{Artigau_2022}
{Artigau}, {\'E}., {Cadieux}, C., {Cook}, N.~J., {et~al.} 2022, \aj, 164, 84, \dodoi{10.3847/1538-3881/ac7ce6}

\bibitem[{{Artigau} {et~al.}(2024){Artigau}, {Cadieux}, {Cook}, {Doyon}, {Dauplaise}, {Arnold}, {Cadieux}, {Donati}, {Cristofari}, {Delfosse}, {Fouqu{\'e}}, {Moutou}, {Larue}, \& {Allart}}]{Artigau_2024}
---. 2024, \aj, 168, 252, \dodoi{10.3847/1538-3881/ad7b30}

\bibitem[{{Astropy Collaboration} {et~al.}(2018){Astropy Collaboration}, {Price-Whelan}, {Sip{\H{o}}cz}, {G{\"u}nther}, {Lim}, {Crawford}, {Conseil}, {Shupe}, {Craig}, {Dencheva}, {Ginsburg}, {VanderPlas}, {Bradley}, {P{\'e}rez-Su{\'a}rez}, {de Val-Borro}, {Aldcroft}, {Cruz}, {Robitaille}, {Tollerud}, {Ardelean}, {Babej}, {Bach}, {Bachetti}, {Bakanov}, {Bamford}, {Barentsen}, {Barmby}, {Baumbach}, {Berry}, {Biscani}, {Boquien}, {Bostroem}, {Bouma}, {Brammer}, {Bray}, {Breytenbach}, {Buddelmeijer}, {Burke}, {Calderone}, {Cano Rodr{\'\i}guez}, {Cara}, {Cardoso}, {Cheedella}, {Copin}, {Corrales}, {Crichton}, {D'Avella}, {Deil}, {Depagne}, {Dietrich}, {Donath}, {Droettboom}, {Earl}, {Erben}, {Fabbro}, {Ferreira}, {Finethy}, {Fox}, {Garrison}, {Gibbons}, {Goldstein}, {Gommers}, {Greco}, {Greenfield}, {Groener}, {Grollier}, {Hagen}, {Hirst}, {Homeier}, {Horton}, {Hosseinzadeh}, {Hu}, {Hunkeler}, {Ivezi{\'c}}, {Jain}, {Jenness}, {Kanarek}, {Kendrew}, {Kern}, {Kerzendorf}, {Khvalko}, {King}, {Kirkby}, {Kulkarni},
  {Kumar}, {Lee}, {Lenz}, {Littlefair}, {Ma}, {Macleod}, {Mastropietro}, {McCully}, {Montagnac}, {Morris}, {Mueller}, {Mumford}, {Muna}, {Murphy}, {Nelson}, {Nguyen}, {Ninan}, {N{\"o}the}, {Ogaz}, {Oh}, {Parejko}, {Parley}, {Pascual}, {Patil}, {Patil}, {Plunkett}, {Prochaska}, {Rastogi}, {Reddy Janga}, {Sabater}, {Sakurikar}, {Seifert}, {Sherbert}, {Sherwood-Taylor}, {Shih}, {Sick}, {Silbiger}, {Singanamalla}, {Singer}, {Sladen}, {Sooley}, {Sornarajah}, {Streicher}, {Teuben}, {Thomas}, {Tremblay}, {Turner}, {Terr{\'o}n}, {van Kerkwijk}, {de la Vega}, {Watkins}, {Weaver}, {Whitmore}, {Woillez}, {Zabalza}, \& {Astropy Contributors}}]{Astropy_2018}
{Astropy Collaboration}, {Price-Whelan}, A.~M., {Sip{\H{o}}cz}, B.~M., {et~al.} 2018, \aj, 156, 123, \dodoi{10.3847/1538-3881/aabc4f}

\bibitem[{{Astudillo-Defru} {et~al.}(2017){Astudillo-Defru}, {D{\'\i}az}, {Bonfils}, {Almenara}, {Delisle}, {Bouchy}, {Delfosse}, {Forveille}, {Lovis}, {Mayor}, {Murgas}, {Pepe}, {Santos}, {S{\'e}gransan}, {Udry}, \& {W{\"u}nsche}}]{Astudillo-Defru_2017}
{Astudillo-Defru}, N., {D{\'\i}az}, R.~F., {Bonfils}, X., {et~al.} 2017, \aap, 605, L11, \dodoi{10.1051/0004-6361/201731581}

\bibitem[{{Banerjee} {et~al.}(2024){Banerjee}, {Barstow}, {Gressier}, {Espinoza}, {Sing}, {Allen}, {Birkmann}, {Challener}, {Crouzet}, {Haswell}, {Lewis}, {Lewis}, \& {Yang}}]{Banerjee_2024}
{Banerjee}, A., {Barstow}, J.~K., {Gressier}, A., {et~al.} 2024, \apjl, 975, L11, \dodoi{10.3847/2041-8213/ad73d0}

\bibitem[{{Barclay} {et~al.}(2023){Barclay}, {Sheppard}, {Latouf}, {Mandell}, {Quintana}, {Gilbert}, {Liuzzi}, {Villanueva}, {Arney}, {Brande}, {Col{\'o}n}, {Covone}, {Crossfield}, {Damiano}, {Domagal-Goldman}, {Fauchez}, {Fiscale}, {Gallo}, {Hedges}, {Hu}, {Kite}, {Koll}, {Kopparapu}, {Kostov}, {Kreidberg}, {Lopez}, {Mang}, {Morley}, {Mullally}, {Mullally}, {Pidhorodetska}, {Schlieder}, {Vega}, {Youngblood}, \& {Zieba}}]{Barclay_2023}
{Barclay}, T., {Sheppard}, K.~B., {Latouf}, N., {et~al.} 2023, arXiv e-prints, arXiv:2301.10866, \dodoi{10.48550/arXiv.2301.10866}

\bibitem[{Barr(2008)}]{barr_mobile_2008}
Barr, A.~C. 2008, Journal of Geophysical Research (Planets), 113, E07009, \dodoi{10.1029/2008JE003114}

\bibitem[{Barr {et~al.}(2018)Barr, Dobos, \& Kiss}]{barr_interior_2018}
Barr, A.~C., Dobos, V., \& Kiss, L.~L. 2018, Astronomy and Astrophysics, 613, A37, \dodoi{10.1051/0004-6361/201731992}

\bibitem[{{Bello-Arufe} {et~al.}(2025){Bello-Arufe}, {Damiano}, {Bennett}, {Hu}, {Welbanks}, {MacDonald}, {Seligman}, {Sing}, {Tokadjian}, {Oza}, \& {Yang}}]{Bello-Arufe_2025}
{Bello-Arufe}, A., {Damiano}, M., {Bennett}, K.~A., {et~al.} 2025, \apjl, 980, L26, \dodoi{10.3847/2041-8213/adaf22}

\bibitem[{{Benneke} \& {Seager}(2012)}]{Benneke_2012}
{Benneke}, B., \& {Seager}, S. 2012, \apj, 753, 100, \dodoi{10.1088/0004-637X/753/2/100}

\bibitem[{Bertaux {et~al.}(2014)Bertaux, Lallement, Ferron, Boonne, \& Bodichon}]{Bertaux_2014}
Bertaux, J.~L., Lallement, R., Ferron, S., Boonne, C., \& Bodichon, R. 2014, Astronomy \& Astrophysics, 564, A46, \dodoi{10.1051/0004-6361/201322383}

\bibitem[{{Bond} {et~al.}(2010){Bond}, {O'Brien}, \& {Lauretta}}]{Bond_2010}
{Bond}, J.~C., {O'Brien}, D.~P., \& {Lauretta}, D.~S. 2010, \apj, 715, 1050, \dodoi{10.1088/0004-637X/715/2/1050}

\bibitem[{{Bonsor} {et~al.}(2021){Bonsor}, {Jofr{\'e}}, {Shorttle}, {Rogers}, {Xu(许偲艺)}, \& {Melis}}]{Bonsor_2021}
{Bonsor}, A., {Jofr{\'e}}, P., {Shorttle}, O., {et~al.} 2021, \mnras, 503, 1877, \dodoi{10.1093/mnras/stab370}

\bibitem[{{Bouchy} {et~al.}(2001){Bouchy}, {Pepe}, \& {Queloz}}]{Bouchy_2001}
{Bouchy}, F., {Pepe}, F., \& {Queloz}, D. 2001, \aap, 374, 733, \dodoi{10.1051/0004-6361:20010730}

\bibitem[{{Brinkman} {et~al.}(2024){Brinkman}, {Polanski}, {Huber}, {Weiss}, {Valencia}, \& {Plotnykov}}]{Brinkman_2024}
{Brinkman}, C.~L., {Polanski}, A.~S., {Huber}, D., {et~al.} 2024, arXiv e-prints, arXiv:2409.08361, \dodoi{10.48550/arXiv.2409.08361}

\bibitem[{{Burn} {et~al.}(2024){Burn}, {Mordasini}, {Mishra}, {Haldemann}, {Venturini}, {Emsenhuber}, \& {Henning}}]{Burn_2024}
{Burn}, R., {Mordasini}, C., {Mishra}, L., {et~al.} 2024, Nature Astronomy, 8, 463, \dodoi{10.1038/s41550-023-02183-7}

\bibitem[{{Cadieux} {et~al.}(2022){Cadieux}, {Doyon}, {Plotnykov}, {H{\'e}brard}, {Jahandar}, {Artigau}, {Valencia}, {Cook}, {Martioli}, {Vandal}, {Donati}, {Cloutier}, {Narita}, {Fukui}, {Hirano}, {Bouchy}, {Cowan}, {Gonzales}, {Ciardi}, {Stassun}, {Arnold}, {Benneke}, {Boisse}, {Bonfils}, {Carmona}, {Cort{\'e}s-Zuleta}, {Delfosse}, {Forveille}, {Fouqu{\'e}}, {Gomes da Silva}, {Jenkins}, {Kiefer}, {K{\'o}sp{\'a}l}, {Lafreni{\`e}re}, {Martins}, {Moutou}, {do Nascimento}, {Ould-Elhkim}, {Pelletier}, {Twicken}, {Bouma}, {Cartwright}, {Darveau-Bernier}, {Grankin}, {Ikoma}, {Kagetani}, {Kawauchi}, {Kodama}, {Kotani}, {Latham}, {Menou}, {Ricker}, {Seager}, {Tamura}, {Vanderspek}, \& {Watanabe}}]{Cadieux_2022}
{Cadieux}, C., {Doyon}, R., {Plotnykov}, M., {et~al.} 2022, \aj, 164, 96, \dodoi{10.3847/1538-3881/ac7cea}

\bibitem[{{Cadieux} {et~al.}(2024{\natexlab{a}}){Cadieux}, {Plotnykov}, {Doyon}, {Valencia}, {Jahandar}, {Dang}, {Turbet}, {Fauchez}, {Cloutier}, {Cherubim}, {Artigau}, {Cook}, {Edwards}, {Hallatt}, {Charnay}, {Bouchy}, {Allart}, {Mignon}, {Baron}, {Barros}, {Benneke}, {Canto Martins}, {Cowan}, {De Medeiros}, {Delfosse}, {Delgado-Mena}, {Dumusque}, {Ehrenreich}, {Frensch}, {Gonz{\'a}lez Hern{\'a}ndez}, {Hara}, {Lafreni{\`e}re}, {Lo Curto}, {Malo}, {Melo}, {Mounzer}, {Passeger}, {Pepe}, {Poulin-Girard}, {Santos}, {Sosnowska}, {Su{\'a}rez Mascare{\~n}o}, {Thibault}, {Vaulato}, {Wade}, \& {Wildi}}]{Cadieux_2024a}
{Cadieux}, C., {Plotnykov}, M., {Doyon}, R., {et~al.} 2024{\natexlab{a}}, \apjl, 960, L3, \dodoi{10.3847/2041-8213/ad1691}

\bibitem[{{Cadieux} {et~al.}(2024{\natexlab{b}}){Cadieux}, {Doyon}, {MacDonald}, {Turbet}, {Artigau}, {Lim}, {Radica}, {Fauchez}, {Salhi}, {Dang}, {Albert}, {Coulombe}, {Cowan}, {Lafreni{\`e}re}, {L'Heureux}, {Piaulet-Ghorayeb}, {Benneke}, {Cloutier}, {Charnay}, {Cook}, {Fournier-Tondreau}, {Plotnykov}, \& {Valencia}}]{Cadieux_2024b}
{Cadieux}, C., {Doyon}, R., {MacDonald}, R.~J., {et~al.} 2024{\natexlab{b}}, \apjl, 970, L2, \dodoi{10.3847/2041-8213/ad5afa}

\bibitem[{{Cherubim} {et~al.}(2023){Cherubim}, {Cloutier}, {Charbonneau}, {Stockdale}, {Stassun}, {Schwarz}, {Safonov}, {Mortier}, {Lewin}, {Latham}, {Horne}, {Haywood}, {Gonzales}, {Goliguzova}, {Collins}, {Ciardi}, {Bieryla}, {Belinski}, {Wohler}, {Watson}, {Vanderspek}, {Udry}, {Sozzetti}, {S{\'e}gransan}, {Sasselov}, {Ricker}, {Rice}, {Poretti}, {Piotto}, {Pepe}, {Molinari}, {Micela}, {Mayor}, {Lovis}, {L{\'o}pez-Morales}, {Jenkins}, {Essack}, {Dumusque}, {Doty}, {Col{\'o}n}, {Cameron}, \& {Buchhave}}]{Cherubim_2023}
{Cherubim}, C., {Cloutier}, R., {Charbonneau}, D., {et~al.} 2023, \aj, 165, 167, \dodoi{10.3847/1538-3881/acbdfd}

\bibitem[{{Cloutier} {et~al.}(2018{\natexlab{a}}){Cloutier}, {Doyon}, {Bouchy}, \& {H{\'e}brard}}]{Cloutier_2018b}
{Cloutier}, R., {Doyon}, R., {Bouchy}, F., \& {H{\'e}brard}, G. 2018{\natexlab{a}}, \aj, 156, 82, \dodoi{10.3847/1538-3881/aacea9}

\bibitem[{{Cloutier} \& {Menou}(2020)}]{Cloutier-Menou_2020}
{Cloutier}, R., \& {Menou}, K. 2020, \aj, 159, 211, \dodoi{10.3847/1538-3881/ab8237}

\bibitem[{{Cloutier} {et~al.}(2018{\natexlab{b}}){Cloutier}, {Artigau}, {Delfosse}, {Malo}, {Moutou}, {Doyon}, {Donati}, {Cumming}, {Dumusque}, {H{\'e}brard}, \& {Menou}}]{Cloutier_2018a}
{Cloutier}, R., {Artigau}, {\'E}., {Delfosse}, X., {et~al.} 2018{\natexlab{b}}, \aj, 155, 93, \dodoi{10.3847/1538-3881/aaa54e}

\bibitem[{{Cloutier} {et~al.}(2019){Cloutier}, {Astudillo-Defru}, {Bonfils}, {Jenkins}, {Berdi{\~n}as}, {Ricker}, {Vanderspek}, {Latham}, {Seager}, {Winn}, {Jenkins}, {Almenara}, {Bouchy}, {Delfosse}, {D{\'\i}az}, {D{\'\i}az}, {Doyon}, {Figueira}, {Forveille}, {Kurtovic}, {Lovis}, {Mayor}, {Menou}, {Morgan}, {Morris}, {Muirhead}, {Murgas}, {Pepe}, {Santos}, {S{\'e}gransan}, {Smith}, {Tenenbaum}, {Torres}, {Udry}, {Vezie}, \& {Villasenor}}]{Cloutier_2019}
{Cloutier}, R., {Astudillo-Defru}, N., {Bonfils}, X., {et~al.} 2019, \aap, 629, A111, \dodoi{10.1051/0004-6361/201935957}

\bibitem[{{Cook} {et~al.}(2022){Cook}, {Artigau}, {Doyon}, {Hobson}, {Martioli}, {Bouchy}, {Moutou}, {Carmona}, {Usher}, {Fouqu{\'e}}, {Arnold}, {Delfosse}, {Boisse}, {Cadieux}, {Vandal}, {Donati}, \& {Desli{\`e}res}}]{Cook_2022}
{Cook}, N.~J., {Artigau}, {\'E}., {Doyon}, R., {et~al.} 2022, \pasp, 134, 114509, \dodoi{10.1088/1538-3873/ac9e74}

\bibitem[{{Damiano} {et~al.}(2024){Damiano}, {Bello-Arufe}, {Yang}, \& {Hu}}]{Damiano_2024}
{Damiano}, M., {Bello-Arufe}, A., {Yang}, J., \& {Hu}, R. 2024, arXiv e-prints, arXiv:2403.13265, \dodoi{10.48550/arXiv.2403.13265}

\bibitem[{{Damiano} {et~al.}(2022){Damiano}, {Hu}, {Barclay}, {Zieba}, {Kreidberg}, {Brande}, {Colon}, {Covone}, {Crossfield}, {Domagal-Goldman}, {Fauchez}, {Fiscale}, {Gallo}, {Gilbert}, {Hedges}, {Kite}, {Kopparapu}, {Kostov}, {Morley}, {Mullally}, {Pidhorodetska}, {Schlieder}, \& {Quintana}}]{Damiano_2022}
{Damiano}, M., {Hu}, R., {Barclay}, T., {et~al.} 2022, \aj, 164, 225, \dodoi{10.3847/1538-3881/ac9472}

\bibitem[{Davies \& Davies(2010)}]{davies_earths_2010}
Davies, J.~H., \& Davies, D.~R. 2010, Solid Earth, 1, 5, \dodoi{10.5194/se-1-5-2010}

\bibitem[{{Dawson} \& {Johnson}(2012)}]{Dawson_2012}
{Dawson}, R.~I., \& {Johnson}, J.~A. 2012, \apj, 756, 122, \dodoi{10.1088/0004-637X/756/2/122}

\bibitem[{Deck {et~al.}(2014)Deck, Agol, Holman, \& Nesvorn{\'y}}]{Deck_2014}
Deck, K.~M., Agol, E., Holman, M.~J., \& Nesvorn{\'y}, D. 2014, The Astrophysical Journal, 787, 132, \dodoi{10.1088/0004-637X/787/2/132}

\bibitem[{{Delmotte} {et~al.}(2006){Delmotte}, {Dolensky}, {Padovani}, {Retzlaff}, {Rit{\'e}}, {Rosati}, {Slijkhuis}, {Wicenec}, {Fernique}, \& {Micol}}]{Delmotte_2006}
{Delmotte}, N., {Dolensky}, M., {Padovani}, P., {et~al.} 2006, in Astronomical Society of the Pacific Conference Series, Vol. 351, Astronomical Data Analysis Software and Systems XV, ed. C.~{Gabriel}, C.~{Arviset}, D.~{Ponz}, \& S.~{Enrique}, 690

\bibitem[{{Demangeon} {et~al.}(2021){Demangeon}, {Zapatero Osorio}, {Alibert}, {Barros}, {Adibekyan}, {Tabernero}, {Antoniadis-Karnavas}, {Camacho}, {Su{\'a}rez Mascare{\~n}o}, {Oshagh}, {Micela}, {Sousa}, {Lovis}, {Pepe}, {Rebolo}, {Cristiani}, {Santos}, {Allart}, {Allende Prieto}, {Bossini}, {Bouchy}, {Cabral}, {Damasso}, {Di Marcantonio}, {D'Odorico}, {Ehrenreich}, {Faria}, {Figueira}, {G{\'e}nova Santos}, {Haldemann}, {Hara}, {Gonz{\'a}lez Hern{\'a}ndez}, {Lavie}, {Lillo-Box}, {Lo Curto}, {Martins}, {M{\'e}gevand}, {Mehner}, {Molaro}, {Nunes}, {Pall{\'e}}, {Pasquini}, {Poretti}, {Sozzetti}, \& {Udry}}]{Demangeon_2021}
{Demangeon}, O.~D.~S., {Zapatero Osorio}, M.~R., {Alibert}, Y., {et~al.} 2021, \aap, 653, A41, \dodoi{10.1051/0004-6361/202140728}

\bibitem[{{Dittmann} {et~al.}(2017){Dittmann}, {Irwin}, {Charbonneau}, {Bonfils}, {Astudillo-Defru}, {Haywood}, {Berta-Thompson}, {Newton}, {Rodriguez}, {Winters}, {Tan}, {Almenara}, {Bouchy}, {Delfosse}, {Forveille}, {Lovis}, {Murgas}, {Pepe}, {Santos}, {Udry}, {W{\"u}nsche}, {Esquerdo}, {Latham}, \& {Dressing}}]{Dittmann_2017}
{Dittmann}, J.~A., {Irwin}, J.~M., {Charbonneau}, D., {et~al.} 2017, \nat, 544, 333

\bibitem[{Dobos \& Turner(2015)}]{dobos_viscoelastic_2015}
Dobos, V., \& Turner, E.~L. 2015, The Astrophysical Journal, 804, 41, \dodoi{10.1088/0004-637X/804/1/41}

\bibitem[{{Donati} {et~al.}(2020){Donati}, {Kouach}, {Moutou}, {Doyon}, {Delfosse}, {Artigau}, {Baratchart}, {Lacombe}, {Barrick}, {H{\'e}brard}, {Bouchy}, {Saddlemyer}, {Par{\`e}s}, {Rabou}, {Micheau}, {Dolon}, {Reshetov}, {Challita}, {Carmona}, {Striebig}, {Thibault}, {Martioli}, {Cook}, {Fouqu{\'e}}, {Vermeulen}, {Wang}, {Arnold}, {Pepe}, {Boisse}, {Figueira}, {Bouvier}, {Ray}, {Feugeade}, {Morin}, {Alencar}, {Hobson}, {Castilho}, {Udry}, {Santos}, {Hernandez}, {Benedict}, {Vall{\'e}e}, {Gallou}, {Dupieux}, {Larrieu}, {Perruchot}, {Sottile}, {Moreau}, {Usher}, {Baril}, {Wildi}, {Chazelas}, {Malo}, {Bonfils}, {Loop}, {Kerley}, {Wevers}, {Dunn}, {Pazder}, {Macdonald}, {Dubois}, {Carri{\'e}}, {Valentin}, {Henault}, {Yan}, \& {Steinmetz}}]{Donati_2020}
{Donati}, J.~F., {Kouach}, D., {Moutou}, C., {et~al.} 2020, \mnras, 498, 5684, \dodoi{10.1093/mnras/staa2569}

\bibitem[{{Donati} {et~al.}(2023){Donati}, {Lehmann}, {Cristofari}, {Fouqu{\'e}}, {Moutou}, {Charpentier}, {Ould-Elhkim}, {Carmona}, {Delfosse}, {Artigau}, {Alencar}, {Cadieux}, {Arnold}, {Petit}, {Morin}, {Forveille}, {Cloutier}, {Doyon}, {H{\'e}brard}, \& {SLS Collaboration}}]{Donati_2023}
{Donati}, J.~F., {Lehmann}, L.~T., {Cristofari}, P.~I., {et~al.} 2023, \mnras, 525, 2015, \dodoi{10.1093/mnras/stad2301}

\bibitem[{{Dorn} {et~al.}(2017){Dorn}, {Hinkel}, \& {Venturini}}]{Dorn_2017}
{Dorn}, C., {Hinkel}, N.~R., \& {Venturini}, J. 2017, \aap, 597, A38, \dodoi{10.1051/0004-6361/201628749}

\bibitem[{{Ducrot} {et~al.}(2024){Ducrot}, {Lagage}, {Min}, {Gillon}, {Bell}, {Tremblin}, {Greene}, {Dyrek}, {Bouwman}, {Waters}, {G{\"u}del}, {Henning}, {Vandenbussche}, {Absil}, {Barrado}, {Boccaletti}, {Coulais}, {Decin}, {Edwards}, {Gastaud}, {Glasse}, {Kendrew}, {Olofsson}, {Patapis}, {Pye}, {Rouan}, {Whiteford}, {Argyriou}, {Cossou}, {Glauser}, {Krause}, {Lahuis}, {Royer}, {Scheithauer}, {Colina}, {van Dishoeck}, {Ostlin}, {Ray}, \& {Wright}}]{Ducrot_2024}
{Ducrot}, E., {Lagage}, P.-O., {Min}, M., {et~al.} 2024, Nature Astronomy, \dodoi{10.1038/s41550-024-02428-z}

\bibitem[{{Engle} \& {Guinan}(2023)}]{Engle_2023}
{Engle}, S.~G., \& {Guinan}, E.~F. 2023, \apjl, 954, L50, \dodoi{10.3847/2041-8213/acf472}

\bibitem[{{Espinoza}(2018)}]{Espinoza_2018}
{Espinoza}, N. 2018, Research Notes of the American Astronomical Society, 2, 209, \dodoi{10.3847/2515-5172/aaef38}

\bibitem[{{Espinoza} {et~al.}(2019){Espinoza}, {Kossakowski}, \& {Brahm}}]{Espinoza_2019}
{Espinoza}, N., {Kossakowski}, D., \& {Brahm}, R. 2019, \mnras, 490, 2262, \dodoi{10.1093/mnras/stz2688}

\bibitem[{{Farhat} {et~al.}(2025){Farhat}, {Auclair-Desrotour}, {Bou{\'e}}, {Lichtenberg}, \& {Laskar}}]{farhat_tides_2025}
{Farhat}, M., {Auclair-Desrotour}, P., {Bou{\'e}}, G., {Lichtenberg}, T., \& {Laskar}, J. 2025, \apj, 979, 133, \dodoi{10.3847/1538-4357/ad9b93}

\bibitem[{{Feroz} {et~al.}(2019){Feroz}, {Hobson}, {Cameron}, \& {Pettitt}}]{Feroz_2019}
{Feroz}, F., {Hobson}, M.~P., {Cameron}, E., \& {Pettitt}, A.~N. 2019, The Open Journal of Astrophysics, 2, 10, \dodoi{10.21105/astro.1306.2144}

\bibitem[{Fischer \& Spohn(1990)}]{fischer_thermal-orbital_1990}
Fischer, H.-J., \& Spohn, T. 1990, Icarus, 83, 39, \dodoi{10.1016/0019-1035(90)90005-T}

\bibitem[{{Foreman-Mackey}(2018)}]{celerite2_2018}
{Foreman-Mackey}, D. 2018, Research Notes of the American Astronomical Society, 2, 31, \dodoi{10.3847/2515-5172/aaaf6c}

\bibitem[{{Foreman-Mackey} {et~al.}(2017){Foreman-Mackey}, {Agol}, {Ambikasaran}, \& {Angus}}]{celerite1_2017}
{Foreman-Mackey}, D., {Agol}, E., {Ambikasaran}, S., \& {Angus}, R. 2017, \aj, 154, 220, \dodoi{10.3847/1538-3881/aa9332}

\bibitem[{{Foreman-Mackey} {et~al.}(2013){Foreman-Mackey}, {Hogg}, {Lang}, \& {Goodman}}]{Foreman-Mackey_2013}
{Foreman-Mackey}, D., {Hogg}, D.~W., {Lang}, D., \& {Goodman}, J. 2013, \pasp, 125, 306, \dodoi{10.1086/670067}

\bibitem[{{Fromont} {et~al.}(2024){Fromont}, {Ahlers}, {do Amaral}, {Barnes}, {Gilbert}, {Quintana}, {Peacock}, {Barclay}, \& {Youngblood}}]{Fromont_2024}
{Fromont}, E.~F., {Ahlers}, J.~P., {do Amaral}, L. N.~R., {et~al.} 2024, \apj, 961, 115, \dodoi{10.3847/1538-4357/ad0e0e}

\bibitem[{{Fulton} {et~al.}(2018){Fulton}, {Petigura}, {Blunt}, \& {Sinukoff}}]{Fulton_2018}
{Fulton}, B.~J., {Petigura}, E.~A., {Blunt}, S., \& {Sinukoff}, E. 2018, \pasp, 130, 044504, \dodoi{10.1088/1538-3873/aaaaa8}

\bibitem[{{Fulton} {et~al.}(2017){Fulton}, {Petigura}, {Howard}, {Isaacson}, {Marcy}, {Cargile}, {Hebb}, {Weiss}, {Johnson}, {Morton}, {Sinukoff}, {Crossfield}, \& {Hirsch}}]{Fulton_2017}
{Fulton}, B.~J., {Petigura}, E.~A., {Howard}, A.~W., {et~al.} 2017, \aj, 154, 109, \dodoi{10.3847/1538-3881/aa80eb}

\bibitem[{{Gaia Collaboration} {et~al.}(2023){Gaia Collaboration}, {Vallenari}, {Brown}, {Prusti}, {de Bruijne}, {Arenou}, {Babusiaux}, {Biermann}, {Creevey}, {Ducourant}, {Evans}, {Eyer}, {Guerra}, {Hutton}, {Jordi}, {Klioner}, {Lammers}, {Lindegren}, {Luri}, {Mignard}, {Panem}, {Pourbaix}, {Randich}, {Sartoretti}, {Soubiran}, {Tanga}, {Walton}, {Bailer-Jones}, {Bastian}, {Drimmel}, {Jansen}, {Katz}, {Lattanzi}, {van Leeuwen}, {Bakker}, {Cacciari}, {Casta{\~n}eda}, {De Angeli}, {Fabricius}, {Fouesneau}, {Fr{\'e}mat}, {Galluccio}, {Guerrier}, {Heiter}, {Masana}, {Messineo}, {Mowlavi}, {Nicolas}, {Nienartowicz}, {Pailler}, {Panuzzo}, {Riclet}, {Roux}, {Seabroke}, {Sordo}, {Th{\'e}venin}, {Gracia-Abril}, {Portell}, {Teyssier}, {Altmann}, {Andrae}, {Audard}, {Bellas-Velidis}, {Benson}, {Berthier}, {Blomme}, {Burgess}, {Busonero}, {Busso}, {C{\'a}novas}, {Carry}, {Cellino}, {Cheek}, {Clementini}, {Damerdji}, {Davidson}, {de Teodoro}, {Nu{\~n}ez Campos}, {Delchambre}, {Dell'Oro}, {Esquej},
  {Fern{\'a}ndez-Hern{\'a}ndez}, {Fraile}, {Garabato}, {Garc{\'\i}a-Lario}, {Gosset}, {Haigron}, {Halbwachs}, {Hambly}, {Harrison}, {Hern{\'a}ndez}, {Hestroffer}, {Hodgkin}, {Holl}, {Jan{\ss}en}, {Jevardat de Fombelle}, {Jordan}, {Krone-Martins}, {Lanzafame}, {L{\"o}ffler}, {Marchal}, {Marrese}, {Moitinho}, {Muinonen}, {Osborne}, {Pancino}, {Pauwels}, {Recio-Blanco}, {Reyl{\'e}}, {Riello}, {Rimoldini}, {Roegiers}, {Rybizki}, {Sarro}, {Siopis}, {Smith}, {Sozzetti}, {Utrilla}, {van Leeuwen}, {Abbas}, {{\'A}brah{\'a}m}, {Abreu Aramburu}, {Aerts}, {Aguado}, {Ajaj}, {Aldea-Montero}, {Altavilla}, {{\'A}lvarez}, {Alves}, {Anders}, {Anderson}, {Anglada Varela}, {Antoja}, {Baines}, {Baker}, {Balaguer-N{\'u}{\~n}ez}, {Balbinot}, {Balog}, {Barache}, {Barbato}, {Barros}, {Barstow}, {Bartolom{\'e}}, {Bassilana}, {Bauchet}, {Becciani}, {Bellazzini}, {Berihuete}, {Bernet}, {Bertone}, {Bianchi}, {Binnenfeld}, {Blanco-Cuaresma}, {Blazere}, {Boch}, {Bombrun}, {Bossini}, {Bouquillon}, {Bragaglia}, {Bramante}, {Breedt},
  {Bressan}, {Brouillet}, {Brugaletta}, {Bucciarelli}, {Burlacu}, {Butkevich}, {Buzzi}, {Caffau}, {Cancelliere}, {Cantat-Gaudin}, {Carballo}, {Carlucci}, {Carnerero}, {Carrasco}, {Casamiquela}, {Castellani}, {Castro-Ginard}, {Chaoul}, {Charlot}, {Chemin}, {Chiaramida}, {Chiavassa}, {Chornay}, {Comoretto}, {Contursi}, {Cooper}, {Cornez}, {Cowell}, {Crifo}, {Cropper}, {Crosta}, {Crowley}, {Dafonte}, {Dapergolas}, {David}, {David}, {de Laverny}, {De Luise}, {De March}, {De Ridder}, {de Souza}, {de Torres}, {del Peloso}, {del Pozo}, {Delbo}, {Delgado}, {Delisle}, {Demouchy}, {Dharmawardena}, {Di Matteo}, {Diakite}, {Diener}, {Distefano}, {Dolding}, {Edvardsson}, {Enke}, {Fabre}, {Fabrizio}, {Faigler}, {Fedorets}, {Fernique}, {Fienga}, {Figueras}, {Fournier}, {Fouron}, {Fragkoudi}, {Gai}, {Garcia-Gutierrez}, {Garcia-Reinaldos}, {Garc{\'\i}a-Torres}, {Garofalo}, {Gavel}, {Gavras}, {Gerlach}, {Geyer}, {Giacobbe}, {Gilmore}, {Girona}, {Giuffrida}, {Gomel}, {Gomez}, {Gonz{\'a}lez-N{\'u}{\~n}ez},
  {Gonz{\'a}lez-Santamar{\'\i}a}, {Gonz{\'a}lez-Vidal}, {Granvik}, {Guillout}, {Guiraud}, {Guti{\'e}rrez-S{\'a}nchez}, {Guy}, {Hatzidimitriou}, {Hauser}, {Haywood}, {Helmer}, {Helmi}, {Sarmiento}, {Hidalgo}, {Hilger}, {H{\l}adczuk}, {Hobbs}, {Holland}, {Huckle}, {Jardine}, {Jasniewicz}, {Jean-Antoine Piccolo}, {Jim{\'e}nez-Arranz}, {Jorissen}, {Juaristi Campillo}, {Julbe}, {Karbevska}, {Kervella}, {Khanna}, {Kontizas}, {Kordopatis}, {Korn}, {K{\'o}sp{\'a}l}, {Kostrzewa-Rutkowska}, {Kruszy{\'n}ska}, {Kun}, {Laizeau}, {Lambert}, {Lanza}, {Lasne}, {Le Campion}, {Lebreton}, {Lebzelter}, {Leccia}, {Leclerc}, {Lecoeur-Taibi}, {Liao}, {Licata}, {Lindstr{\o}m}, {Lister}, {Livanou}, {Lobel}, {Lorca}, {Loup}, {Madrero Pardo}, {Magdaleno Romeo}, {Managau}, {Mann}, {Manteiga}, {Marchant}, {Marconi}, {Marcos}, {Marcos Santos}, {Mar{\'\i}n Pina}, {Marinoni}, {Marocco}, {Marshall}, {Martin Polo}, {Mart{\'\i}n-Fleitas}, {Marton}, {Mary}, {Masip}, {Massari}, {Mastrobuono-Battisti}, {Mazeh}, {McMillan}, {Messina}, {Michalik},
  {Millar}, {Mints}, {Molina}, {Molinaro}, {Moln{\'a}r}, {Monari}, {Mongui{\'o}}, {Montegriffo}, {Montero}, {Mor}, {Mora}, {Morbidelli}, {Morel}, {Morris}, {Muraveva}, {Murphy}, {Musella}, {Nagy}, {Noval}, {Oca{\~n}a}, {Ogden}, {Ordenovic}, {Osinde}, {Pagani}, {Pagano}, {Palaversa}, {Palicio}, {Pallas-Quintela}, {Panahi}, {Payne-Wardenaar}, {Pe{\~n}alosa Esteller}, {Penttil{\"a}}, {Pichon}, {Piersimoni}, {Pineau}, {Plachy}, {Plum}, {Poggio}, {Pr{\v{s}}a}, {Pulone}, {Racero}, {Ragaini}, {Rainer}, {Raiteri}, {Rambaux}, {Ramos}, {Ramos-Lerate}, {Re Fiorentin}, {Regibo}, {Richards}, {Rios Diaz}, {Ripepi}, {Riva}, {Rix}, {Rixon}, {Robichon}, {Robin}, {Robin}, {Roelens}, {Rogues}, {Rohrbasser}, {Romero-G{\'o}mez}, {Rowell}, {Royer}, {Ruz Mieres}, {Rybicki}, {Sadowski}, {S{\'a}ez N{\'u}{\~n}ez}, {Sagrist{\`a} Sell{\'e}s}, {Sahlmann}, {Salguero}, {Samaras}, {Sanchez Gimenez}, {Sanna}, {Santove{\~n}a}, {Sarasso}, {Schultheis}, {Sciacca}, {Segol}, {Segovia}, {S{\'e}gransan}, {Semeux}, {Shahaf}, {Siddiqui}, {Siebert},
  {Siltala}, {Silvelo}, {Slezak}, {Slezak}, {Smart}, {Snaith}, {Solano}, {Solitro}, {Souami}, {Souchay}, {Spagna}, {Spina}, {Spoto}, {Steele}, {Steidelm{\"u}ller}, {Stephenson}, {S{\"u}veges}, {Surdej}, {Szabados}, {Szegedi-Elek}, {Taris}, {Taylor}, {Teixeira}, {Tolomei}, {Tonello}, {Torra}, {Torra}, {Torralba Elipe}, {Trabucchi}, {Tsounis}, {Turon}, {Ulla}, {Unger}, {Vaillant}, {van Dillen}, {van Reeven}, {Vanel}, {Vecchiato}, {Viala}, {Vicente}, {Voutsinas}, {Weiler}, {Wevers}, {Wyrzykowski}, {Yoldas}, {Yvard}, {Zhao}, {Zorec}, {Zucker}, \& {Zwitter}}]{Gaia_2023}
{Gaia Collaboration}, {Vallenari}, A., {Brown}, A.~G.~A., {et~al.} 2023, \aap, 674, A1, \dodoi{10.1051/0004-6361/202243940}

\bibitem[{{Gardner} {et~al.}(2023){Gardner}, {Mather}, {Abbott}, {Abell}, {Abernathy}, {Abney}, {Abraham}, {Abraham}, {Abul-Huda}, {Acton}, {Adams}, {Adams}, {Adler}, {Adriaensen}, {Aguilar}, {Ahmed}, {Ahmed}, {Ahmed}, {Albat}, {Albert}, {Alberts}, {Aldridge}, {Allen}, {Allen}, {Altenburg}, {Altunc}, {Alvarez}, {{\'A}lvarez-M{\'a}rquez}, {Alves de Oliveira}, {Ambrose}, {Anandakrishnan}, {Andersen}, {Anderson}, {Anderson}, {Anderson}, {Anderson}, {Aprea}, {Archer}, {Arenberg}, {Argyriou}, {Arribas}, {Artigau}, {Arvai}, {Atcheson}, {Atkinson}, {Averbukh}, {Aymergen}, {Bacinski}, {Baggett}, {Bagnasco}, {Baker}, {Balzano}, {Banks}, {Baran}, {Barker}, {Barrett}, {Barringer}, {Barto}, {Bast}, {Baudoz}, {Baum}, {Beatty}, {Beaulieu}, {Bechtold}, {Beck}, {Beddard}, {Beichman}, {Bellagama}, {Bely}, {Berger}, {Bergeron}, {Bernier}, {Bertch}, {Beskow}, {Betz}, {Biagetti}, {Birkmann}, {Bjorklund}, {Blackwood}, {Blazek}, {Blossfeld}, {Bluth}, {Boccaletti}, {Boegner}, {Bohlin}, {Boia}, {B{\"o}ker}, {Bonaventura}, {Bond},
  {Bosley}, {Boucarut}, {Bouchet}, {Bouwman}, {Bower}, {Bowers}, {Bowers}, {Boyce}, {Boyer}, {Boyer}, {Boyer}, {Boyer}, {Bradley}, {Brady}, {Brandl}, {Brannen}, {Breda}, {Bremmer}, {Brennan}, {Bresnahan}, {Bright}, {Broiles}, {Bromenschenkel}, {Brooks}, {Brooks}, {Brown}, {Brown}, {Brown}, {Bruce}, {Bryson}, {Bujanda}, {Bullock}, {Bunker}, {Bureo}, {Burt}, {Bush}, {Bushouse}, {Bussman}, {Cabaud}, {Cale}, {Calhoon}, {Calvani}, {Canipe}, {Caputo}, {Cara}, {Carey}, {Case}, {Cesari}, {Cetorelli}, {Chance}, {Chandler}, {Chaney}, {Chapman}, {Charlot}, {Chayer}, {Cheezum}, {Chen}, {Chen}, {Cherinka}, {Chichester}, {Chilton}, {Chittiraibalan}, {Clampin}, {Clark}, {Clark}, {Clark}, {Claybrooks}, {Cleveland}, {Cohen}, {Cohen}, {Col{\'o}n}, {Coleman}, {Colina}, {Comber}, {Comeau}, {Comer}, {Conde Reis}, {Connolly}, {Conroy}, {Contos}, {Contreras}, {Cook}, {Cooper}, {Cooper}, {Correia}, {Correnti}, {Cossou}, {Costanza}, {Coulais}, {Cox}, {Coyle}, {Cracraft}, {Crew}, {Curtis}, {Cusveller}, {Da Costa Maciel}, {Dailey},
  {Daugeron}, {Davidson}, {Davies}, {Davis}, {Davis}, {Day}, {de Chambure}, {de Jong}, {De Marchi}, {Dean}, {Decker}, {Delisa}, {Dell}, {Dellagatta}, {Dembinska}, {Demosthenes}, {Dencheva}, {Deneu}, {DePriest}, {Deschenes}, {Dethienne}, {Detre}, {Diaz}, {Dicken}, {DiFelice}, {Dillman}, {Disharoon}, {Dixon}, {Doggett}, {Dominguez}, {Donaldson}, {Doria-Warner}, {Santos}, {Doty}, {Douglas}, {Doyon}, {Dressler}, {Driggers}, {Driggers}, {Dunn}, {DuPrie}, {Dupuis}, {Durning}, {Dutta}, {Earl}, {Eccleston}, {Ecobichon}, {Egami}, {Ehrenwinkler}, {Eisenhamer}, {Eisenhower}, {Eisenstein}, {El Hamel}, {Elie}, {Elliott}, {Elliott}, {Engesser}, {Espinoza}, {Etienne}, {Etxaluze}, {Evans}, {Fabreguettes}, {Falcolini}, {Falini}, {Fatig}, {Feeney}, {Feinberg}, {Fels}, {Ferdous}, {Ferguson}, {Ferrarese}, {Ferreira}, {Ferruit}, {Ferry}, {Filippazzo}, {Firre}, {Fix}, {Flagey}, {Flanagan}, {Fleming}, {Florian}, {Flynn}, {Foiadelli}, {Fontaine}, {Fontanella}, {Forshay}, {Fortner}, {Fox}, {Framarini}, {Francisco}, {Franck}, {Franx},
  {Franz}, {Friedman}, {Friend}, {Frost}, {Fu}, {Fullerton}, {Gaillard}, {Galkin}, {Gallagher}, {Galyer}, {Garc{\'\i}a Mar{\'\i}n}, {Gardner}, {Garland}, {Garrett}, {Gasman}, {G{\'a}sp{\'a}r}, {Gastaud}, {Gaudreau}, {Gauthier}, {Geers}, {Geithner}, {Gennaro}, {Gerber}, {Gereau}, {Giampaoli}, {Giardino}, {Gibbons}, {Gilbert}, {Gilman}, {Girard}, {Giuliano}, {Gkountis}, {Glasse}, {Glassmire}, {Glauser}, {Glazer}, {Goldberg}, {Golimowski}, {Gonzaga}, {Gordon}, {Gordon}, {Goudfrooij}, {Gough}, {Graham}, {Grau}, {Green}, {Greene}, {Greene}, {Greenfield}, {Greenhouse}, {Greve}, {Greville}, {Grimaldi}, {Groe}, {Groebner}, {Grumm}, {Grundy}, {G{\"u}del}, {Guillard}, {Guldalian}, {Gunn}, {Gurule}, {Gutman}, {Guy}, {Guyot}, {Hack}, {Haderlein}, {Hagan}, {Hagedorn}, {Hainline}, {Haley}, {Hami}, {Hamilton}, {Hammann}, {Hammel}, {Hanley}, {Hansen}, {Hardy}, {Harnisch}, {Harr}, {Harris}, {Hart}, {Hartig}, {Hasan}, {Hashim}, {Hashimoto}, {Haskins}, {Hawkins}, {Hayden}, {Hayden}, {Healy}, {Hecht}, {Heeg}, {Hejal}, {Helm},
  {Hengemihle}, {Henning}, {Henry}, {Henry}, {Henshaw}, {Hernandez}, {Herrington}, {Heske}, {Hesman}, {Hickey}, {Hilbert}, {Hines}, {Hinz}, {Hirsch}, {Hitcho}, {Hodapp}, {Hodge}, {Hoffman}, {Holfeltz}, {Holler}, {Hoppa}, {Horner}, {Howard}, {Howard}, {Huber}, {Hunkeler}, {Hunter}, {Hunter}, {Hurd}, {Hurst}, {Hutchings}, {Hylan}, {Ignat}, {Illingworth}, {Irish}, {Isaacs}, {Jackson}, {Jaffe}, {Jahic}, {Jahromi}, {Jakobsen}, {James}, {James}, {James}, {Jamieson}, {Jandra}, {Jayawardhana}, {Jedrzejewski}, {Jeffers}, {Jensen}, {Joanne}, {Johns}, {Johnson}, {Johnson}, {Johnson}, {Johnson}, {Johnson}, {Johnson}, {Johnstone}, {Jollet}, {Jones}, {Jones}, {Jones}, {Jones}, {Jones}, {Jordan}, {Jordan}, {Jue}, {Jurkowski}, {Justis}, {Justtanont}, {Kaleida}, {Kalirai}, {Kalmanson}, {Kaltenegger}, {Kammerer}, {Kan}, {Kanarek}, {Kao}, {Karakla}, {Karl}, {Kassin}, {Kauffman}, {Kavanagh}, {Kelley}, {Kelly}, {Kendrew}, {Kennedy}, {Kenny}, {Keski-Kuha}, {Keyes}, {Khan}, {Kidwell}, {Kimble}, {King}, {King}, {Kinzel}, {Kirk},
  {Kirkpatrick}, {Klaassen}, {Klingemann}, {Klintworth}, {Knapp}, {Knight}, {Knollenberg}, {Knutsen}, {Koehler}, {Koekemoer}, {Kofler}, {Kontson}, {Kovacs}, {Kozhurina-Platais}, {Krause}, {Kriss}, {Krist}, {Kristoffersen}, {Krogel}, {Krueger}, {Kulp}, {Kumari}, {Kwan}, {Kyprianou}, {Labador}, {Labiano}, {Lafreni{\`e}re}, {Lagage}, {Laidler}, {Laine}, {Laird}, {Lajoie}, {Lallo}, {Lam}, {LaMassa}, {Lambros}, {Lampenfield}, {Lander}, {Langston}, {Larson}, {Larson}, {LaVerghetta}, {Law}, {Lawrence}, {Lee}, {Lee}, {Lee}, {Leisenring}, {Leveille}, {Levenson}, {Levi}, {Levine}, {Lewis}, {Lewis}, {Lewis}, {Libralato}, {Lidon}, {Liebrecht}, {Lightsey}, {Lilly}, {Lim}, {Lim}, {Ling}, {Link}, {Link}, {Lipinski}, {Liu}, {Lo}, {Lobmeyer}, {Logue}, {Long}, {Long}, {Long}, {Long}, {L{\'o}pez-Caniego}, {Lotz}, {Love-Pruitt}, {Lubskiy}, {Luers}, {Luetgens}, {Luevano}, {Lui}, {Lund}, {Lundquist}, {Lunine}, {L{\"u}tzgendorf}, {Lynch}, {MacDonald}, {MacDonald}, {Macias}, {Macklis}, {Maghami}, {Maharaja}, {Maiolino},
  {Makrygiannis}, {Malla}, {Malumuth}, {Manjavacas}, {Marini}, {Marrione}, {Marston}, {Martel}, {Martin}, {Martin}, {Martinez}, {Maschmann}, {Masci}, {Masetti}, {Maszkiewicz}, {Matthews}, {Matuskey}, {McBrayer}, {McCarthy}, {McCaughrean}, {McClare}, {McClare}, {McCloskey}, {McClurg}, {McCoy}, {McElwain}, {McGregor}, {McGuffey}, {McKay}, {McKenzie}, {McLean}, {McMaster}, {McNeil}, {De Meester}, {Mehalick}, {Meixner}, {Mel{\'e}ndez}, {Menzel}, {Menzel}, {Merz}, {Mesterharm}, {Meyer}, {Meyett}, {Meza}, {Midwinter}, {Milam}, {Miller}, {Miller}, {Miskey}, {Misselt}, {Mitchell}, {Mohan}, {Montoya}, {Moran}, {Morishita}, {Moro-Mart{\'\i}n}, {Morrison}, {Morrison}, {Morse}, {Moschos}, {Moseley}, {Mosier}, {Mosner}, {Mountain}, {Muckenthaler}, {Mueller}, {Mueller}, {Muhiem}, {M{\"u}hlmann}, {Mullally}, {Mullen}, {Munger}, {Murphy}, {Murray}, {Muzerolle}, {Mycroft}, {Myers}, {Myers}, {Myers}, {Myers}, {Myrick}, {Nagle}, {Nayak}, {Naylor}, {Neff}, {Nelan}, {Nella}, {Nguyen}, {Nguyen}, {Nickson}, {Nidhiry}, {Niedner},
  {Nieto-Santisteban}, {Nikolov}, {Nishisaka}, {Noriega-Crespo}, {Nota}, {O'Mara}, {Oboryshko}, {O'Brien}, {Ochs}, {Offenberg}, {Ogle}, {Ohl}, {Olmsted}, {Osborne}, {O'Shaughnessy}, {{\"O}stlin}, {O'Sullivan}, {Otor}, {Ottens}, {Ouellette}, {Outlaw}, {Owens}, {Pacifici}, {Page}, {Paranilam}, {Park}, {Parrish}, {Paschal}, {Patapis}, {Patel}, {Patrick}, {Pattishall}, {Paul}, {Paul}, {Pauly}, {Pavlovsky}, {Pe{\~n}a-Guerrero}, {Pedder}, {Peek}, {Pelham}, {Penanen}, {Perriello}, {Perrin}, {Perrine}, {Perrygo}, {Peslier}, {Petach}, {Peterson}, {Pfarr}, {Pierson}, {Pietraszkiewicz}, {Pilchen}, {Pipher}, {Pirzkal}, {Pitman}, {Player}, {Plesha}, {Plitzke}, {Pohner}, {Poletis}, {Pollizzi}, {Polster}, {Pontius}, {Pontoppidan}, {Porges}, {Potter}, {Prescott}, {Proffitt}, {Pueyo}, {Quispe Neira}, {Radich}, {Rager}, {Rameau}, {Ramey}, {Ramos Alarcon}, {Rampini}, {Rapp}, {Rashford}, {Rauscher}, {Ravindranath}, {Rawle}, {Rawlings}, {Ray}, {Regan}, {Rehm}, {Rehm}, {Reid}, {Reis}, {Renk}, {Reoch}, {Ressler}, {Rest},
  {Reynolds}, {Richon}, {Richon}, {Ridgaway}, {Riedel}, {Rieke}, {Rieke}, {Rifelli}, {Rigby}, {Riggs}, {Ringel}, {Ritchie}, {Rix}, {Robberto}, {Robinson}, {Robinson}, {Robinson}, {Rock}, {Rodriguez}, {Rodr{\'\i}guez del Pino}, {Roellig}, {Rohrbach}, {Roman}, {Romelfanger}, {Romo}, {Rosales}, {Rose}, {Roteliuk}, {Roth}, {Rothwell}, {Rouzaud}, {Rowe}, {Rowlands}, {Roy}, {Royer}, {Rui}, {Rumler}, {Rumpl}, {Russ}, {Ryan}, {Ryan}, {Saad}, {Sabata}, {Sabatino}, {Sabbi}, {Sabelhaus}, {Sabia}, {Sahu}, {Saif}, {Salvignol}, {Samara-Ratna}, {Samuelson}, {Sanders}, {Sappington}, {Sargent}, {Sauer}, {Savadkin}, {Sawicki}, {Schappell}, {Scheffer}, {Scheithauer}, {Scherer}, {Schiff}, {Schlawin}, {Schmeitzky}, {Schmitz}, {Schmude}, {Schneider}, {Schreiber}, {Schroeven-Deceuninck}, {Schultz}, {Schwab}, {Schwartz}, {Scoccimarro}, {Scott}, {Scott}, {Seaton}, {Seely}, {Seery}, {Seidleck}, {Sembach}, {Shanahan}, {Shaughnessy}, {Shaw}, {Shay}, {Sheehan}, {Sheth}, {Shih}, {Shivaei}, {Siegel}, {Sienkiewicz}, {Simmons}, {Simon},
  {Sirianni}, {Sivaramakrishnan}, {Slade}, {Sloan}, {Slocum}, {Slowinski}, {Smith}, {Smith}, {Smith}, {Smith}, {Smith}, {Smith}, {Smolik}, {Soderblom}, {Sohn}, {Sokol}, {Sonneborn}, {Sontag}, {Sooy}, {Soummer}, {Southwood}, {Spain}, {Sparmo}, {Speer}, {Spencer}, {Sprofera}, {Stallcup}, {Stanley}, {Stansberry}, {Stark}, {Starr}, {Stassi}, {Steck}, {Steeley}, {Stephens}, {Stephenson}, {Stewart}, {Stiavelli}, {}, {Strada}, {Straughn}, {Streetman}, {Strickland}, {Strobele}, {Stuhlinger}, {Stys}, {Such}, {Sukhatme}, {Sullivan}, {Sullivan}, {Sumner}, {Sun}, {Sunnquist}, {Swade}, {Swam}, {Swenton}, {Swoish}, {Tam Litten}, {Tamas}, {Tao}, {Taylor}, {Taylor}, {te Plate}, {Van Tea}, {Teague}, {Telfer}, {Temim}, {Texter}, {Thatte}, {Thompson}, {Thompson}, {Thomson}, {Thronson}, {Tierney}, {Tikkanen}, {Tinnin}, {Tippet}, {Todd}, {Tran}, {Trauger}, {Trejo}, {Vinh Truong}, {Tsukamoto}, {Tufail}, {Tumlinson}, {Tustain}, {Tyra}, {Ubeda}, {Underwood}, {Uzzo}, {Vaclavik}, {Valenduc}, {Valenti}, {Van Campen}, {van de Wetering},
  {Van Der Marel}, {van Haarlem}, {Vandenbussche}, {van Dishoeck}, {Vanterpool}, {Vernoy}, {Vila Costas}, {Volk}, {Voorzaat}, {Voyton}, {Vydra}, {Waddy}, {Waelkens}, {Wahlgren}, {Walker}, {Wander}, {Warfield}, {Warner}, {Wasiak}, {Wasiak}, {Wehner}, {Weiler}, {Weilert}, {Weiss}, {Wells}, {Welty}, {Wheate}, {Wheeler}, {White}, {Whitehouse}, {Whiteleather}, {Whitman}, {Williams}, {Willmer}, {Willott}, {Willoughby}, {Wilson}, {Wilson}, {Wilson}, {Windhorst}, {Wislowski}, {Wolfe}, {Wolfe}, {Wolff}, {Wondel}, {Woo}, {Woods}, {Worden}, {Workman}, {Wright}, {Wu}, {Wu}, {Wun}, {Wymer}, {Yadetie}, {Yan}, {Yang}, {Yates}, {Yeager}, {Yerger}, {Young}, {Young}, {Yu}, {Yu}, {Zak}, {Zeidler}, {Zepp}, {Zhou}, {Zincke}, {Zonak}, \& {Zondag}}]{Gardner_2023}
{Gardner}, J.~P., {Mather}, J.~C., {Abbott}, R., {et~al.} 2023, \pasp, 135, 068001, \dodoi{10.1088/1538-3873/acd1b5}

\bibitem[{{Gaudi} {et~al.}(2020){Gaudi}, {Seager}, {Mennesson}, {Kiessling}, {Warfield}, {Cahoy}, {Clarke}, {Domagal-Goldman}, {Feinberg}, {Guyon}, {Kasdin}, {Mawet}, {Plavchan}, {Robinson}, {Rogers}, {Scowen}, {Somerville}, {Stapelfeldt}, {Stark}, {Stern}, {Turnbull}, {Amini}, {Kuan}, {Martin}, {Morgan}, {Redding}, {Stahl}, {Webb}, {Alvarez-Salazar}, {Arnold}, {Arya}, {Balasubramanian}, {Baysinger}, {Bell}, {Below}, {Benson}, {Blais}, {Booth}, {Bourgeois}, {Bradford}, {Brewer}, {Brooks}, {Cady}, {Caldwell}, {Calvet}, {Carr}, {Chan}, {Cormarkovic}, {Coste}, {Cox}, {Danner}, {Davis}, {Dewell}, {Dorsett}, {Dunn}, {East}, {Effinger}, {Eng}, {Freebury}, {Garcia}, {Gaskin}, {Greene}, {Hennessy}, {Hilgemann}, {Hood}, {Holota}, {Howe}, {Huang}, {Hull}, {Hunt}, {Hurd}, {Johnson}, {Kissil}, {Knight}, {Kolenz}, {Kraus}, {Krist}, {Li}, {Lisman}, {Mandic}, {Mann}, {Marchen}, {Marrese-Reading}, {McCready}, {McGown}, {Missun}, {Miyaguchi}, {Moore}, {Nemati}, {Nikzad}, {Nissen}, {Novicki}, {Perrine}, {Pineda}, {Polanco},
  {Putnam}, {Qureshi}, {Richards}, {Eldorado Riggs}, {Rodgers}, {Rud}, {Saini}, {Scalisi}, {Scharf}, {Schulz}, {Serabyn}, {Sigrist}, {Sikkia}, {Singleton}, {Shaklan}, {Smith}, {Southerd}, {Stahl}, {Steeves}, {Sturges}, {Sullivan}, {Tang}, {Taras}, {Tesch}, {Therrell}, {Tseng}, {Valente}, {Van Buren}, {Villalvazo}, {Warwick}, {Webb}, {Westerhoff}, {Wofford}, {Wu}, {Woo}, {Wood}, {Ziemer}, {Arney}, {Anderson}, {Ma{\'\i}z-Apell{\'a}niz}, {Bartlett}, {Belikov}, {Bendek}, {Cenko}, {Douglas}, {Dulz}, {Evans}, {Faramaz}, {Feng}, {Ferguson}, {Follette}, {Ford}, {Garc{\'\i}a}, {Geha}, {Gelino}, {G{\"o}tberg}, {Hildebrandt}, {Hu}, {Jahnke}, {Kennedy}, {Kreidberg}, {Isella}, {Lopez}, {Marchis}, {Macri}, {Marley}, {Matzko}, {Mazoyer}, {McCandliss}, {Meshkat}, {Mordasini}, {Morris}, {Nielsen}, {Newman}, {Petigura}, {Postman}, {Reines}, {Roberge}, {Roederer}, {Ruane}, {Schwieterman}, {Sirbu}, {Spalding}, {Teplitz}, {Tumlinson}, {Turner}, {Werk}, {Wofford}, {Wyatt}, {Young}, \& {Zellem}}]{Gaudi_2020}
{Gaudi}, B.~S., {Seager}, S., {Mennesson}, B., {et~al.} 2020, arXiv e-prints, arXiv:2001.06683.
\newblock \doarXiv{2001.06683}

\bibitem[{{Gilbert} {et~al.}(2022){Gilbert}, {MacDougall}, \& {Petigura}}]{Gilbert_2022}
{Gilbert}, G.~J., {MacDougall}, M.~G., \& {Petigura}, E.~A. 2022, \aj, 164, 92, \dodoi{10.3847/1538-3881/ac7f2f}

\bibitem[{{Gillon} {et~al.}(2017){Gillon}, {Triaud}, {Demory}, {Jehin}, {Agol}, {Deck}, {Lederer}, {de Wit}, {Burdanov}, {Ingalls}, {Bolmont}, {Leconte}, {Raymond}, {Selsis}, {Turbet}, {Barkaoui}, {Burgasser}, {Burleigh}, {Carey}, {Chaushev}, {Copperwheat}, {Delrez}, {Fernandes}, {Holdsworth}, {Kotze}, {Van Grootel}, {Almleaky}, {Benkhaldoun}, {Magain}, \& {Queloz}}]{Gillon_2017}
{Gillon}, M., {Triaud}, A. H.~M.~J., {Demory}, B.-O., {et~al.} 2017, \nat, 542, 456, \dodoi{10.1038/nature21360}

\bibitem[{{Ginzburg} {et~al.}(2018){Ginzburg}, {Schlichting}, \& {Sari}}]{Ginzburg_2018}
{Ginzburg}, S., {Schlichting}, H.~E., \& {Sari}, R. 2018, \mnras, 476, 759, \dodoi{10.1093/mnras/sty290}

\bibitem[{{Gkouvelis} {et~al.}(2025){Gkouvelis}, {Pozuelos}, {Drant}, {Farhat}, {Tian}, \& {Ak{\i}n}}]{Gkouvelis_2025}
{Gkouvelis}, L., {Pozuelos}, F.~J., {Drant}, T., {et~al.} 2025, arXiv e-prints, arXiv:2506.02188, \dodoi{10.48550/arXiv.2506.02188}

\bibitem[{{Gonz{\'a}lez Hern{\'a}ndez} {et~al.}(2024){Gonz{\'a}lez Hern{\'a}ndez}, {Su{\'a}rez Mascare{\~n}o}, {Silva}, {Stefanov}, {Faria}, {Tabernero}, {Sozzetti}, {Rebolo}, {Pepe}, {Santos}, {Cristiani}, {Lovis}, {Dumusque}, {Figueira}, {Lillo-Box}, {Nari}, {Benatti}, {Hobson}, {Castro-Gonz{\'a}lez}, {Allart}, {Passegger}, {Zapatero Osorio}, {Adibekyan}, {Alibert}, {Allende Prieto}, {Bouchy}, {Damasso}, {D'Odorico}, {Di Marcantonio}, {Ehrenreich}, {Lo Curto}, {Santos}, {Martins}, {Mehner}, {Micela}, {Molaro}, {Nunes}, {Palle}, {Sousa}, \& {Udry}}]{Gonzalez-Hernandez_2024}
{Gonz{\'a}lez Hern{\'a}ndez}, J.~I., {Su{\'a}rez Mascare{\~n}o}, A., {Silva}, A.~M., {et~al.} 2024, \aap, 690, A79, \dodoi{10.1051/0004-6361/202451311}

\bibitem[{{Greene} {et~al.}(2023){Greene}, {Bell}, {Ducrot}, {Dyrek}, {Lagage}, \& {Fortney}}]{Greene_2023}
{Greene}, T.~P., {Bell}, T.~J., {Ducrot}, E., {et~al.} 2023, \nat, 618, 39, \dodoi{10.1038/s41586-023-05951-7}

\bibitem[{{Gressier} {et~al.}(2024){Gressier}, {Espinoza}, {Allen}, {Sing}, {Banerjee}, {Barstow}, {Valenti}, {Lewis}, {Birkmann}, {Challener}, {Manjavacas}, {Alves de Oliveira}, {Crouzet}, \& {Beck}}]{Gressier_2024}
{Gressier}, A., {Espinoza}, N., {Allen}, N.~H., {et~al.} 2024, \apjl, 975, L10, \dodoi{10.3847/2041-8213/ad73d1}

\bibitem[{{Harris} {et~al.}(2020){Harris}, {Millman}, {van der Walt}, {Gommers}, {Virtanen}, {Cournapeau}, {Wieser}, {Taylor}, {Berg}, {Smith}, {Kern}, {Picus}, {Hoyer}, {van Kerkwijk}, {Brett}, {Haldane}, {del R{\'\i}o}, {Wiebe}, {Peterson}, {G{\'e}rard-Marchant}, {Sheppard}, {Reddy}, {Weckesser}, {Abbasi}, {Gohlke}, \& {Oliphant}}]{Harris_2020}
{Harris}, C.~R., {Millman}, K.~J., {van der Walt}, S.~J., {et~al.} 2020, \nat, 585, 357, \dodoi{10.1038/s41586-020-2649-2}

\bibitem[{{Haywood} {et~al.}(2014){Haywood}, {Collier Cameron}, {Queloz}, {Barros}, {Deleuil}, {Fares}, {Gillon}, {Lanza}, {Lovis}, {Moutou}, {Pepe}, {Pollacco}, {Santerne}, {S{\'e}gransan}, \& {Unruh}}]{Haywood_2014}
{Haywood}, R.~D., {Collier Cameron}, A., {Queloz}, D., {et~al.} 2014, \mnras, 443, 2517, \dodoi{10.1093/mnras/stu1320}

\bibitem[{Henning {et~al.}(2009)Henning, O'Connell, \& Sasselov}]{henning_tidally_2009}
Henning, W.~G., O'Connell, R.~J., \& Sasselov, D.~D. 2009, The Astrophysical Journal, 707, 1000, \dodoi{10.1088/0004-637X/707/2/1000}

\bibitem[{{Higson} {et~al.}(2019){Higson}, {Handley}, {Hobson}, \& {Lasenby}}]{Higson_2019}
{Higson}, E., {Handley}, W., {Hobson}, M., \& {Lasenby}, A. 2019, Statistics and Computing, 29, 891, \dodoi{10.1007/s11222-018-9844-0}

\bibitem[{{Huber} {et~al.}(2017){Huber}, {Czesla}, \& {Schmitt}}]{Huber_2017}
{Huber}, K.~F., {Czesla}, S., \& {Schmitt}, J.~H.~M.~M. 2017, \aap, 597, A113, \dodoi{10.1051/0004-6361/201629699}

\bibitem[{{Hunter}(2007)}]{Hunter_2007}
{Hunter}, J.~D. 2007, Computing in Science and Engineering, 9, 90, \dodoi{10.1109/MCSE.2007.55}

\bibitem[{{Jenkins} {et~al.}(2016){Jenkins}, {Twicken}, {McCauliff}, {Campbell}, {Sanderfer}, {Lung}, {Mansouri-Samani}, {Girouard}, {Tenenbaum}, {Klaus}, {Smith}, {Caldwell}, {Chacon}, {Henze}, {Heiges}, {Latham}, {Morgan}, {Swade}, {Rinehart}, \& {Vanderspek}}]{Jenkins_2016}
{Jenkins}, J.~M., {Twicken}, J.~D., {McCauliff}, S., {et~al.} 2016, in Society of Photo-Optical Instrumentation Engineers (SPIE) Conference Series, Vol. 9913, Software and Cyberinfrastructure for Astronomy IV, ed. G.~{Chiozzi} \& J.~C. {Guzman}, 99133E, \dodoi{10.1117/12.2233418}

\bibitem[{{Kipping}(2010)}]{Kipping_2010}
{Kipping}, D.~M. 2010, \mnras, 407, 301, \dodoi{10.1111/j.1365-2966.2010.16894.x}

\bibitem[{{Kipping}(2013)}]{Kipping_2013}
---. 2013, \mnras, 435, 2152, \dodoi{10.1093/mnras/stt1435}

\bibitem[{{Kipping}(2014)}]{Kipping_2014}
---. 2014, \mnras, 440, 2164, \dodoi{10.1093/mnras/stu318}

\bibitem[{{Kopparapu} {et~al.}(2014){Kopparapu}, {Ramirez}, {SchottelKotte}, {Kasting}, {Domagal-Goldman}, \& {Eymet}}]{Kopparapu_2014}
{Kopparapu}, R.~K., {Ramirez}, R.~M., {SchottelKotte}, J., {et~al.} 2014, \apjl, 787, L29, \dodoi{10.1088/2041-8205/787/2/L29}

\bibitem[{{Kopparapu} {et~al.}(2016){Kopparapu}, {Wolf}, {Haqq-Misra}, {Yang}, {Kasting}, {Meadows}, {Terrien}, \& {Mahadevan}}]{Kopparapu_2016}
{Kopparapu}, R.~k., {Wolf}, E.~T., {Haqq-Misra}, J., {et~al.} 2016, \apj, 819, 84, \dodoi{10.3847/0004-637X/819/1/84}

\bibitem[{{Kopparapu} {et~al.}(2013){Kopparapu}, {Ramirez}, {Kasting}, {Eymet}, {Robinson}, {Mahadevan}, {Terrien}, {Domagal-Goldman}, {Meadows}, \& {Deshpande}}]{Kopparapu_2013}
{Kopparapu}, R.~K., {Ramirez}, R., {Kasting}, J.~F., {et~al.} 2013, \apj, 765, 131, \dodoi{10.1088/0004-637X/765/2/131}

\bibitem[{{Kostov} {et~al.}(2019){Kostov}, {Schlieder}, {Barclay}, {Quintana}, {Col{\'o}n}, {Brande}, {Collins}, {Feinstein}, {Hadden}, {Kane}, {Kreidberg}, {Kruse}, {Lam}, {Matthews}, {Montet}, {Pozuelos}, {Stassun}, {Winters}, {Ricker}, {Vanderspek}, {Latham}, {Seager}, {Winn}, {Jenkins}, {Afanasev}, {Armstrong}, {Arney}, {Boyd}, {Barentsen}, {Barkaoui}, {Batalha}, {Beichman}, {Bayliss}, {Burke}, {Burdanov}, {Cacciapuoti}, {Carson}, {Charbonneau}, {Christiansen}, {Ciardi}, {Clampin}, {Collins}, {Conti}, {Coughlin}, {Covone}, {Crossfield}, {Delrez}, {Domagal-Goldman}, {Dressing}, {Ducrot}, {Essack}, {Everett}, {Fauchez}, {Foreman-Mackey}, {Gan}, {Gilbert}, {Gillon}, {Gonzales}, {Hamann}, {Hedges}, {Hocutt}, {Hoffman}, {Horch}, {Horne}, {Howell}, {Hynes}, {Ireland}, {Irwin}, {Isopi}, {Jensen}, {Jehin}, {Kaltenegger}, {Kielkopf}, {Kopparapu}, {Lewis}, {Lopez}, {Lissauer}, {Mann}, {Mallia}, {Mandell}, {Matson}, {Mazeh}, {Monsue}, {Moran}, {Moran}, {Morley}, {Morris}, {Muirhead}, {Mukai}, {Mullally}, {Mullally},
  {Murray}, {Narita}, {Palle}, {Pidhorodetska}, {Quinn}, {Relles}, {Rinehart}, {Ritsko}, {Rodriguez}, {Rowden}, {Rowe}, {Sebastian}, {Sefako}, {Shahaf}, {Shporer}, {Ta{\~n}{\'o}n Reyes}, {Tenenbaum}, {Ting}, {Twicken}, {van Belle}, {Vega}, {Volosin}, {Walkowicz}, \& {Youngblood}}]{Kostov_2019}
{Kostov}, V.~B., {Schlieder}, J.~E., {Barclay}, T., {et~al.} 2019, \aj, 158, 32, \dodoi{10.3847/1538-3881/ab2459}

\bibitem[{{Kreidberg}(2015)}]{Kreidberg_2015}
{Kreidberg}, L. 2015, \pasp, 127, 1161, \dodoi{10.1086/683602}

\bibitem[{{Lange}(2023)}]{Lange_2023}
{Lange}, J.~U. 2023, \mnras, 525, 3181, \dodoi{10.1093/mnras/stad2441}

\bibitem[{{Lillo-Box} {et~al.}(2020){Lillo-Box}, {Figueira}, {Leleu}, {Acu{\~n}a}, {Faria}, {Hara}, {Santos}, {Correia}, {Robutel}, {Deleuil}, {Barrado}, {Sousa}, {Bonfils}, {Mousis}, {Almenara}, {Astudillo-Defru}, {Marcq}, {Udry}, {Lovis}, \& {Pepe}}]{Lillo-Box2020}
{Lillo-Box}, J., {Figueira}, P., {Leleu}, A., {et~al.} 2020, \aap, 642, A121

\bibitem[{{Lim} {et~al.}(2023){Lim}, {Benneke}, {Doyon}, {MacDonald}, {Piaulet}, {Artigau}, {Coulombe}, {Radica}, {L'Heureux}, {Albert}, {Rackham}, {de Wit}, {Salhi}, {Roy}, {Flagg}, {Fournier-Tondreau}, {Taylor}, {Cook}, {Lafreni{\`e}re}, {Cowan}, {Kaltenegger}, {Rowe}, {Espinoza}, {Dang}, \& {Darveau-Bernier}}]{Lim_2023}
{Lim}, O., {Benneke}, B., {Doyon}, R., {et~al.} 2023, \apjl, 955, L22, \dodoi{10.3847/2041-8213/acf7c4}

\bibitem[{{Louca} {et~al.}(2023){Louca}, {Miguel}, {Tsai}, {Froning}, {Loyd}, \& {France}}]{Louca_2023}
{Louca}, A.~J., {Miguel}, Y., {Tsai}, S.-M., {et~al.} 2023, \mnras, 521, 3333, \dodoi{10.1093/mnras/stac1220}

\bibitem[{{Lovis} \& {Fischer}(2010)}]{Lovis_2010}
{Lovis}, C., \& {Fischer}, D. 2010, in Exoplanets, ed. S.~{Seager}, 27--53

\bibitem[{{Luger} \& {Barnes}(2015)}]{Luger-Barnes_2015}
{Luger}, R., \& {Barnes}, R. 2015, Astrobiology, 15, 119, \dodoi{10.1089/ast.2014.1231}

\bibitem[{{Luo} {et~al.}(2024){Luo}, {Dorn}, \& {Deng}}]{Luo_2024}
{Luo}, H., {Dorn}, C., \& {Deng}, J. 2024, arXiv e-prints, arXiv:2401.16394, \dodoi{10.48550/arXiv.2401.16394}

\bibitem[{{Luque} {et~al.}(2021){Luque}, {Serrano}, {Molaverdikhani}, {Nixon}, {Livingston}, {Guenther}, {Pall{\'e}}, {Madhusudhan}, {Nowak}, {Korth}, {Cochran}, {Hirano}, {Chaturvedi}, {Goffo}, {Albrecht}, {Barrag{\'a}n}, {Brice{\~n}o}, {Cabrera}, {Charbonneau}, {Cloutier}, {Collins}, {Collins}, {Col{\'o}n}, {Crossfield}, {Csizmadia}, {Dai}, {Deeg}, {Esposito}, {Fridlund}, {Gandolfi}, {Georgieva}, {Glidden}, {Goeke}, {Grziwa}, {Hatzes}, {Henze}, {Howell}, {Irwin}, {Jenkins}, {Jensen}, {K{\'a}bath}, {Kidwell}, {Kielkopf}, {Knudstrup}, {Lam}, {Latham}, {Lissauer}, {Mann}, {Matthews}, {Mireles}, {Narita}, {Paegert}, {Persson}, {Redfield}, {Ricker}, {Rodler}, {Schlieder}, {Scott}, {Seager}, {{\v{S}}ubjak}, {Tan}, {Ting}, {Vanderspek}, {Van Eylen}, {Winn}, \& {Ziegler}}]{Luque_2021}
{Luque}, R., {Serrano}, L.~M., {Molaverdikhani}, K., {et~al.} 2021, \aap, 645, A41, \dodoi{10.1051/0004-6361/202039455}

\bibitem[{{Luque} {et~al.}(2022){Luque}, {Nowak}, {Hirano}, {Kossakowski}, {Pall{\'e}}, {Nixon}, {Morello}, {Amado}, {Albrecht}, {Caballero}, {Cifuentes}, {Cochran}, {Deeg}, {Dreizler}, {Esparza-Borges}, {Fukui}, {Gandolfi}, {Goffo}, {Guenther}, {Hatzes}, {Henning}, {Kabath}, {Kawauchi}, {Korth}, {Kotani}, {Kudo}, {Kuzuhara}, {Lafarga}, {Lam}, {Livingston}, {Morales}, {Muresan}, {Murgas}, {Narita}, {Osborne}, {Parviainen}, {Passegger}, {Persson}, {Quirrenbach}, {Redfield}, {Reffert}, {Reiners}, {Ribas}, {Serrano}, {Tamura}, {Van Eylen}, {Watanabe}, \& {Zapatero Osorio}}]{Luque_2022}
{Luque}, R., {Nowak}, G., {Hirano}, T., {et~al.} 2022, \aap, 666, A154, \dodoi{10.1051/0004-6361/202244426}

\bibitem[{{Mann} {et~al.}(2015){Mann}, {Feiden}, {Gaidos}, {Boyajian}, \& {von Braun}}]{Mann_2015}
{Mann}, A.~W., {Feiden}, G.~A., {Gaidos}, E., {Boyajian}, T., \& {von Braun}, K. 2015, \apj, 804, 64, \dodoi{10.1088/0004-637X/804/1/64}

\bibitem[{{Mann} {et~al.}(2019){Mann}, {Dupuy}, {Kraus}, {Gaidos}, {Ansdell}, {Ireland}, {Rizzuto}, {Hung}, {Dittmann}, {Factor}, {Feiden}, {Martinez}, {Ru{\'\i}z-Rodr{\'\i}guez}, \& {Thao}}]{Mann_2019}
{Mann}, A.~W., {Dupuy}, T., {Kraus}, A.~L., {et~al.} 2019, \apj, 871, 63, \dodoi{10.3847/1538-4357/aaf3bc}

\bibitem[{{Marconi} {et~al.}(2024){Marconi}, {Abreu}, {Adibekyan}, {Alberti}, {Albrecht}, {Alcaniz}, {Aliverti}, {Allende Prieto}, {Alvarado-Gomez}, {Alves}, {Amado}, {Amate}, {Andersen}, {Antoniucci}, {Artigau}, {Bailet}, {Baker}, {Baldini}, {Balestra}, {Barnes}, {Baron}, {Barros}, {Bauer}, {Beaulieu}, {Bellido-Tirado}, {Benneke}, {Bensby}, {Bergin}, {Berio}, {Biazzo}, {Bigot}, {Bik}, {Birkby}, {Blind}, {Boebion}, {Boisse}, {Bolmont}, {Bolton}, {Bonaglia}, {Bonfils}, {Bonhomme}, {Borsa}, {Bouret}, {Brandeker}, {Brandner}, {Broeg}, {Brogi}, {Brousseau}, {Brucalassi}, {Brynnel}, {Buchhave}, {Buscher}, {Cabona}, {Cabral}, {Calderone}, {Calvo-Ortega}, {Cantalloube}, {Canto Martins}, {Carbonaro}, {Caujolle}, {Chauvin}, {Chazelas}, {Cheffot}, {Cheng}, {Chiavassa}, {Christensen}, {Cirami}, {Cirasuolo}, {Cook}, {Cooke}, {Coretti}, {Covino}, {Cowan}, {Cresci}, {Cristiani}, {Cunha Parro}, {Cupani}, {D'Odorico}, {Dadi}, {de Castro Le{\~a}o}, {De Cia}, {De Medeiros}, {Debras}, {Debus}, {Delorme}, {Demangeon}, {Derie},
  {Dessauges-Zavadsky}, {Di Marcantonio}, {Di Stefano}, {Dionies}, {Domiciano de Souza}, {Doyon}, {Dunn}, {Egner}, {Ehrenreich}, {Faria}, {Ferruzzi}, {Feruglio}, {Fisher}, {Fontana}, {Frank}, {Fuesslein}, {Fumagalli}, {Fusco}, {Fynbo}, {Gabella}, {Gaessler}, {Gallo}, {Gao}, {Genolet}, {Genoni}, {Giacobbe}, {Giro}, {Gon{\c{c}}alves}, {Gonzalez}, {Gonz{\'a}lez-Hern{\'a}ndez}, {Gouvret}, {Gracia T{\'e}mich}, {Haehnelt}, {Haniff}, {Hatzes}, {Helled}, {Hoeijmakers}, {Hughes}, {Huke}, {Ivanisenko}, {J{\"a}rvinen}, {J{\"a}rvinen}, {Kaminski}, {Kern}, {Knoche}, {Kordt}, {Korhonen}, {Korn}, {Kouach}, {Kowzan}, {Kreidberg}, {Landoni}, {Lanotte}, {Lavail}, {Lavie}, {Lee}, {Lehmitz}, {Li}, {Li}, {Liske}, {Lovis}, {Lucatello}, {Lunney}, {MacIntosh}, {Madhusudhan}, {Magrini}, {Maiolino}, {Maldonado}, {Malo}, {Man}, {Marquart}, {Marques}, {Marques}, {Martinez}, {Martins}, {Martins}, {Martins}, {Maslowski}, {Mason}, {Mason}, {McCracken}, {Melo e Sousa}, {Mergo}, {Micela}, {Milakovi{\'c}}, {Molli{\`e}re}, {Monteiro},
  {Montgomery}, {Mordasini}, {Morin}, {Mucciarelli}, {Murphy}, {N'Diaye}, {Nardetto}, {Neichel}, {Neri}, {Niedzielski}, {Niemczura}, {Nisini}, {Nortmann}, {Noterdaeme}, {Nunes}, {Oggioni}, {Olchewsky}, {Oliva}, {{\"O}nel}, {Origlia}, {{\"O}stlin}, {Ouellette}, {Pall{\'e}}, {Papaderos}, {Pariani}, \& {Pasquini}}]{Marconi_2024}
{Marconi}, A., {Abreu}, M., {Adibekyan}, V., {et~al.} 2024, in Society of Photo-Optical Instrumentation Engineers (SPIE) Conference Series, Vol. 13096, Ground-based and Airborne Instrumentation for Astronomy X, ed. J.~J. {Bryant}, K.~{Motohara}, \& J.~R.~D. {Vernet}, 1309613, \dodoi{10.1117/12.3017966}

\bibitem[{{Mayor} {et~al.}(2003){Mayor}, {Pepe}, {Queloz}, {Bouchy}, {Rupprecht}, {Lo Curto}, {Avila}, {Benz}, {Bertaux}, {Bonfils}, {Dall}, {Dekker}, {Delabre}, {Eckert}, {Fleury}, {Gilliotte}, {Gojak}, {Guzman}, {Kohler}, {Lizon}, {Longinotti}, {Lovis}, {Megevand}, {Pasquini}, {Reyes}, {Sivan}, {Sosnowska}, {Soto}, {Udry}, {van Kesteren}, {Weber}, \& {Weilenmann}}]{Mayor_2003}
{Mayor}, M., {Pepe}, F., {Queloz}, D., {et~al.} 2003, The Messenger, 114, 20

\bibitem[{{Ment} {et~al.}(2019){Ment}, {Dittmann}, {Astudillo-Defru}, {Charbonneau}, {Irwin}, {Bonfils}, {Murgas}, {Almenara}, {Forveille}, {Agol}, {Ballard}, {Berta-Thompson}, {Bouchy}, {Cloutier}, {Delfosse}, {Doyon}, {Dressing}, {Esquerdo}, {Haywood}, {Kipping}, {Latham}, {Lovis}, {Newton}, {Pepe}, {Rodriguez}, {Santos}, {Tan}, {Udry}, {Winters}, \& {W{\"u}nsche}}]{Ment_2019}
{Ment}, K., {Dittmann}, J.~A., {Astudillo-Defru}, N., {et~al.} 2019, \aj, 157, 32, \dodoi{10.3847/1538-3881/aaf1b110.48550/arXiv.1808.00485}

\bibitem[{Moore(2003)}]{moore_tidal_2003}
Moore, W.~B. 2003, Journal of Geophysical Research (Planets), 108, 5096, \dodoi{10.1029/2002JE001943}

\bibitem[{{Nicholls} {et~al.}(2025){Nicholls}, {Guimond}, {Hay}, {Chatterjee}, {Lichtenberg}, \& {Pierrehumbert}}]{Nicholls_2025}
{Nicholls}, H., {Guimond}, C.~M., {Hay}, H. C.~F.~C., {et~al.} 2025, arXiv e-prints, arXiv:2505.03604, \dodoi{10.48550/arXiv.2505.03604}

\bibitem[{{Osborne} {et~al.}(2024){Osborne}, {Van Eylen}, {Goffo}, {Gandolfi}, {Nowak}, {Persson}, {Livingston}, {Weeks}, {Pall{\'e}}, {Luque}, {Hellier}, {Carleo}, {Redfield}, {Hirano}, {Garbaccio Gili}, {Alarcon}, {Barrag{\'a}n}, {Casasayas-Barris}, {D{\'\i}az}, {Esposito}, {Knudstrup}, {Jenkins}, {Murgas}, {Orell-Miquel}, {Rodler}, {Serrano}, {Stangret}, {Albrecht}, {Alqasim}, {Cochran}, {Deeg}, {Fridlund}, {Hatzes}, {Korth}, \& {Lam}}]{Osborne_2024}
{Osborne}, H.~L.~M., {Van Eylen}, V., {Goffo}, E., {et~al.} 2024, \mnras, 527, 11138, \dodoi{10.1093/mnras/stad3837}

\bibitem[{{Ostberg} {et~al.}(2023){Ostberg}, {Kane}, {Li}, {Schwieterman}, {Hill}, {Bott}, {Dalba}, {Fetherolf}, {Head}, \& {Unterborn}}]{Ostberg_2023}
{Ostberg}, C., {Kane}, S.~R., {Li}, Z., {et~al.} 2023, \aj, 165, 168, \dodoi{10.3847/1538-3881/acbfaf}

\bibitem[{{Owen} \& {Campos Estrada}(2020)}]{Owen_2020}
{Owen}, J.~E., \& {Campos Estrada}, B. 2020, \mnras, 491, 5287, \dodoi{10.1093/mnras/stz3435}

\bibitem[{{Owen} \& {Wu}(2017)}]{Owen_2017}
{Owen}, J.~E., \& {Wu}, Y. 2017, \apj, 847, 29, \dodoi{10.3847/1538-4357/aa890a}

\bibitem[{{Palle} {et~al.}(2023){Palle}, {Biazzo}, {Bolmont}, {Molliere}, {Poppenhaeger}, {Birkby}, {Brogi}, {Chauvin}, {Chiavassa}, {Hoeijmakers}, {Lellouch}, {Lovis}, {Maiolino}, {Nortmann}, {Parviainen}, {Pino}, {Turbet}, {Wender}, {Albrecht}, {Antoniucci}, {Barros}, {Beaudoin}, {Benneke}, {Boisse}, {Bonomo}, {Borsa}, {Brandeker}, {Brandner}, {Buchhave}, {Cheffot}, {Deborde}, {Debras}, {Doyon}, {Di Marcantonio}, {Giacobbe}, {Gonzalez Hernandez}, {Helled}, {Kreidberg}, {Machado}, {Maldonado}, {Marconi}, {Canto Martins}, {Miceli}, {Mordasini}, {N'Diaye}, {Niedzielski}, {Nisini}, {Origlia}, {Peroux}, {Pietrow}, {Pinna}, {Rauscher}, {Reffert}, {Rousselot}, {Sanna}, {Simonnin}, {Suarez Mascareno}, {Zanutta}, \& {Zechmeister}}]{Palle_2023}
{Palle}, E., {Biazzo}, K., {Bolmont}, E., {et~al.} 2023, arXiv e-prints, arXiv:2311.17075, \dodoi{10.48550/arXiv.2311.17075}

\bibitem[{{Parviainen} {et~al.}(2024){Parviainen}, {Luque}, \& {Palle}}]{Parviainen_2024}
{Parviainen}, H., {Luque}, R., \& {Palle}, E. 2024, \mnras, 527, 5693, \dodoi{10.1093/mnras/stad3504}

\bibitem[{{Pepe} {et~al.}(2021){Pepe}, {Cristiani}, {Rebolo}, {Santos}, {Dekker}, {Cabral}, {Di Marcantonio}, {Figueira}, {Lo Curto}, {Lovis}, {Mayor}, {M{\'e}gevand}, {Molaro}, {Riva}, {Zapatero Osorio}, {Amate}, {Manescau}, {Pasquini}, {Zerbi}, {Adibekyan}, {Abreu}, {Affolter}, {Alibert}, {Aliverti}, {Allart}, {Allende Prieto}, {{\'A}lvarez}, {Alves}, {Avila}, {Baldini}, {Bandy}, {Barros}, {Benz}, {Bianco}, {Borsa}, {Bourrier}, {Bouchy}, {Broeg}, {Calderone}, {Cirami}, {Coelho}, {Conconi}, {Coretti}, {Cumani}, {Cupani}, {D'Odorico}, {Damasso}, {Deiries}, {Delabre}, {Demangeon}, {Dumusque}, {Ehrenreich}, {Faria}, {Fragoso}, {Genolet}, {Genoni}, {G{\'e}nova Santos}, {Gonz{\'a}lez Hern{\'a}ndez}, {Hughes}, {Iwert}, {Kerber}, {Knudstrup}, {Landoni}, {Lavie}, {Lillo-Box}, {Lizon}, {Maire}, {Martins}, {Mehner}, {Micela}, {Modigliani}, {Monteiro}, {Monteiro}, {Moschetti}, {Murphy}, {Nunes}, {Oggioni}, {Oliveira}, {Oshagh}, {Pall{\'e}}, {Pariani}, {Poretti}, {Rasilla}, {Rebord{\~a}o}, {Redaelli}, {Santana Tschudi},
  {Santin}, {Santos}, {S{\'e}gransan}, {Schmidt}, {Segovia}, {Sosnowska}, {Sozzetti}, {Sousa}, {Span{\`o}}, {Su{\'a}rez Mascare{\~n}o}, {Tabernero}, {Tenegi}, {Udry}, \& {Zanutta}}]{Pepe_2021}
{Pepe}, F., {Cristiani}, S., {Rebolo}, R., {et~al.} 2021, \aap, 645, A96, \dodoi{10.1051/0004-6361/202038306}

\bibitem[{{Peterson} {et~al.}(2023){Peterson}, {Benneke}, {Collins}, {Piaulet}, {Crossfield}, {Ali-Dib}, {Christiansen}, {Gagn{\'e}}, {Faherty}, {Kite}, {Dressing}, {Charbonneau}, {Murgas}, {Cointepas}, {Almenara}, {Bonfils}, {Kane}, {Werner}, {Gorjian}, {Roy}, {Shporer}, {Pozuelos}, {Socia}, {Cloutier}, {Dietrich}, {Irwin}, {Weiss}, {Waalkes}, {Berta-Thomson}, {Evans}, {Apai}, {Parviainen}, {Pall{\'e}}, {Narita}, {Howard}, {Dragomir}, {Barkaoui}, {Gillon}, {Jehin}, {Ducrot}, {Benkhaldoun}, {Fukui}, {Mori}, {Nishiumi}, {Kawauchi}, {Ricker}, {Latham}, {Winn}, {Seager}, {Isaacson}, {Bixel}, {Gibbs}, {Jenkins}, {Smith}, {Chavez}, {Rackham}, {Henning}, {Gabor}, {Chen}, {Espinoza}, {Jensen}, {Collins}, {Schwarz}, {Conti}, {Wang}, {Kielkopf}, {Mao}, {Horne}, {Sefako}, {Quinn}, {Moldovan}, {Fausnaugh}, {F{\.z}{\.z}r{\'e}sz}, \& {Barclay}}]{peterson_temperate_2023}
{Peterson}, M.~S., {Benneke}, B., {Collins}, K., {et~al.} 2023, \nat, 617, 701, \dodoi{10.1038/s41586-023-05934-8}

\bibitem[{{Piaulet} {et~al.}(2023){Piaulet}, {Benneke}, {Almenara}, {Dragomir}, {Knutson}, {Thorngren}, {Peterson}, {Crossfield}, {Kempton}, {Kubyshkina}, {Howard}, {Angus}, {Isaacson}, {Weiss}, {Beichman}, {Fortney}, {Fossati}, {Lammer}, {McCullough}, {Morley}, \& {Wong}}]{Piaulet_2023}
{Piaulet}, C., {Benneke}, B., {Almenara}, J.~M., {et~al.} 2023, Nature Astronomy, 7, 206, \dodoi{10.1038/s41550-022-01835-4}

\bibitem[{{Pidhorodetska} {et~al.}(2021){Pidhorodetska}, {Moran}, {Schwieterman}, {Barclay}, {Fauchez}, {Lewis}, {Quintana}, {Villanueva}, {Domagal-Goldman}, {Schlieder}, {Gilbert}, {Kane}, \& {Kostov}}]{Pidhorodetska_2021}
{Pidhorodetska}, D., {Moran}, S.~E., {Schwieterman}, E.~W., {et~al.} 2021, \aj, 162, 169, \dodoi{10.3847/1538-3881/ac1171}

\bibitem[{{Plotnykov} \& {Valencia}(2020)}]{Plotnykov_2020}
{Plotnykov}, M., \& {Valencia}, D. 2020, \mnras, 499, 932, \dodoi{10.1093/mnras/staa2615}

\bibitem[{{Plotnykov} \& {Valencia}(2024)}]{Plotnykov_2024}
---. 2024, \mnras, 530, 3488, \dodoi{10.1093/mnras/stae993}

\bibitem[{{Quanz} {et~al.}(2022){Quanz}, {Ottiger}, {Fontanet}, {Kammerer}, {Menti}, {Dannert}, {Gheorghe}, {Absil}, {Airapetian}, {Alei}, {Allart}, {Angerhausen}, {Blumenthal}, {Buchhave}, {Cabrera}, {Carri{\'o}n-Gonz{\'a}lez}, {Chauvin}, {Danchi}, {Dandumont}, {Defr{\'e}re}, {Dorn}, {Ehrenreich}, {Ertel}, {Fridlund}, {Garc{\'\i}a Mu{\~n}oz}, {Gasc{\'o}n}, {Girard}, {Glauser}, {Grenfell}, {Guidi}, {Hagelberg}, {Helled}, {Ireland}, {Janson}, {Kopparapu}, {Korth}, {Kozakis}, {Kraus}, {L{\'e}ger}, {Leedj{\"a}rv}, {Lichtenberg}, {Lillo-Box}, {Linz}, {Liseau}, {Loicq}, {Mahendra}, {Malbet}, {Mathew}, {Mennesson}, {Meyer}, {Mishra}, {Molaverdikhani}, {Noack}, {Oza}, {Pall{\'e}}, {Parviainen}, {Quirrenbach}, {Rauer}, {Ribas}, {Rice}, {Romagnolo}, {Rugheimer}, {Schwieterman}, {Serabyn}, {Sharma}, {Stassun}, {Szul{\'a}gyi}, {Wang}, {Wunderlich}, {Wyatt}, \& {LIFE Collaboration}}]{Quanz_2022}
{Quanz}, S.~P., {Ottiger}, M., {Fontanet}, E., {et~al.} 2022, \aap, 664, A21, \dodoi{10.1051/0004-6361/202140366}

\bibitem[{{Quick} {et~al.}(2020){Quick}, {Roberge}, {Mlinar}, \& {Hedman}}]{Quick_2020}
{Quick}, L.~C., {Roberge}, A., {Mlinar}, A.~B., \& {Hedman}, M.~M. 2020, \pasp, 132, 084402, \dodoi{10.1088/1538-3873/ab9504}

\bibitem[{{Quirrenbach} {et~al.}(2018){Quirrenbach}, {Amado}, {Ribas}, {Reiners}, {Caballero}, {Seifert}, {Aceituno}, {Azzaro}, {Baroch}, {Barrado}, {Bauer}, {Becerril}, {B{\`e}jar}, {Ben{\'\i}tez}, {Brinkm{\"o}ller}, {Cardona Guill{\'e}n}, {Cifuentes}, {Colom{\'e}}, {Cort{\'e}s-Contreras}, {Czesla}, {Dreizler}, {Fr{\"o}lich}, {Fuhrmeister}, {Galad{\'\i}-Enr{\'\i}quez}, {Gonz{\'a}lez Hern{\'a}ndez}, {Gonz{\'a}lez Peinado}, {Guenther}, {de Guindos}, {Hagen}, {Hatzes}, {Hauschildt}, {Helmling}, {Henning}, {Herbort}, {Hern{\'a}ndez Casta{\~n}o}, {Herrero}, {Hintz}, {Jeffers}, {Johnson}, {de Juan}, {Kaminski}, {Klahr}, {K{\"u}rster}, {Lafarga}, {Sairam}, {Lamp{\'o}n}, {Lara}, {Launhardt}, {L{\'o}pez del Fresno}, {L{\'o}pez-Puertas}, {Luque}, {Mandel}, {Marfil}, {Mart{\'\i}n}, {Mart{\'\i}n-Ruiz}, {Mathar}, {Montes}, {Morales}, {Nagel}, {Nortmann}, {Nowak}, {Pall{\'e}}, {Passegger}, {Pavlov}, {Pedraz}, {P{\'e}rez-Medialdea}, {Perger}, {Rebolo}, {Reffert}, {Rodr{\'\i}guez}, {Rodr{\'\i}guez L{\'o}pez}, {Rosich},
  {Sabotta}, {Sadegi}, {Salz}, {S{\'a}nchez-L{\'o}pez}, {Sanz-Forcada}, {Sarkis}, {Sch{\"a}fer}, {Schiller}, {Schmitt}, {Sch{\"o}fer}, {Schweitzer}, {Shulyak}, {Solano}, {Stahl}, {Tala Pinto}, {Trifonov}, {Zapatero Osorio}, {Yan}, {Zechmeister}, {Abell{\'a}n}, {Abril}, {Alonso-Floriano}, {Ammler-von Eiff}, {Anglada-Escud{\'e}}, {Anwand-Heerwart}, {Arroyo-Torres}, {Berdi{\~n}as}, {Bergondy}, {Bl{\"u}mcke}, {del Burgo}, {Cano}, {Carro}, {C{\'a}rdenas}, {Casal}, {Claret}, {D{\'\i}ez-Alonso}, {Doellinger}, {Dorda}, {Feiz}, {Fern{\'a}ndez}, {Ferro}, {Gaisn{\'e}}, {Gallardo}, {G{\'a}lvez-Ortiz}, {Garc{\'\i}a-Piquer}, {Garc{\'\i}a-Vargas}, {Garrido}, {Gesa}, {G{\'o}mez Galera}, {Gonz{\'a}lez-{\'A}lvarez}, {Gonz{\'a}lez-Cuesta}, {Grohnert}, {Gr{\"o}zinger}, {Gu{\`a}rdia}, {Guijarro}, {Hedrosa}, {Hermann}, {Hermelo}, {Hern{\'a}ndez Arab{\'\i}}, {Hern{\'a}ndez Hernando}, {Hidalgo}, {Holgado}, {Huber}, {Huber}, {Huke}, {Kehr}, {Kim}, {Klein}, {Kl{\"u}ter}, {Klutsch}, {Labarga}, {Labiche}, {Lamert}, {Laun}, {L{\'a}zaro},
  {Lemke}, {Lenzen}, {Llamas}, {Lizon}, {Lodieu}, {L{\'o}pez Gonz{\'a}lez}, {L{\'o}pez-Morales}, {L{\'o}pez Salas}, {L{\'o}pez-Santiago}, {Mag{\'a}n Madinabeitia}, {Mall}, {Mancini}, {Mar{\'\i}n Molina}, {Mart{\'\i}nez-Rodr{\'\i}guez}, {Maroto Fern{\'a}ndez}, {Marvin}, {Mirabet}, {Moreno-Raya}, {Moya}, {Mundt}, {Naranjo}, {Panduro}, {Pascual}, {P{\'e}rez-Calpena}, {Perryman}, {Pluto}, {Ram{\'o}n}, {Redondo}, {Reinhart}, {Rhode}, {Rix}, {Rodler}, {Rohloff}, {S{\'a}nchez-Blanco}, {S{\'a}nchez Carrasco}, {Sarmiento}, {Schmidt}, {Storz}, {Strachan}, {St{\"u}rmer}, {Su{\'a}rez}, {Tabernero}, {Tal-Or}, {Tulloch}, {Ulbrich}, {Veredas}, {Vico Linares}, {Vidal-Dasilva}, {Vilardell}, {Wagner}, {Winkler}, {Wolthoff}, \& {Xu}}]{Quirrenbach_2018}
{Quirrenbach}, A., {Amado}, P.~J., {Ribas}, I., {et~al.} 2018, in Society of Photo-Optical Instrumentation Engineers (SPIE) Conference Series, Vol. 10702, Ground-based and Airborne Instrumentation for Astronomy VII, ed. C.~J. {Evans}, L.~{Simard}, \& H.~{Takami}, 107020W, \dodoi{10.1117/12.2313689}

\bibitem[{{Radica} {et~al.}(2022){Radica}, {Artigau}, {Lafreni{\'e}re}, {Cadieux}, {Cook}, {Doyon}, {Amado}, {Caballero}, {Henning}, {Quirrenbach}, {Reiners}, \& {Ribas}}]{Radica_2022}
{Radica}, M., {Artigau}, {\'E}., {Lafreni{\'e}re}, D., {et~al.} 2022, \mnras, 517, 5050, \dodoi{10.1093/mnras/stac3024}

\bibitem[{{Radica} {et~al.}(2025){Radica}, {Piaulet-Ghorayeb}, {Taylor}, {Coulombe}, {Benneke}, {Albert}, {Artigau}, {Cowan}, {Doyon}, {Lafreni{\`e}re}, {L'Heureux}, \& {Lim}}]{Radica_2025}
{Radica}, M., {Piaulet-Ghorayeb}, C., {Taylor}, J., {et~al.} 2025, \apjl, 979, L5, \dodoi{10.3847/2041-8213/ada381}

\bibitem[{{Rajpaul} {et~al.}(2015){Rajpaul}, {Aigrain}, {Osborne}, {Reece}, \& {Roberts}}]{Rajpaul_2015}
{Rajpaul}, V., {Aigrain}, S., {Osborne}, M.~A., {Reece}, S., \& {Roberts}, S. 2015, \mnras, 452, 2269, \dodoi{10.1093/mnras/stv1428}

\bibitem[{{Rajpaul} {et~al.}(2024){Rajpaul}, {Barrag{\'a}n}, \& {Zicher}}]{Rajpaul_2024}
{Rajpaul}, V.~M., {Barrag{\'a}n}, O., \& {Zicher}, N. 2024, \mnras, 530, 4665, \dodoi{10.1093/mnras/stae778}

\bibitem[{Renner {et~al.}(2000)Renner, Evans, \& Hirth}]{renner_rheologically_2000}
Renner, J., Evans, B., \& Hirth, G. 2000, Earth and Planetary Science Letters, 181, 585, \dodoi{10.1016/S0012-821X(00)00222-3}

\bibitem[{{Ricker} {et~al.}(2015){Ricker}, {Winn}, {Vanderspek}, {Latham}, {Bakos}, {Bean}, {Berta-Thompson}, {Brown}, {Buchhave}, {Butler}, {Butler}, {Chaplin}, {Charbonneau}, {Christensen-Dalsgaard}, {Clampin}, {Deming}, {Doty}, {De Lee}, {Dressing}, {Dunham}, {Endl}, {Fressin}, {Ge}, {Henning}, {Holman}, {Howard}, {Ida}, {Jenkins}, {Jernigan}, {Johnson}, {Kaltenegger}, {Kawai}, {Kjeldsen}, {Laughlin}, {Levine}, {Lin}, {Lissauer}, {MacQueen}, {Marcy}, {McCullough}, {Morton}, {Narita}, {Paegert}, {Palle}, {Pepe}, {Pepper}, {Quirrenbach}, {Rinehart}, {Sasselov}, {Sato}, {Seager}, {Sozzetti}, {Stassun}, {Sullivan}, {Szentgyorgyi}, {Torres}, {Udry}, \& {Villasenor}}]{Ricker_2015}
{Ricker}, G.~R., {Winn}, J.~N., {Vanderspek}, R., {et~al.} 2015, Journal of Astronomical Telescopes, Instruments, and Systems, 1, 014003, \dodoi{10.1117/1.JATIS.1.1.014003}

\bibitem[{{Ridgway} {et~al.}(2023){Ridgway}, {Zamyatina}, {Mayne}, {Manners}, {Lambert}, {Braam}, {Drummond}, {H{\'e}brard}, {Palmer}, \& {Kohary}}]{Ridgway_2023}
{Ridgway}, R.~J., {Zamyatina}, M., {Mayne}, N.~J., {et~al.} 2023, \mnras, 518, 2472, \dodoi{10.1093/mnras/stac3105}

\bibitem[{{Rodr{\'\i}guez Mart{\'\i}nez} {et~al.}(2023){Rodr{\'\i}guez Mart{\'\i}nez}, {Martin}, {Gaudi}, {Schulze}, {Asnodkar}, {Boley}, \& {Ballard}}]{Rodriguez_2023}
{Rodr{\'\i}guez Mart{\'\i}nez}, R., {Martin}, D.~V., {Gaudi}, B.~S., {et~al.} 2023, \aj, 166, 137, \dodoi{10.3847/1538-3881/aced9a}

\bibitem[{{Rogers} {et~al.}(2023){Rogers}, {Schlichting}, \& {Owen}}]{Rogers_2023}
{Rogers}, J.~G., {Schlichting}, H.~E., \& {Owen}, J.~E. 2023, \apjl, 947, L19, \dodoi{10.3847/2041-8213/acc86f}

\bibitem[{{Scarsdale} {et~al.}(2024){Scarsdale}, {Wogan}, {Wakeford}, {Wallack}, {Batalha}, {Alderson}, {Aguichine}, {Wolfgang}, {Teske}, {Moran}, {L{\'o}pez-Morales}, {Kirk}, {Gordon}, {Gao}, {Batalha}, {Alam}, \& {Adams Redai}}]{Scarsdale_2024}
{Scarsdale}, N., {Wogan}, N., {Wakeford}, H.~R., {et~al.} 2024, \aj, 168, 276, \dodoi{10.3847/1538-3881/ad73cf}

\bibitem[{Segatz {et~al.}(1988)Segatz, Spohn, Ross, \& Schubert}]{segatz_tidal_1988}
Segatz, M., Spohn, T., Ross, M.~N., \& Schubert, G. 1988, Icarus, 75, 187, \dodoi{10.1016/0019-1035(88)90001-2}

\bibitem[{{Segura} {et~al.}(2010){Segura}, {Walkowicz}, {Meadows}, {Kasting}, \& {Hawley}}]{Segura_2010}
{Segura}, A., {Walkowicz}, L.~M., {Meadows}, V., {Kasting}, J., \& {Hawley}, S. 2010, Astrobiology, 10, 751, \dodoi{10.1089/ast.2009.0376}

\bibitem[{{Seligman} {et~al.}(2024){Seligman}, {Feinstein}, {Lai}, {Welbanks}, {Taylor}, {Becker}, {Adams}, {Morgan}, \& {Bergner}}]{Seligman_2024}
{Seligman}, D.~Z., {Feinstein}, A.~D., {Lai}, D., {et~al.} 2024, \apj, 961, 22, \dodoi{10.3847/1538-4357/ad0b82}

\bibitem[{{Silva} {et~al.}(2022){Silva}, {Faria}, {Santos}, {Sousa}, {Viana}, {Martins}, {Figueira}, {Lovis}, {Pepe}, {Cristiani}, {Rebolo}, {Allart}, {Cabral}, {Mehner}, {Sozzetti}, {Su{\'a}rez Mascare{\~n}o}, {Martins}, {Ehrenreich}, {M{\'e}gevand}, {Palle}, {Lo Curto}, {Tabernero}, {Lillo-Box}, {Gonz{\'a}lez Hern{\'a}ndez}, {Zapatero Osorio}, {Hara}, {Nunes}, {Di Marcantonio}, {Udry}, {Adibekyan}, \& {Dumusque}}]{Silva_2022}
{Silva}, A.~M., {Faria}, J.~P., {Santos}, N.~C., {et~al.} 2022, \aap, 663, A143, \dodoi{10.1051/0004-6361/202142262}

\bibitem[{Skilling(2006)}]{Skilling_2006}
Skilling, J. 2006, Bayesian Analysis, 1, 833 , \dodoi{10.1214/06-BA127}

\bibitem[{{Skrutskie} {et~al.}(2006){Skrutskie}, {Cutri}, {Stiening}, {Weinberg}, {Schneider}, {Carpenter}, {Beichman}, {Capps}, {Chester}, {Elias}, {Huchra}, {Liebert}, {Lonsdale}, {Monet}, {Price}, {Seitzer}, {Jarrett}, {Kirkpatrick}, {Gizis}, {Howard}, {Evans}, {Fowler}, {Fullmer}, {Hurt}, {Light}, {Kopan}, {Marsh}, {McCallon}, {Tam}, {Van Dyk}, \& {Wheelock}}]{Skrutskie_2006}
{Skrutskie}, M.~F., {Cutri}, R.~M., {Stiening}, R., {et~al.} 2006, \aj, 131, 1163, \dodoi{10.1086/498708}

\bibitem[{{Smith} {et~al.}(2012){Smith}, {Stumpe}, {Van Cleve}, {Jenkins}, {Barclay}, {Fanelli}, {Girouard}, {Kolodziejczak}, {McCauliff}, {Morris}, \& {Twicken}}]{Smith_2012}
{Smith}, J.~C., {Stumpe}, M.~C., {Van Cleve}, J.~E., {et~al.} 2012, \pasp, 124, 1000, \dodoi{10.1086/667697}

\bibitem[{Solomatov \& Moresi(2000)}]{solomatov_scaling_2000}
Solomatov, V.~S., \& Moresi, L.-N. 2000, Journal of Geophysical Research, 105, 21795, \dodoi{10.1029/2000JB900197}

\bibitem[{{Stock} {et~al.}(2023){Stock}, {Kemmer}, {Kossakowski}, {Sabotta}, {Reffert}, \& {Quirrenbach}}]{Stock_2023}
{Stock}, S., {Kemmer}, J., {Kossakowski}, D., {et~al.} 2023, \aap, 674, A108, \dodoi{10.1051/0004-6361/202244629}

\bibitem[{{Stumpe} {et~al.}(2014){Stumpe}, {Smith}, {Catanzarite}, {Van Cleve}, {Jenkins}, {Twicken}, \& {Girouard}}]{Stumpe_2014}
{Stumpe}, M.~C., {Smith}, J.~C., {Catanzarite}, J.~H., {et~al.} 2014, \pasp, 126, 100, \dodoi{10.1086/674989}

\bibitem[{{Stumpe} {et~al.}(2012){Stumpe}, {Smith}, {Van Cleve}, {Twicken}, {Barclay}, {Fanelli}, {Girouard}, {Jenkins}, {Kolodziejczak}, {McCauliff}, \& {Morris}}]{Stumpe_2012}
{Stumpe}, M.~C., {Smith}, J.~C., {Van Cleve}, J.~E., {et~al.} 2012, \pasp, 124, 985, \dodoi{10.1086/667698}

\bibitem[{{Su{\'a}rez Mascare{\~n}o} {et~al.}(2023){Su{\'a}rez Mascare{\~n}o}, {Gonz{\'a}lez-{\'A}lvarez}, {Zapatero Osorio}, {Lillo-Box}, {Faria}, {Passegger}, {Gonz{\'a}lez Hern{\'a}ndez}, {Figueira}, {Sozzetti}, {Rebolo}, {Pepe}, {Santos}, {Cristiani}, {Lovis}, {Silva}, {Ribas}, {Amado}, {Caballero}, {Quirrenbach}, {Reiners}, {Zechmeister}, {Adibekyan}, {Alibert}, {B{\'e}jar}, {Benatti}, {D'Odorico}, {Damasso}, {Delisle}, {Di Marcantonio}, {Dreizler}, {Ehrenreich}, {Hatzes}, {Hara}, {Henning}, {Kaminski}, {L{\'o}pez-Gonz{\'a}lez}, {Martins}, {Micela}, {Montes}, {Pall{\'e}}, {Pedraz}, {Rodr{\'\i}guez}, {Rodr{\'\i}guez-L{\'o}pez}, {Tal-Or}, {Sousa}, \& {Udry}}]{Suarez-Mascarno_2023}
{Su{\'a}rez Mascare{\~n}o}, A., {Gonz{\'a}lez-{\'A}lvarez}, E., {Zapatero Osorio}, M.~R., {et~al.} 2023, \aap, 670, A5, \dodoi{10.1051/0004-6361/202244991}

\bibitem[{{Thiabaud} {et~al.}(2015){Thiabaud}, {Marboeuf}, {Alibert}, {Leya}, \& {Mezger}}]{Thiabaud_2015}
{Thiabaud}, A., {Marboeuf}, U., {Alibert}, Y., {Leya}, I., \& {Mezger}, K. 2015, \aap, 580, A30, \dodoi{10.1051/0004-6361/201525963}

\bibitem[{{Trotta}(2008)}]{Trotta2008}
{Trotta}, R. 2008, Contemporary Physics, 49, 71, \dodoi{10.1080/00107510802066753}

\bibitem[{{Turbet} {et~al.}(2023){Turbet}, {Fauchez}, {Leconte}, {Bolmont}, {Chaverot}, {Forget}, {Millour}, {Selsis}, {Charnay}, {Ducrot}, {Gillon}, {Maurel}, \& {Villanueva}}]{Turbet_2023}
{Turbet}, M., {Fauchez}, T.~J., {Leconte}, J., {et~al.} 2023, \aap, 679, A126, \dodoi{10.1051/0004-6361/202347539}

\bibitem[{{Unterborn} {et~al.}(2016){Unterborn}, {Dismukes}, \& {Panero}}]{Unterborn_2016}
{Unterborn}, C.~T., {Dismukes}, E.~E., \& {Panero}, W.~R. 2016, \apj, 819, 32, \dodoi{10.3847/0004-637X/819/1/32}

\bibitem[{{Van Eylen} \& {Albrecht}(2015)}]{Van-Eylen_2015}
{Van Eylen}, V., \& {Albrecht}, S. 2015, \apj, 808, 126, \dodoi{10.1088/0004-637X/808/2/126}

\bibitem[{{VanWyngarden} \& {Cloutier}(2024)}]{VanWyngarden_2024}
{VanWyngarden}, M., \& {Cloutier}, R. 2024, \aj, 168, 154, \dodoi{10.3847/1538-3881/ad6903}

\bibitem[{{Veeder} {et~al.}(2012){Veeder}, {Davies}, {Matson}, {Johnson}, {Williams}, \& {Radebaugh}}]{veeder_io_2012}
{Veeder}, G.~J., {Davies}, A.~G., {Matson}, D.~L., {et~al.} 2012, \icarus, 219, 701, \dodoi{10.1016/j.icarus.2012.04.004}

\bibitem[{{Venturini} {et~al.}(2024){Venturini}, {Ronco}, {Guilera}, {Haldemann}, {Mordasini}, \& {Miller Bertolami}}]{Venturini_2024}
{Venturini}, J., {Ronco}, M.~P., {Guilera}, O.~M., {et~al.} 2024, \aap, 686, L9, \dodoi{10.1051/0004-6361/202349088}

\bibitem[{{Virtanen} {et~al.}(2020){Virtanen}, {Gommers}, {Oliphant}, {Haberland}, {Reddy}, {Cournapeau}, {Burovski}, {Peterson}, {Weckesser}, {Bright}, {van der Walt}, {Brett}, {Wilson}, {Millman}, {Mayorov}, {Nelson}, {Jones}, {Kern}, {Larson}, {Carey}, {Polat}, {Feng}, {Moore}, {VanderPlas}, {Laxalde}, {Perktold}, {Cimrman}, {Henriksen}, {Quintero}, {Harris}, {Archibald}, {Ribeiro}, {Pedregosa}, {van Mulbregt}, \& {SciPy 1. 0 Contributors}}]{Virtanen_2020}
{Virtanen}, P., {Gommers}, R., {Oliphant}, T.~E., {et~al.} 2020, Nature Methods, 17, 261, \dodoi{10.1038/s41592-019-0686-2}

\bibitem[{{Wang} {et~al.}(2018){Wang}, {Lineweaver}, \& {Ireland}}]{Wang_2018}
{Wang}, H.~S., {Lineweaver}, C.~H., \& {Ireland}, T.~R. 2018, \icarus, 299, 460, \dodoi{10.1016/j.icarus.2017.08.024}

\bibitem[{{Winn}(2010)}]{Winn_2010}
{Winn}, J.~N. 2010, in Exoplanets, ed. S.~{Seager}, 55--77, \dodoi{10.48550/arXiv.1001.2010}

\bibitem[{{Yang} {et~al.}(2014){Yang}, {Bou{\'e}}, {Fabrycky}, \& {Abbot}}]{Yang_2014}
{Yang}, J., {Bou{\'e}}, G., {Fabrycky}, D.~C., \& {Abbot}, D.~S. 2014, \apjl, 787, L2, \dodoi{10.1088/2041-8205/787/1/L2}

\bibitem[{{Yang} {et~al.}(2013){Yang}, {Cowan}, \& {Abbot}}]{Yang_2013}
{Yang}, J., {Cowan}, N.~B., \& {Abbot}, D.~S. 2013, \apjl, 771, L45, \dodoi{10.1088/2041-8205/771/2/L45}

\bibitem[{{Yang} \& {Hu}(2024)}]{Yang_2024}
{Yang}, J., \& {Hu}, R. 2024, \apj, 966, 189, \dodoi{10.3847/1538-4357/ad35c8}

\bibitem[{{Zacharias} {et~al.}(2013){Zacharias}, {Finch}, {Girard}, {Henden}, {Bartlett}, {Monet}, \& {Zacharias}}]{Zacharias_2013}
{Zacharias}, N., {Finch}, C.~T., {Girard}, T.~M., {et~al.} 2013, \aj, 145, 44, \dodoi{10.1088/0004-6256/145/2/44}

\bibitem[{{Zahnle} \& {Catling}(2017)}]{Zahnle_2017}
{Zahnle}, K.~J., \& {Catling}, D.~C. 2017, \apj, 843, 122, \dodoi{10.3847/1538-4357/aa7846}

\bibitem[{{Zechmeister} {et~al.}(2018){Zechmeister}, {Reiners}, {Amado}, {Azzaro}, {Bauer}, {B{\'e}jar}, {Caballero}, {Guenther}, {Hagen}, {Jeffers}, {Kaminski}, {K{\"u}rster}, {Launhardt}, {Montes}, {Morales}, {Quirrenbach}, {Reffert}, {Ribas}, {Seifert}, {Tal-Or}, \& {Wolthoff}}]{Zechmeister_2018}
{Zechmeister}, M., {Reiners}, A., {Amado}, P.~J., {et~al.} 2018, \aap, 609, A12, \dodoi{10.1051/0004-6361/201731483}

\bibitem[{{Zechmeister} {et~al.}(2019){Zechmeister}, {Dreizler}, {Ribas}, {Reiners}, {Caballero}, {Bauer}, {B{\'e}jar}, {Gonz{\'a}lez-Cuesta}, {Herrero}, {Lalitha}, {L{\'o}pez-Gonz{\'a}lez}, {Luque}, {Morales}, {Pall{\'e}}, {Rodr{\'\i}guez}, {Rodr{\'\i}guez L{\'o}pez}, {Tal-Or}, {Anglada-Escud{\'e}}, {Quirrenbach}, {Amado}, {Abril}, {Aceituno}, {Aceituno}, {Alonso-Floriano}, {Ammler-von Eiff}, {Antona Jim{\'e}nez}, {Anwand-Heerwart}, {Arroyo-Torres}, {Azzaro}, {Baroch}, {Barrado}, {Becerril}, {Ben{\'\i}tez}, {Berdi{\~n}as}, {Bergond}, {Bluhm}, {Brinkm{\"o}ller}, {del Burgo}, {Calvo Ortega}, {Cano}, {Cardona Guill{\'e}n}, {Carro}, {C{\'a}rdenas V{\'a}zquez}, {Casal}, {Casasayas-Barris}, {Casanova}, {Chaturvedi}, {Cifuentes}, {Claret}, {Colom{\'e}}, {Cort{\'e}s-Contreras}, {Czesla}, {D{\'\i}ez-Alonso}, {Dorda}, {Fern{\'a}ndez}, {Fern{\'a}ndez-Mart{\'\i}n}, {Fuhrmeister}, {Fukui}, {Galad{\'\i}-Enr{\'\i}quez}, {Gallardo Cava}, {Garcia de la Fuente}, {Garcia-Piquer}, {Garc{\'\i}a Vargas}, {Gesa}, {G{\'o}ngora
  Rueda}, {Gonz{\'a}lez-{\'A}lvarez}, {Gonz{\'a}lez Hern{\'a}ndez}, {Gonz{\'a}lez-Peinado}, {Gr{\"o}zinger}, {Gu{\`a}rdia}, {Guijarro}, {de Guindos}, {Hatzes}, {Hauschildt}, {Hedrosa}, {Helmling}, {Henning}, {Hermelo}, {Hern{\'a}ndez Arabi}, {Hern{\'a}ndez Casta{\~n}o}, {Hern{\'a}ndez Otero}, {Hintz}, {Huke}, {Huber}, {Jeffers}, {Johnson}, {de Juan}, {Kaminski}, {Kemmer}, {Kim}, {Klahr}, {Klein}, {Kl{\"u}ter}, {Klutsch}, {Kossakowski}, {K{\"u}rster}, {Labarga}, {Lafarga}, {Llamas}, {Lamp{\'o}n}, {Lara}, {Launhardt}, {L{\'a}zaro}, {Lodieu}, {L{\'o}pez del Fresno}, {L{\'o}pez-Puertas}, {L{\'o}pez Salas}, {L{\'o}pez-Santiago}, {Mag{\'a}n Madinabeitia}, {Mall}, {Mancini}, {Mandel}, {Marfil}, {Mar{\'\i}n Molina}, {Maroto Fern{\'a}ndez}, {Mart{\'\i}n}, {Mart{\'\i}n-Fern{\'a}ndez}, {Mart{\'\i}n-Ruiz}, {Marvin}, {Mirabet}, {Monta{\~n}{\'e}s-Rodr{\'\i}guez}, {Montes}, {Moreno-Raya}, {Nagel}, {Naranjo}, {Narita}, {Nortmann}, {Nowak}, {Ofir}, {Oshagh}, {Panduro}, {Parviainen}, {Pascual}, {Passegger}, {Pavlov}, {Pedraz},
  {P{\'e}rez-Calpena}, {P{\'e}rez Medialdea}, {Perger}, {Perryman}, {Rabaza}, {Ram{\'o}n Ballesta}, {Rebolo}, {Redondo}, {Reffert}, {Reinhardt}, {Rhode}, {Rix}, {Rodler}, {Rodr{\'\i}guez Trinidad}, {Rosich}, {Sadegi}, {S{\'a}nchez-Blanco}, {S{\'a}nchez Carrasco}, {S{\'a}nchez-L{\'o}pez}, {Sanz-Forcada}, {Sarkis}, {Sarmiento}, {Sch{\"a}fer}, {Schmitt}, {Sch{\"o}fer}, {Schweitzer}, {Seifert}, {Shulyak}, {Solano}, {Sota}, {Stahl}, {Stock}, {Strachan}, {Stuber}, {St{\"u}rmer}, {Su{\'a}rez}, {Tabernero}, {Tala Pinto}, {Trifonov}, {Veredas}, {Vico Linares}, {Vilardell}, {Wagner}, {Wolthoff}, {Xu}, {Yan}, \& {Zapatero Osorio}}]{Zechmeister_2019}
{Zechmeister}, M., {Dreizler}, S., {Ribas}, I., {et~al.} 2019, \aap, 627, A49, \dodoi{10.1051/0004-6361/201935460}

\bibitem[{{Zeng} {et~al.}(2016){Zeng}, {Sasselov}, \& {Jacobsen}}]{Zeng_2016}
{Zeng}, L., {Sasselov}, D.~D., \& {Jacobsen}, S.~B. 2016, \apj, 819, 127, \dodoi{10.3847/0004-637X/819/2/127}

\bibitem[{{Zeng} {et~al.}(2019){Zeng}, {Jacobsen}, {Sasselov}, {Petaev}, {Vanderburg}, {Lopez-Morales}, {Perez-Mercader}, {Mattsson}, {Li}, {Heising}, {Bonomo}, {Damasso}, {Berger}, {Cao}, {Levi}, \& {Wordsworth}}]{Zeng_2019}
{Zeng}, L., {Jacobsen}, S.~B., {Sasselov}, D.~D., {et~al.} 2019, Proceedings of the National Academy of Science, 116, 9723, \dodoi{10.1073/pnas.1812905116}

\bibitem[{{Zhou} {et~al.}(2022){Zhou}, {Ma}, {Wang}, \& {Zhu}}]{Zhou_2022}
{Zhou}, L., {Ma}, B., {Wang}, Y., \& {Zhu}, Y. 2022, \aj, 164, 203, \dodoi{10.3847/1538-3881/ac8fe9}

\bibitem[{{Zhou} {et~al.}(2023){Zhou}, {Ma}, {Wang}, \& {Zhu}}]{Zhou_2023}
{Zhou}, L., {Ma}, B., {Wang}, Y.-H., \& {Zhu}, Y.-N. 2023, Research in Astronomy and Astrophysics, 23, 025011, \dodoi{10.1088/1674-4527/acaceb}

\bibitem[{{Zieba} {et~al.}(2023){Zieba}, {Kreidberg}, {Ducrot}, {Gillon}, {Morley}, {Schaefer}, {Tamburo}, {Koll}, {Lyu}, {Acu{\~n}a}, {Agol}, {Iyer}, {Hu}, {Lincowski}, {Meadows}, {Selsis}, {Bolmont}, {Mandell}, \& {Suissa}}]{Zieba_2023}
{Zieba}, S., {Kreidberg}, L., {Ducrot}, E., {et~al.} 2023, \nat, 620, 746, \dodoi{10.1038/s41586-023-06232-z}

\end{thebibliography}
\bibliographystyle{aasjournal}

\appendix
\counterwithin{table}{section}
\counterwithin{figure}{section}

\section{TESS Light Curves and Transit Timing Measurements}

The complete TESS light curves covering 25 sectors is presented in Figure~\ref{fig:tess_gp}. Transit timing measurements described in Section~\ref{sec:ttv} are reported in Table~\ref{table:ttv}.

\begin{figure*}[h!]
  
  \minipage{0.5\textwidth}
  \centering
  \includegraphics[width=1\linewidth]{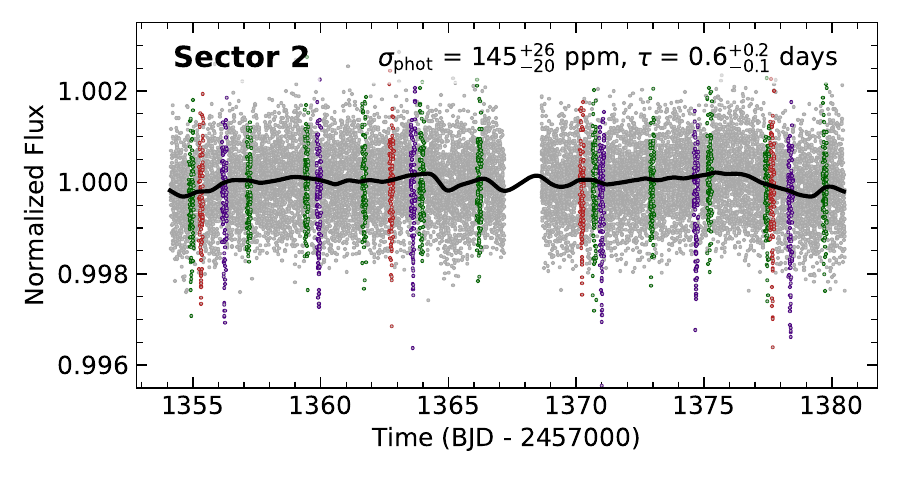}
  \endminipage\hfill
  \minipage{0.5\textwidth}
  \centering
  \includegraphics[width=1\linewidth]{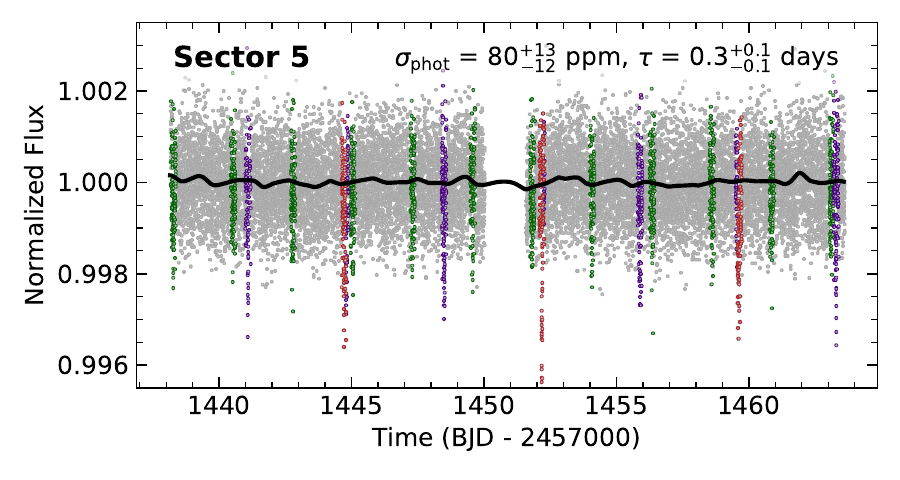}
  \endminipage\hfill\\

  \vspace{-0.3cm}
  
  \minipage{0.5\textwidth}
  \centering
  \includegraphics[width=1\linewidth]{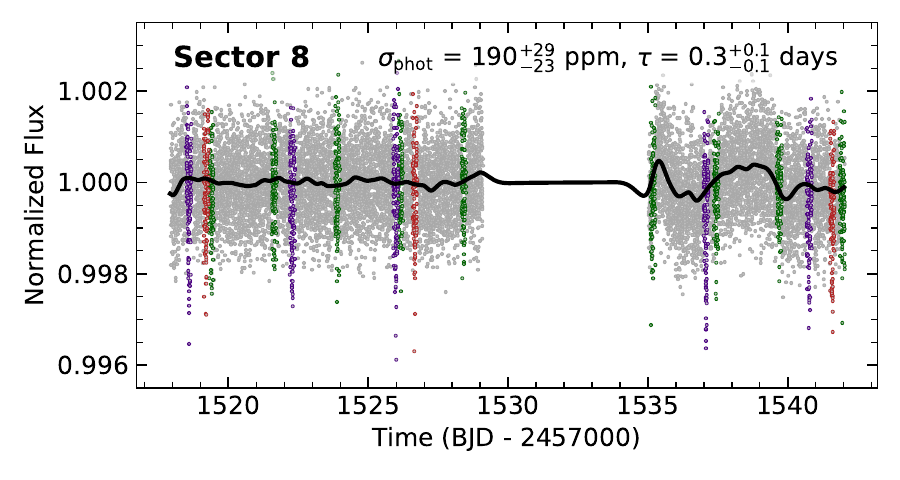}
  \endminipage\hfill
  \minipage{0.5\textwidth}
  \centering
  \includegraphics[width=1\linewidth]{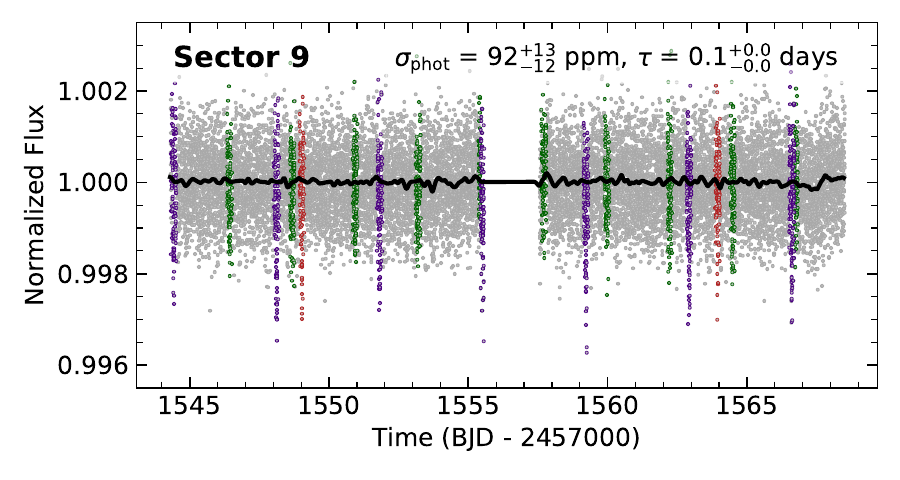}
  \endminipage\hfill\\

  \vspace{-0.3cm}
  
  \minipage{0.5\textwidth}
  \centering
  \includegraphics[width=1\linewidth]{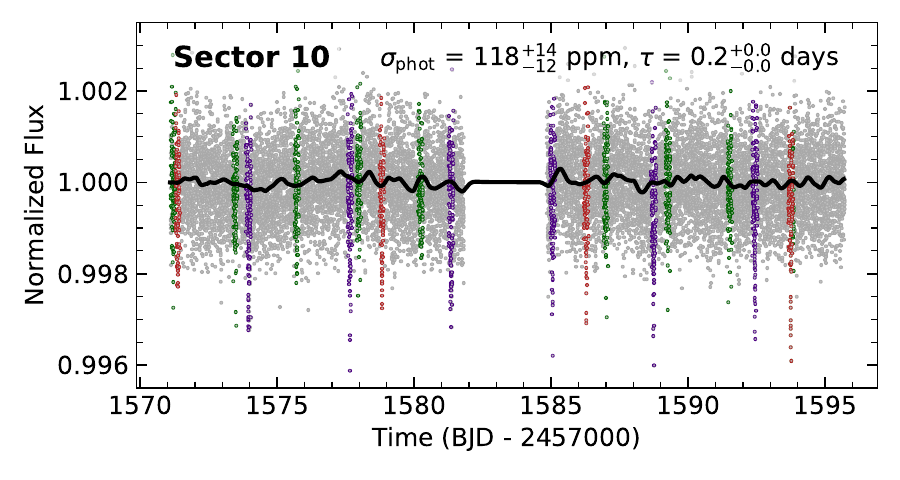}
  \endminipage\hfill
  \minipage{0.5\textwidth}
  \centering
  \includegraphics[width=1\linewidth]{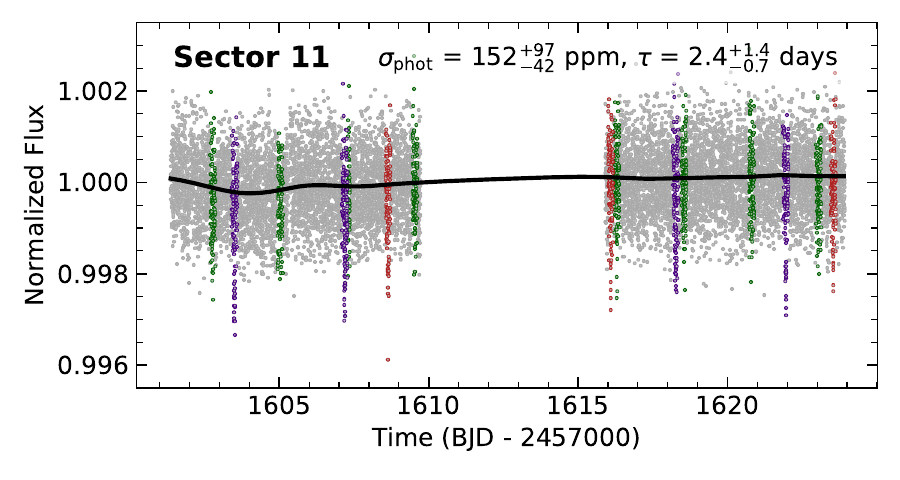}
  \endminipage\hfill\\

  \vspace{-0.3cm}
  
  \minipage{0.5\textwidth}
  \centering
  \includegraphics[width=1\linewidth]{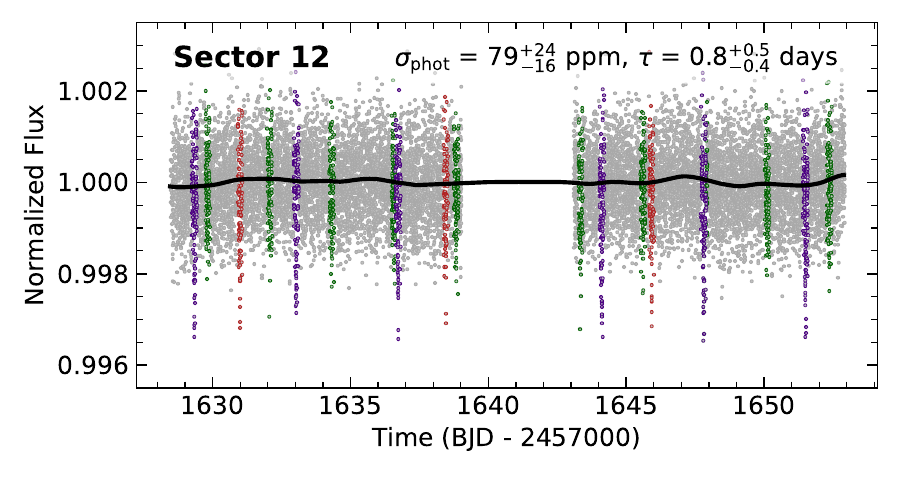}
  \endminipage\hfill
  \minipage{0.5\textwidth}
  \centering
  \includegraphics[width=1\linewidth]{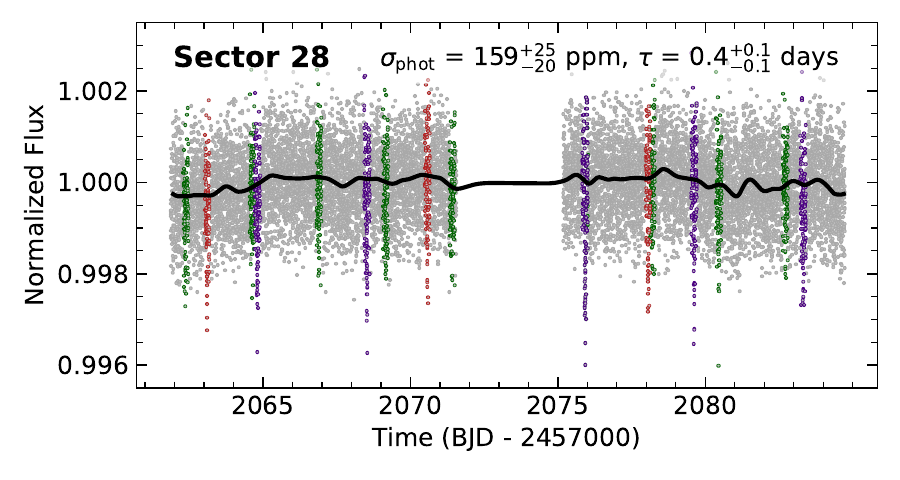}
  \endminipage\hfill\\

  \vspace{-0.3cm}

  \minipage{0.5\textwidth}
  \centering
  \includegraphics[width=1\linewidth]{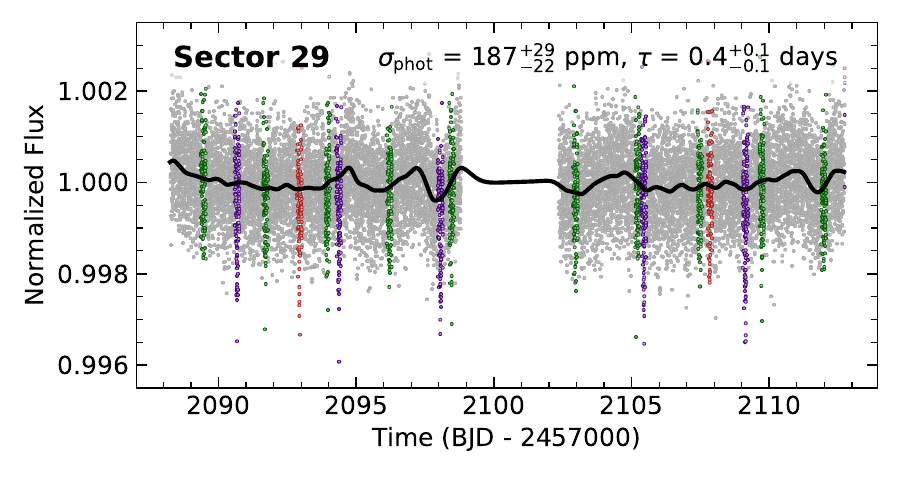}
  \endminipage\hfill
  \minipage{0.5\textwidth}
  \centering
  \includegraphics[width=1\linewidth]{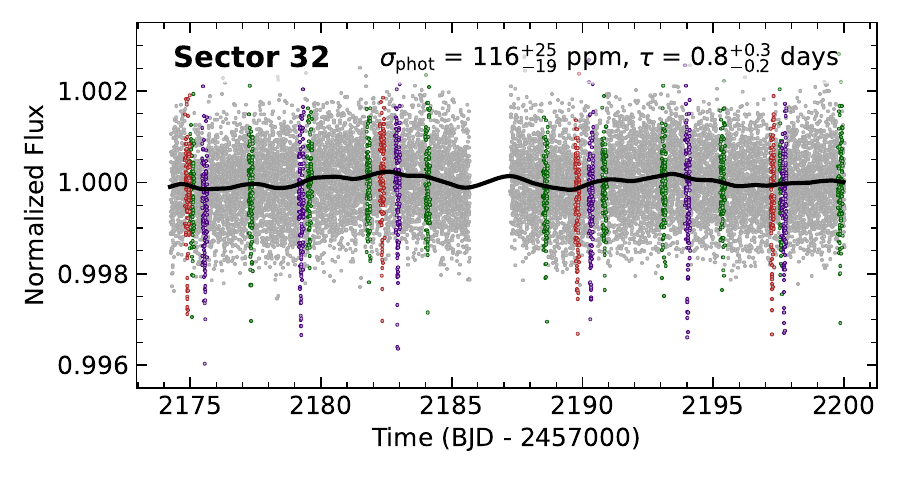}
  \endminipage\hfill\\
  
  \caption{\textit{continue on the next page...}}
  \end{figure*}
  \setcounter{figure}{0}
  \begin{figure*}[h!] 
  
  \minipage{0.5\textwidth}
  \centering
  \includegraphics[width=1\linewidth]{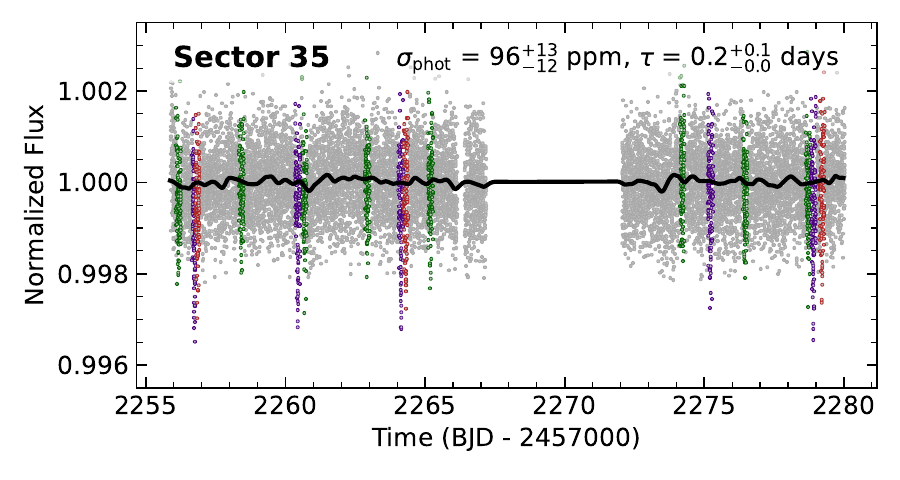}
  \endminipage\hfill
  \minipage{0.5\textwidth}
  \centering
  \includegraphics[width=1\linewidth]{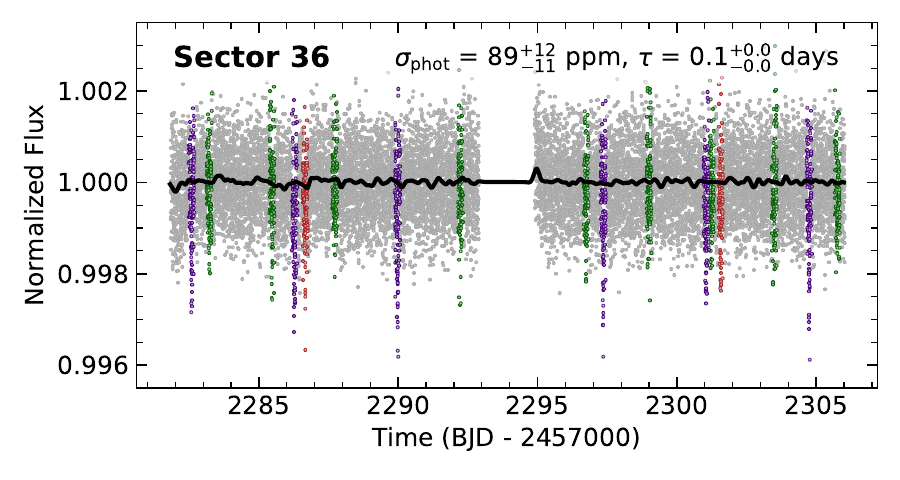}
  \endminipage\hfill\\

  \vspace{-0.3cm}

  \minipage{0.5\textwidth}
  \centering
  \includegraphics[width=1\linewidth]{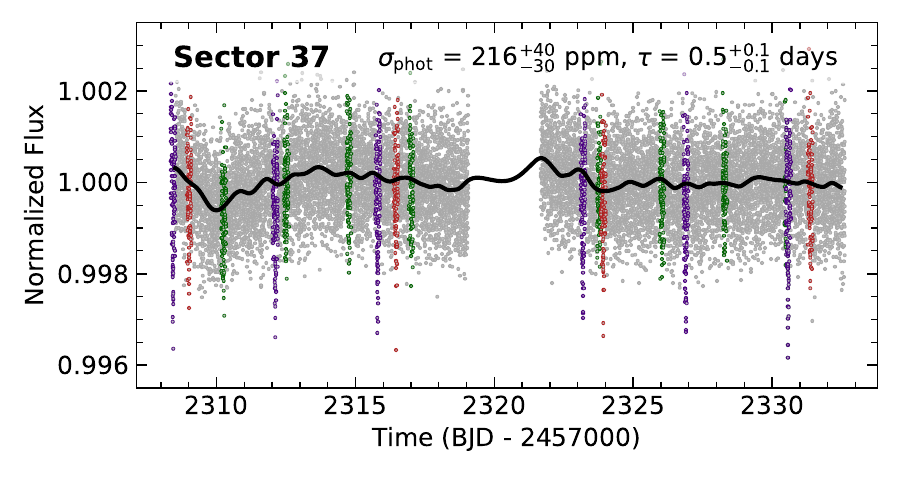}
  \endminipage\hfill
  \minipage{0.5\textwidth}
  \centering
  \includegraphics[width=1\linewidth]{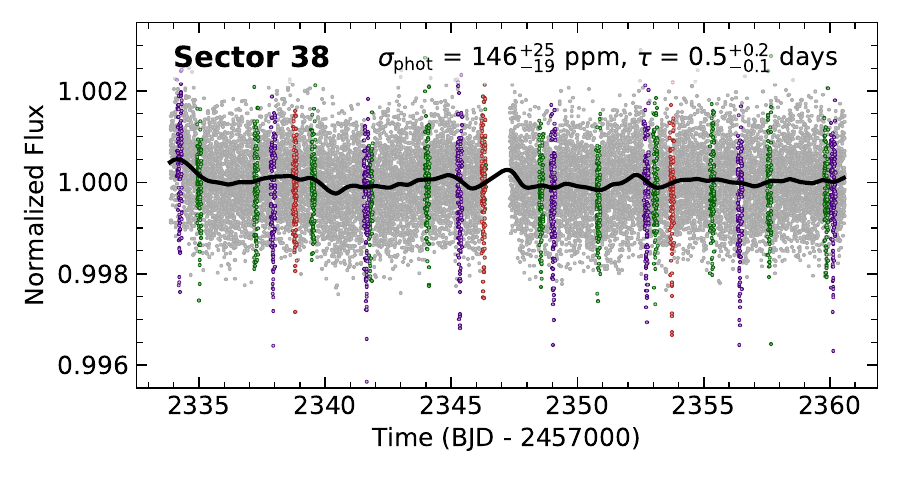}
  \endminipage\hfill\\

  \vspace{-0.3cm}

  \minipage{0.5\textwidth}
  \centering
  \includegraphics[width=1\linewidth]{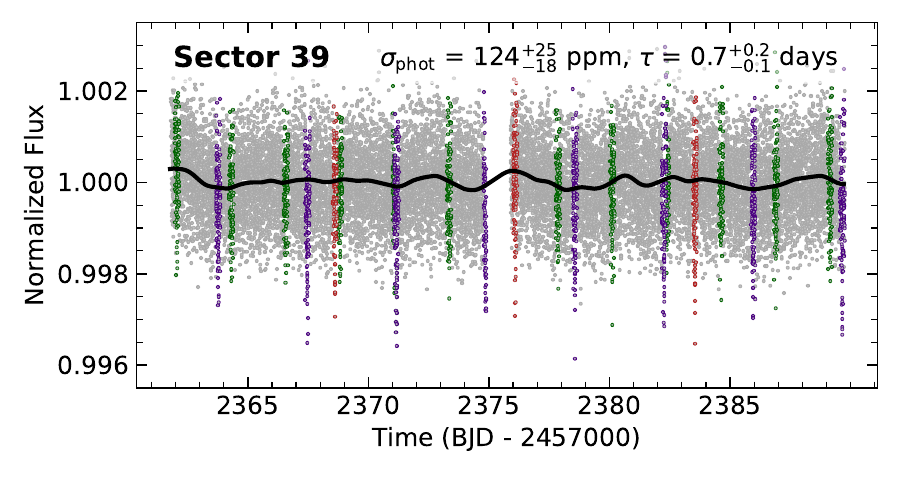}
  \endminipage\hfill
  \minipage{0.5\textwidth}
  \centering
  \includegraphics[width=1\linewidth]{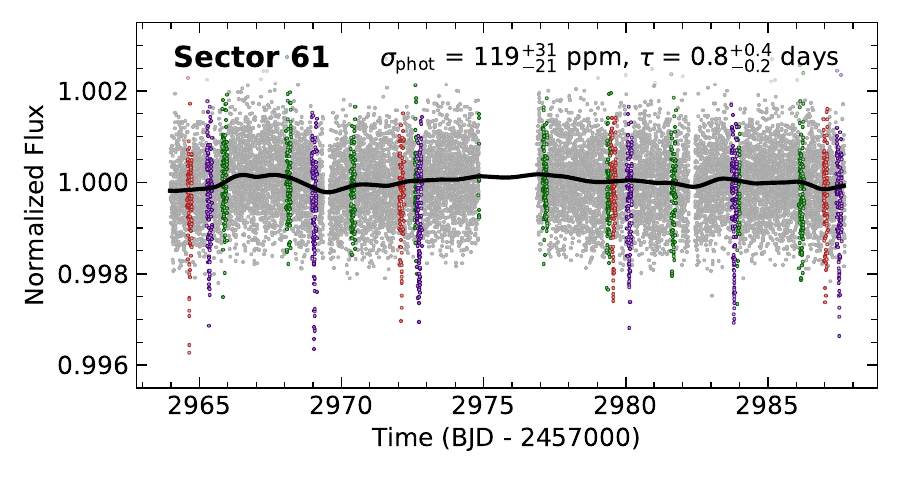}
  \endminipage\hfill\\

  \vspace{-0.3cm}

  \minipage{0.5\textwidth}
  \centering
  \includegraphics[width=1\linewidth]{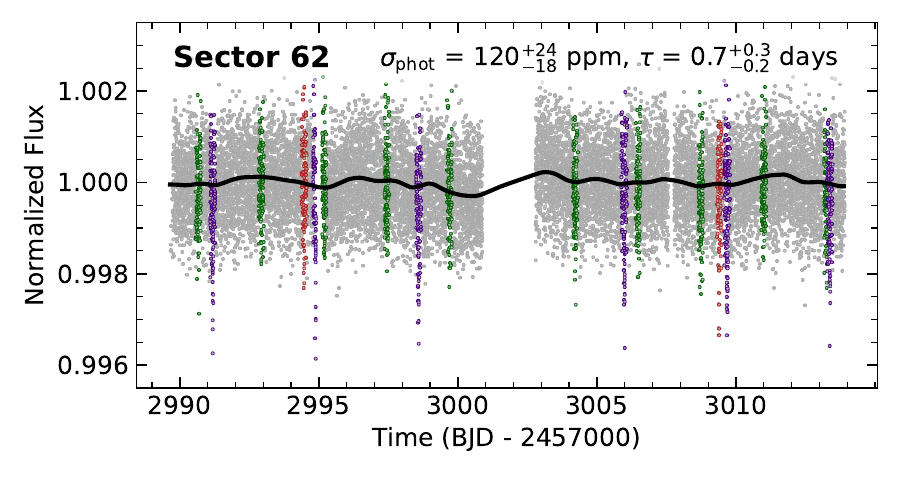}
  \endminipage\hfill
  \minipage{0.5\textwidth}
  \centering
  \includegraphics[width=1\linewidth]{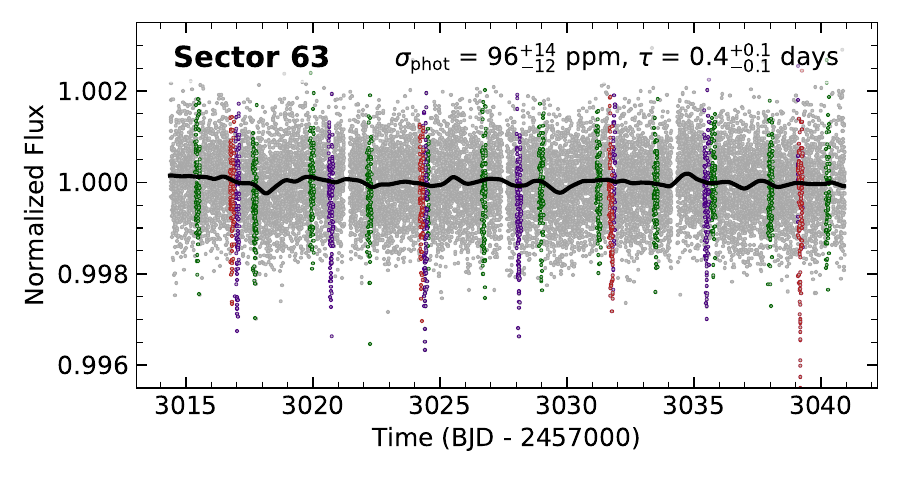}
  \endminipage\hfill\\

  \vspace{-0.3cm}

  \minipage{0.5\textwidth}
  \centering
  \includegraphics[width=1\linewidth]{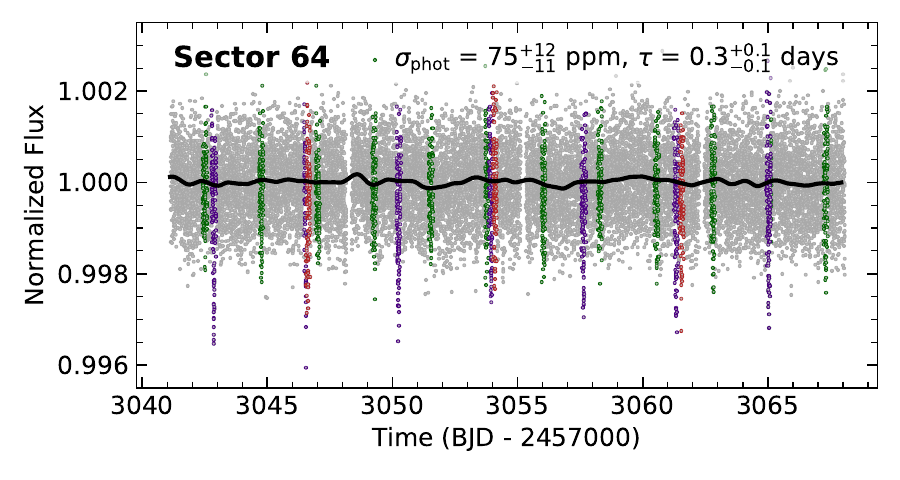}
  \endminipage\hfill
  \minipage{0.5\textwidth}
  \centering
  \includegraphics[width=1\linewidth]{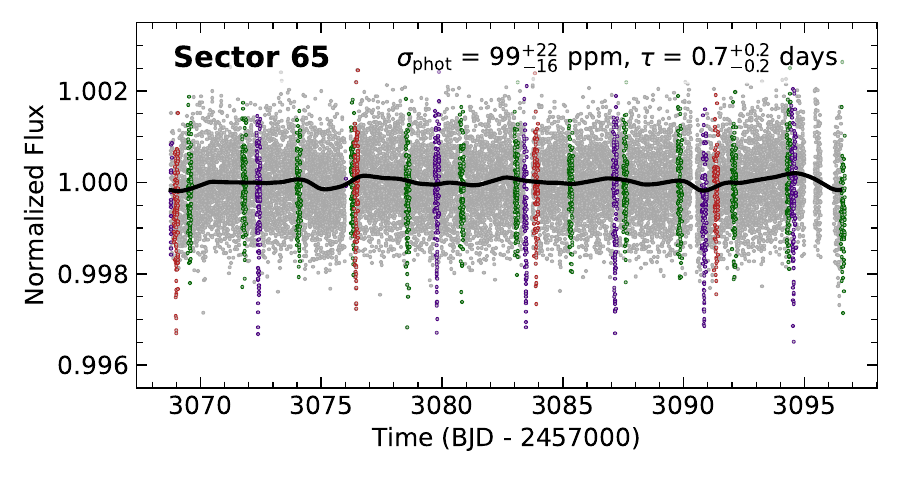}
  \endminipage\hfill\\

  \caption{\textit{continue on the next page...}}
  \end{figure*}
  \setcounter{figure}{0}
  \begin{figure*}[h!]

  \minipage{0.5\textwidth}
  \centering
  \includegraphics[width=1\linewidth]{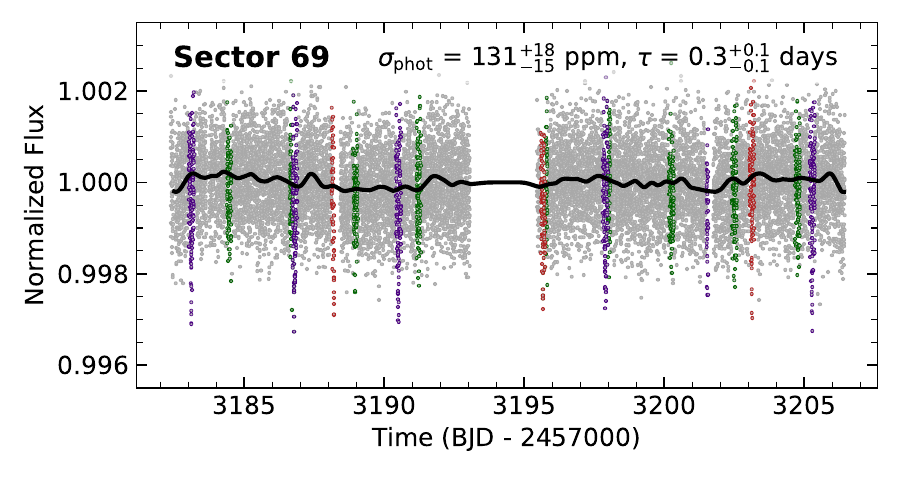}
  \endminipage\hfill
  \minipage{0.5\textwidth}
  \centering
  \includegraphics[width=1\linewidth]{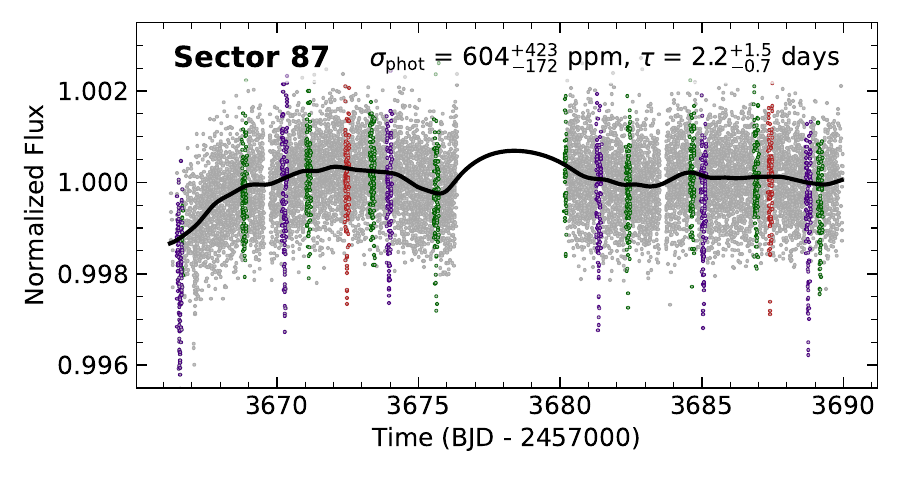}
  \endminipage\hfill\\

  \vspace{-0.3cm}

  \minipage{0.5\textwidth}
  \centering
  \includegraphics[width=1\linewidth]{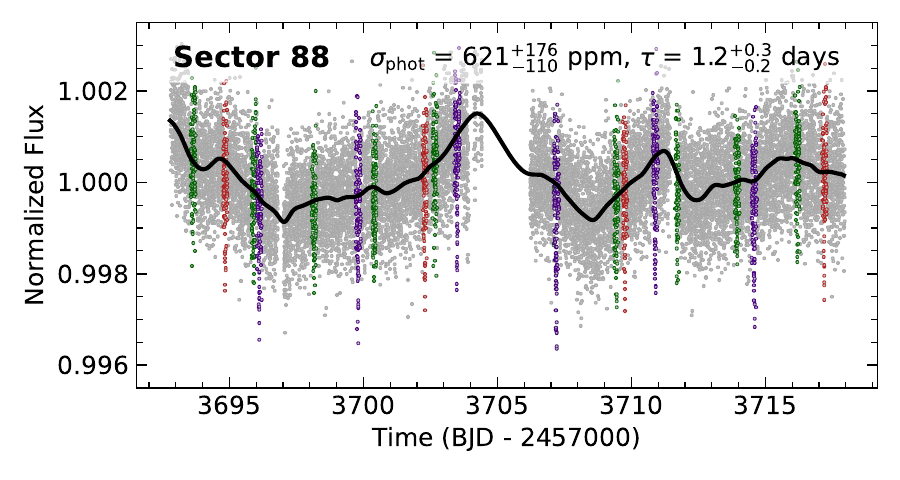}
  \endminipage\hfill
  \minipage{0.5\textwidth}
  \centering
  \includegraphics[width=1\linewidth]{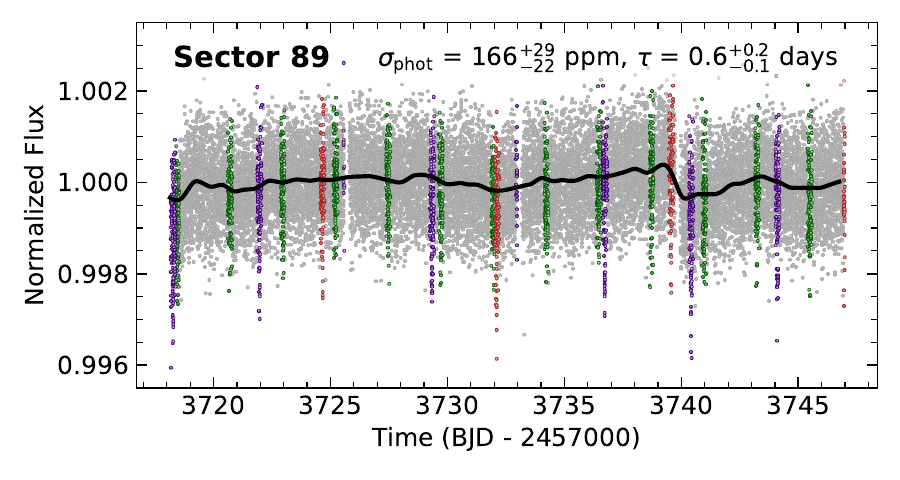}
  \endminipage\hfill\\

  \vspace{-0.3cm}

  \minipage{0.5\textwidth}
  \centering
  \includegraphics[width=1\linewidth]{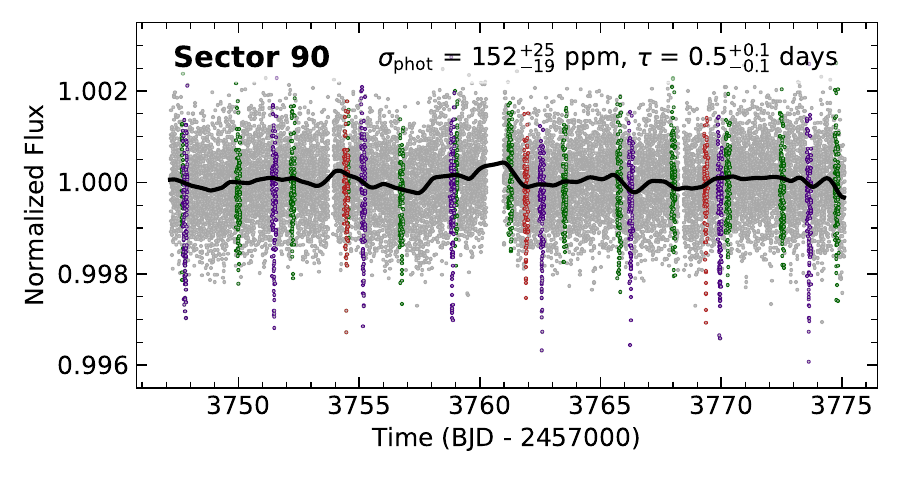}
  \endminipage\hfill

  \caption{Normalized \texttt{PDCSAP} light curve of L~98-59 from TESS between August 23, 2018 (sector 2) and April 9, 2025 (sector 90). Photometry taken during the transit of L~98-59\,b, c, and d are highlighted in green, purple, and red, respectively. The black line is the median GP model of the out-of-transit light curve (details in Sect.~\ref{sec:tess}). The GP amplitude ($\sigma_{\rm phot}$) and coherence timescale ($\tau$) are indicated for each sector. Our GP models were compatible with a baseline flux $f_0 = 1$ and no additional jitter ($\sigma_{\rm jitter} = 0$\,ppm) for all sectors.}
\label{fig:tess_gp}
\end{figure*}

\begin{deluxetable}{cccccc}
\tablecaption{Transit timing of L~98-59\,c and d from TESS\label{table:ttv}}
\tablehead{\colhead{$E_{\rm c}$} &
\colhead{$t_{\rm t,c}$ (BJD)} & \colhead{$\sigma_{t_{\rm t,c}}$ (days)} & \colhead{$E_{\rm d}$} & \colhead{$t_{\rm t,d}$ (BJD)}
 & \colhead{$\sigma_{t_{\rm t,d}}$ (days)}
 }
\startdata
$-4$ & 2458356.20333 & 0.00113 & $-1$ & 2458355.28653 & 0.00121 \\
$-2$ & 2458363.58666 & 0.00125 & 0 & 2458362.73644 & 0.00155 \\
0 & 2458370.96366 & 0.00197 & 1 & 2458370.19064 & 0.00126 \\
2 & 2458378.34446 & 0.00142 & 2 & 2458377.63978 & 0.00110 \\
19 & 2458441.08692 & 0.00113 & 21 & 2458519.20447 & 0.00157 \\
21 & 2458448.47014 & 0.00094 & 25 & 2458549.00914 & 0.00126 \\
23 & 2458455.85064 & 0.00094 & 27 & 2458563.91139 & 0.00114 \\
\ldots & \ldots & \ldots & \ldots & \ldots \\
\enddata
\tablecomments{$E$ refers to the transit epoch number, i.e., the number of transits since a reference epoch $t_0$. The full content of Table~\ref{table:ttv} is available in machine-readable format.}
\end{deluxetable}

\section{Radial Velocity and \dtemp\ Measurements}

The ESPRESSO and HARPS radial velocities and \dtemp\ measurements used in this work are published in Table~\ref{table:rv}.

\begin{deluxetable}{cccccc}
\tablecaption{Radial velocity and \dtemp\ measurements of L~98-59 with HARPS and ESPRESSO\label{table:rv}}
\tablehead{
\colhead{BJD - 2\,400\,000} & \colhead{RV (m\,s$^{-1}$)} & \colhead{$\sigma_{\rm RV}$ (m\,s$^{-1}$)}
 & \colhead{\dtemp\ (K)} & \colhead{$\sigma_{dTemp}$ (K)} & \colhead{Instrument}
 }
\startdata
58408.853661 & $-5886.615$ & 1.419 & $-0.020$ & 0.203 & HARPS\\
58409.844622 & $-5888.186$ & 1.509 & 0.350 & 0.219 & HARPS\\
58412.858886 & $-5887.482$ & 1.383 & $-0.145$ & 0.200 & HARPS\\
\ldots & \ldots & \ldots & \ldots & \ldots \\
58436.805804 & $-5789.985$ & 0.287 & $-0.076$ & 0.064 & ESPRESSOpre\\
58444.839304 & $-5794.605$ & 0.282 & 0.255  & 0.062 & ESPRESSOpre\\
58463.825434 & $-5796.563$ & 0.242 & 0.028  & 0.053 & ESPRESSOpre\\
\ldots & \ldots & \ldots & \ldots & \ldots \\
58754.863546 & $-5786.180$ & 0.248 & $-0.138$ & 0.054 & ESPRESSOpost\\
58792.826929 & $-5792.131$ & 0.237 & 0.277 & 0.053 & ESPRESSOpost\\
58803.811531 & $-5791.298$ & 0.207 & $-0.156$ & 0.045 & ESPRESSOpost\\
\ldots & \ldots & \ldots & \ldots & \ldots \\
\enddata
\tablecomments{The full content of Table~\ref{table:rv} is available in machine-readable format.}
\end{deluxetable}

\clearpage

\section{Supplementary Material of the Data Analysis}

\subsection{ RV Sensitivity Analysis} \label{sec:sensitivity_map}

\begin{figure}[ht!]
\centering
\includegraphics[width=0.7\linewidth]{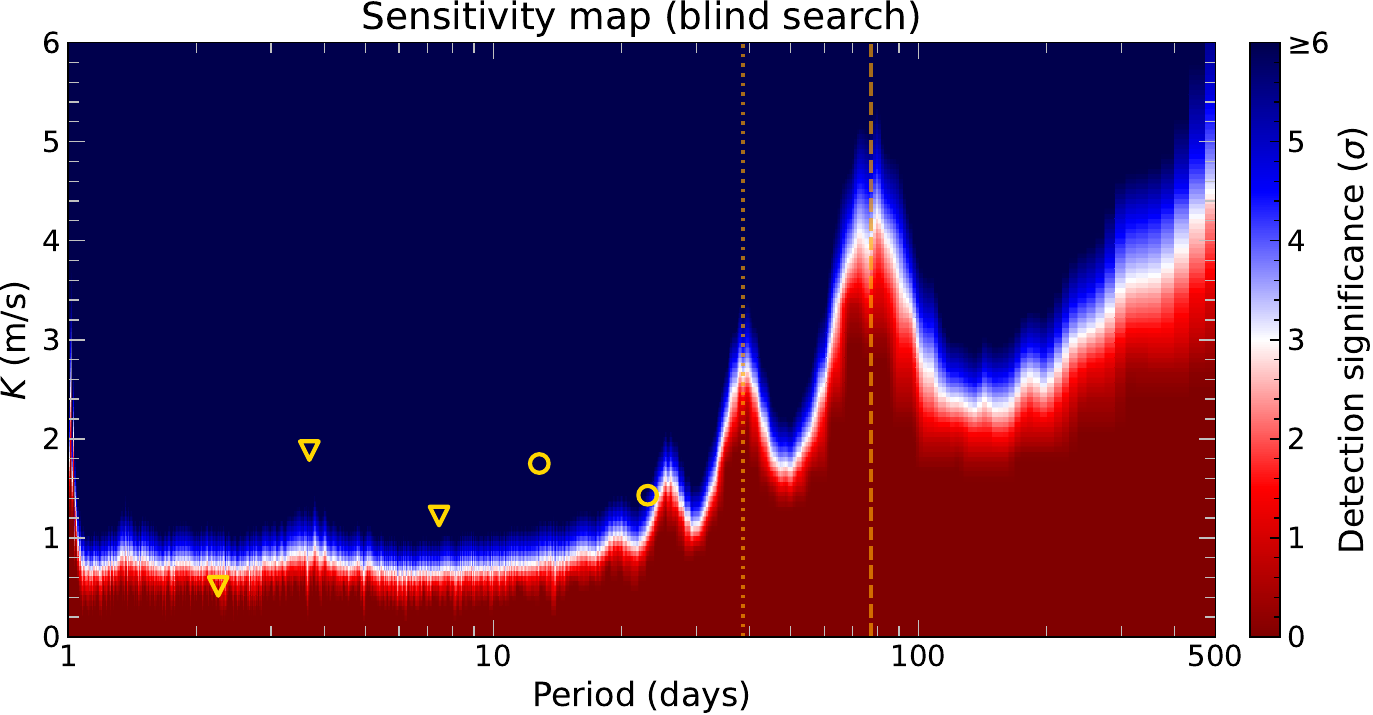}
\caption{Detection sensitivity map of Keplerian signals in the ESPRESSO and HARPS data. The semi-amplitudes of L~98-59\,b--f are depicted with inverted triangles for transiting and circles for non-transiting planets. The detection significance is calculated from the false alarm probability in a Lomb-Scargle periodogram, corresponding to a blind search for periodic signals in a data set. Orange dashed and dotted vertical lines highlight the stellar rotational period ($P_{\rm rot}$) and its first harmonic ($P_{\rm rot}/2$). Our activity modeling with a quasi-periodic GP suppresses small injected signals ($K\approx1$--2\,m\,s$^{-1}$) at long periods and close to the stellar rotation period and harmonics. The ESPRESSO and HARPS RVs are sensitive to planet c through f in a blind search, and to planet b in an informed search using the period and phase constrained by transit.}
\label{fig:rv_detection_map}
\end{figure}

We performed injection-recovery simulations of Keplerian signals in the ESPRESSO+HARPS RVs following a methodology similar to that of \citealt{Gonzalez-Hernandez_2024} (Fig.~16 therein). The injected signals had semi-amplitudes $K$ between 0 and 6\,m\,s$^{-1}$ (step of 0.05\,m\,s$^{-1}$) and orbital periods ranging from 1 to 500\,days, sampled on a logarithmic grid. A quasi-periodic GP using the hyperparameters listed in Table~\ref{table:full_model_params} was subtracted from each simulated dataset to account for stellar activity and assess potential signal suppression. A false alarm probability was then evaluated (Lomb–Scargle periodogram) at the injected period to determine the detection significance. The results are presented in Figure~\ref{fig:rv_detection_map}. This sensitivity map should be interpreted as a blind search for Keplerian signals with no a priori knowledge of period and phase (non-transiting planets). Also, the simulations are valid given that our activity model is true. A more complex modeling of stellar activity in the RV of L~98-59 (e.g., multi-dimensional GP, \citealt{Rajpaul_2024}) could enable the detection of smaller $K$ closer to $P_{\rm rot}$ and its harmonics. For a $P_{\rm rot} = 77.5$\,days, the activity-related RV jitter $\sigma_{\rm act}$ is estimated to be 1.82\,m\,s$^{-1}$ in the optical following \cite{Cloutier_2018b} based on a sample of 21 M dwarfs observed with HARPS \citep{Cloutier_2018a}. For L~98-59, this estimate of $\sigma_{\rm act} = 1.82$\,m\,s$^{-1}$ is similar to the RMS of our median activity GP model in the joint RV--TTV fit (1.79\,m\,s$^{-1}$ RMS). In this exercise, we consider that a Keplerian signal at $P = P_{\rm rot}$ with a variance $\sigma_{\rm Kep}^2 = 3\sigma_{\rm act}^2$ would be detected with a significance of 3$\sigma$. This corresponds to a semi-amplitude $K = 4.46$\,m\,s$^{-1}$.

\subsection{Full Posterior Results of the Joint RV--TTV Fit}

\begin{figure}[ht!]
\centering
\includegraphics[width=0.8\linewidth]{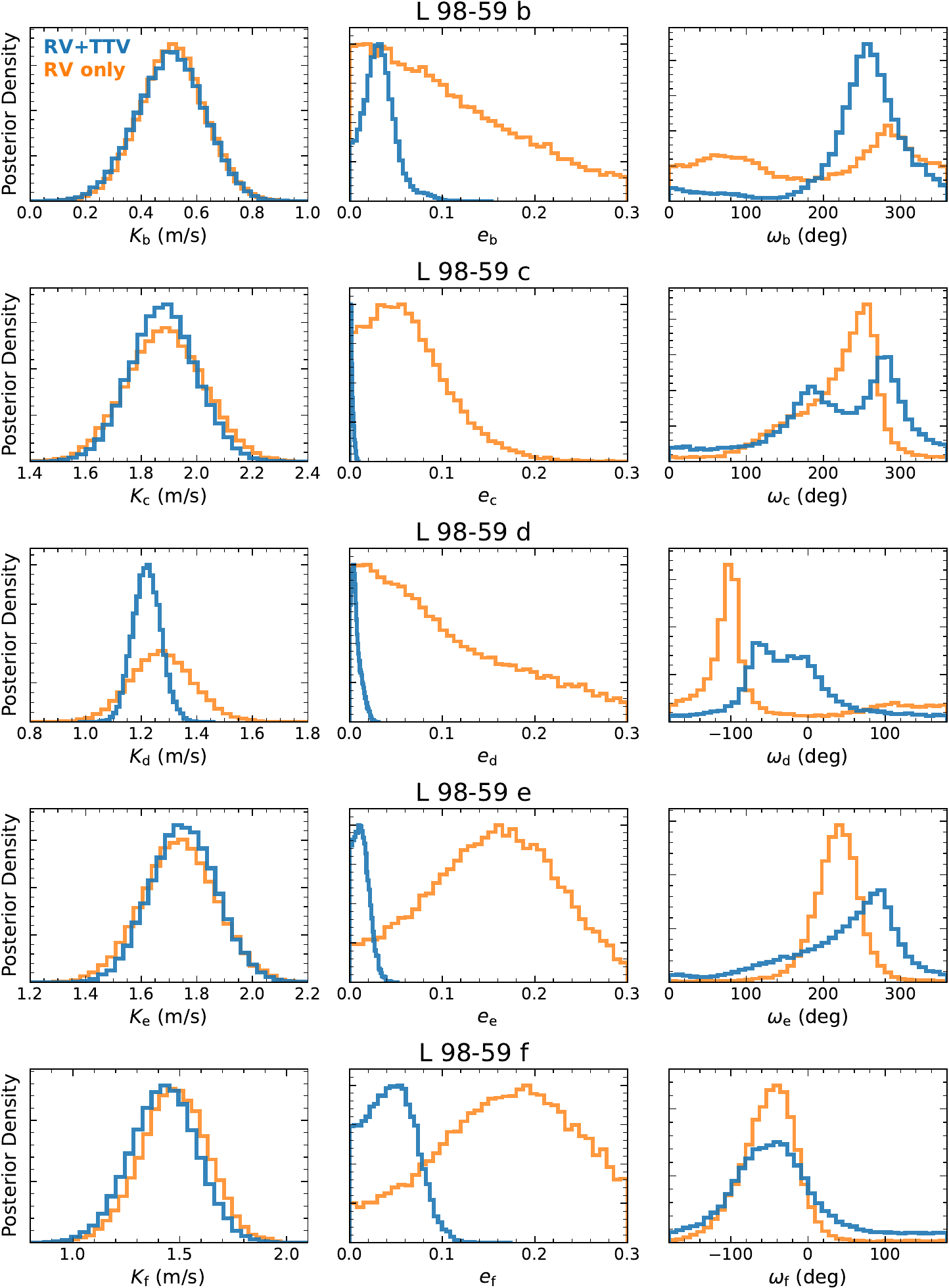}
\caption{Posterior distributions of $K_k$ (left), $e_k$ (center), and $\omega_k$ (right) for the L~98-59 planets from an RV-only fit and the joint RV--TTV fit (fixed-inclination model). The eccentricity distributions are normalized so that their heights are equal to improve visibility. The small TTVs detected on L~98-59\,c and d provide significant constraints on the orbital architecture of the system.}
\label{fig:dist_rv_vs_rvttv}
\end{figure}

The prior and posterior results of the fixed-inclination joint RV--TTV fit are reported in Table~\ref{table:full_model_params}. The full posterior distributions for $K_k$, $e_k$, and $\omega_k$ are shown in Figure~\ref{fig:dist_rv_vs_rvttv}, alongside those obtained from an RV-only fit.

\begin{deluxetable}{lcc}\label{table:full_model_params}
\tablecaption{Results of the joint RV--TTV analysis}
\tablehead{
\colhead{Parameter} & \colhead{Prior} & \colhead{Posterior}
}
\startdata
\multicolumn{2}{c}{\textit{Stellar parameter}} & \\
Stellar mass, $M_{\star}$ (M$_{\odot}$) & $\mathcal{N}\left(0.2923, 0.0067^2\right)$ & 0.2924 $\pm$ 0.0064 \\
\multicolumn{2}{c}{\textit{L~98-59\,b}} & \\
Orbital period, $P_{\rm b}$ (days) & (fixed) & 2.2531140 \\
Time of inferior conjunction, $t_{\rm 0,b}$ (BJD\,-\,2457000) & (fixed) & 1366.17056 \\
Semi-amplitude, $K_{\rm b}$ (m\,s$^{-1}$) & $\mathcal{U}\left(0, 5\right)$ & 0.51 $\pm$ 0.12 \\
Orbit parameterization, $\sqrt{e_{\rm b}} \cos \omega_{\rm b}$ & $\mathcal{U}\left(-\sqrt{0.3}, \sqrt{0.3}\right)$ & $-0.02^{+0.10}_{-0.10}$ \\
Orbit parameterization, $\sqrt{e_{\rm b}} \sin \omega_{\rm b}$ & $\mathcal{U}\left(-\sqrt{0.3}, \sqrt{0.3}\right)$ & $-0.14^{+0.10}_{-0.05}$ \\
Orbital inclination, $i_{\rm b}$ ($^{\circ}$) & (fixed) & 88.08 \\
\multicolumn{2}{c}{\textit{L~98-59\,c}} & \\
Orbital period, $P_{\rm c}$ (days) & (fixed) & 3.6906764 \\
Time of inferior conjunction, $t_{\rm 0,c}$ (BJD\,-\,2457000) & (fixed) & 1367.27303 \\
Semi-amplitude, $K_{\rm c}$ (m\,s$^{-1}$) & $\mathcal{U}\left(0, 5\right)$ & 1.88 $\pm$ 0.12 \\
Orbit parameterization, $\sqrt{e_{\rm c}} \cos \omega_{\rm c}$ & $\mathcal{U}\left(-\sqrt{0.3}, \sqrt{0.3}\right)$ & $-0.01^{+0.03}_{-0.03}$ \\
Orbit parameterization, $\sqrt{e_{\rm c}} \sin \omega_{\rm c}$ & $\mathcal{U}\left(-\sqrt{0.3}, \sqrt{0.3}\right)$ & $-0.02^{+0.03}_{-0.04}$ \\
Orbital inclination, $i_{\rm c}$ ($^{\circ}$) & (fixed) & 88.88 \\
\multicolumn{2}{c}{\textit{L~98-59\,d}} & \\
Orbital period, $P_{\rm d}$ (days) & (fixed) & 7.450729 \\
Time of inferior conjunction, $t_{\rm 0,d}$ (BJD\,-\,2457000) & (fixed) & 1362.74002 \\
Semi-amplitude, $K_{\rm d}$ (m\,s$^{-1}$) & $\mathcal{U}\left(0, 5\right)$ & 1.22 $\pm$ 0.05 \\
Orbit parameterization, $\sqrt{e_{\rm d}} \cos \omega_{\rm d}$ & $\mathcal{U}\left(-\sqrt{0.3}, \sqrt{0.3}\right)$ & 0.05$^{+0.03}_{-0.05}$ \\
Orbit parameterization, $\sqrt{e_{\rm d}} \sin \omega_{\rm d}$ & $\mathcal{U}\left(-\sqrt{0.3}, \sqrt{0.3}\right)$ & $-0.03^{+0.05}_{-0.06}$ \\
Orbital inclination, $i_{\rm d}$ ($^{\circ}$) & (fixed) & 88.44 \\
\multicolumn{2}{c}{\textit{L~98-59\,e}} & \\
Orbital period, $P_{\rm e}$ (days) & $\mathcal{U}\left(8, 18\right)$ & 12.8278 $\pm$ 0.0018 \\
Time of inferior conjunction, $t_{\rm 0,e}$ (BJD\,-\,2457000) & $\mathcal{U}\left(1430, 1448\right)$ & 1438.48 $\pm$ 0.15 \\
Semi-amplitude, $K_{\rm e}$ (m\,s$^{-1}$) & $\mathcal{U}\left(0, 5\right)$ & 1.75 $\pm$ 0.12 \\
Orbit parameterization, $\sqrt{e_{\rm e}} \cos \omega_{\rm e}$ & $\mathcal{U}\left(-\sqrt{0.3}, \sqrt{0.3}\right)$ & $-0.03^{+0.06}_{-0.06}$ \\
Orbit parameterization, $\sqrt{e_{\rm e}} \sin \omega_{\rm e}$ & $\mathcal{U}\left(-\sqrt{0.3}, \sqrt{0.3}\right)$ & $-0.07^{+0.09}_{-0.06}$ \\
Orbital inclination, $i_{\rm e}$ ($^{\circ}$) & (fixed) & 88.47 \\
\multicolumn{2}{c}{\textit{L~98-59\,f}} & \\
Orbital period, $P_{\rm f}$ (days) & $\mathcal{U}\left(18, 35\right)$ & 23.064 $\pm$ 0.055 \\
Time of inferior conjunction, $t_{\rm 0,f}$ (BJD\,-\,2457000) & $\mathcal{U}\left(1420, 1455\right)$ & 1437.84 $\pm$ 0.75 \\
Semi-amplitude, $K_{\rm f}$ (m\,s$^{-1}$) & $\mathcal{U}\left(0, 5\right)$ & 1.43 $\pm$ 0.15 \\
Orbit parameterization, $\sqrt{e_{\rm f}} \cos \omega_{\rm f}$ & $\mathcal{U}\left(-\sqrt{0.3}, \sqrt{0.3}\right)$ & $0.10^{+0.11}_{-0.14}$ \\
Orbit parameterization, $\sqrt{e_{\rm f}} \sin \omega_{\rm f}$ & $\mathcal{U}\left(-\sqrt{0.3}, \sqrt{0.3}\right)$ & $-0.11^{+0.15}_{-0.10}$ \\
Orbital inclination, $i_{\rm f}$ ($^{\circ}$) & (fixed) & 88.47 \\
\multicolumn{2}{c}{\textit{RV systematics}} & \\
HARPS offset, $\gamma_{\scalebox{0.6}{\rm HARPS}}$ (m\,s$^{-1}$) & $\mathcal{U}\left(\rm{ median(RV_{\scalebox{0.6}{\rm HARPS}})} \pm 10\right)$ & $-5886.0$ $\pm$ 1.5 \\
HARPS jitter, $\sigma_{\scalebox{0.6}{\rm HARPS}}$ (m\,s$^{-1}$) & $\mathcal{LU}\left(10^{-3}, 2\right)$ & 0.88 $\pm$ 0.18 \\
ESPRESSOpre offset, $\gamma_{\rm pre}$ (m\,s$^{-1}$) & $\mathcal{U}\left(\rm{ median(RV_{pre})} \pm 10\right)$ & $-5793.2$ $\pm$ 1.5 \\
ESPRESSOpre jitter, $\sigma_{\rm pre}$ (m\,s$^{-1}$) & $\mathcal{LU}\left(10^{-3}, 2\right)$ & 0.67 $\pm$ 0.15 \\
ESPRESSOpost offset, $\gamma_{\rm post}$ (m\,s$^{-1}$) & $\mathcal{U}\left(\rm{ median(RV_{post})} \pm 10\right)$ & $-5791.1$ $\pm$ 1.7 \\
ESPRESSOpost jitter, $\sigma_{\rm pre}$ (m\,s$^{-1}$) & $\mathcal{LU}\left(10^{-3}, 2\right)$ & 0.48$^{+0.21}_{-0.17}$ \\
\multicolumn{2}{c}{\textit{RV activity GP}} & \\
Amplitude, $A$ (m\,s$^{-1}$) & $\mathcal{LU}\left(10^{-3}, 5\right)$ & 2.95$^{+0.62}_{-0.49}$ \\
Timescale, $\ell$ (days) & $\mathcal{TN}\left(85, 20^2, 80, 1000\right)$ & 90$^{+10}_{-7}$ \\
Periodic scale, $\Gamma$ & $\mathcal{N}\left(1.71, 0.29^2\right)$ & 1.92 $\pm$ 0.24 \\
Stellar rotation period, $P_{\rm rot}$ (days) & $\mathcal{N}\left(76.7, 1.5^2\right)$ & 77.3 $\pm$ 1.2
\enddata
\end{deluxetable}

\subsection{Tidal Heating Methods} \label{tidal_heating_methods}

We use the open-source code \texttt{melt}\footnote{\href{https://github.com/cpiaulet/melt}{\texttt{github.com/cpiaulet/melt}}} to evaluate the potential for the rocky planets L~98-59 b and c to host a subsurface magma ocean. \texttt{melt} calculates tidal heating using a simplified energy balance model driven by eccentricity forcing. The model was originally developed to study Io’s tidal heating \citep{fischer_thermal-orbital_1990,moore_tidal_2003,henning_tidally_2009,dobos_viscoelastic_2015}, extended to the TRAPPIST-1 planets \citep{barr_interior_2018}, and recently applied to LP~791-18\,d \citep{peterson_temperate_2023} to assess its interior state. For a range of potential mantle temperatures $T_\mathrm{mantle}$, we calculate the tidal energy dissipation flux $F_\mathrm{tidal}$ \citep{segatz_tidal_1988} following Equations (7) and (8) in \citet{peterson_temperate_2023}, adopting the Maxwell model of viscoelastic rheology to compute the imaginary part of the planet's Love number \citep{fischer_thermal-orbital_1990,moore_tidal_2003,henning_tidally_2009,dobos_viscoelastic_2015,barr_interior_2018}. For each temperature, we also calculate the convective flux $F_\mathrm{conv}$ which reflects the energy transported radially through convection toward the surface \citep{solomatov_scaling_2000,barr_mobile_2008,barr_interior_2018}. Within our model, we solve for the mantle temperature as the value of $T_\mathrm{mantle}$ where $F_\mathrm{conv}=F_\mathrm{tidal}$ is achieved (Eq. (9) in \citealp{peterson_temperate_2023}, assuming a pure rock composition \citep{solomatov_scaling_2000}. We prescribe the values for the shear modulus and viscosity with relations that depend on the temperature regime. Below the solidus temperature at $T_s = 1600$\,K, we follow \citet{fischer_thermal-orbital_1990,henning_tidally_2009}. Between the solidus temperature and the ``breakout point'' temperature $T_b=1800$\,K (beyond which shear stiffness can be neglected; \citealp{renner_rheologically_2000}), we follow the prescription of \citet{moore_tidal_2003}. However, in this higher temperature regime, the rock viscosity strongly depends on the melt fraction $f$ as $\propto \exp(-Bf)$, with the proportionality coefficient $B$ only constrained by experiments, within the range of 10 to 40. In order to account for this model uncertainty, we find the equilibria points between the tidal and convective energy transport fluxes for several values of $B$ spanning the experimentally-constrained range.

For each planet, our model takes as input the planet mass, radius, orbital period, and orbital eccentricity as quoted in Table \ref{table:derivedparams}. Given the high sensitivity of the model to eccentricity, we conservatively instead take the maximum a posteriori value of $e_{\rm b} = 0.031$ and $e_{\rm c} = 0.001$ inferred from the RV--TTV analysis, as the median or upper limits would lead to higher mantle temperatures.

\end{document}